\documentclass[prd,aps,letterpaper,twocolumn,superscriptaddress,preprintnumbers,nofootinbib,floatfix]{revtex4-2}
\pdfoutput=1
\usepackage{xcolor}
\usepackage{graphicx}
\usepackage{amsmath}
\usepackage{amsfonts}
\usepackage{amssymb}
\usepackage{rotating}
\usepackage{subfigure}
\usepackage{paralist} 
\usepackage{verbatim}
\usepackage{float}
\usepackage{soul} 
\usepackage{appendix}
\usepackage{natbib}
\usepackage{epsfig}
\usepackage{hyperref}
\hypersetup{colorlinks,linkcolor=red,urlcolor=blue,citecolor=blue}
\usepackage[utf8]{inputenc}
\usepackage[english]{babel}
\usepackage{tabularx}
\newcolumntype{C}{>{\centering\arraybackslash}X}
\usepackage{mathpazo} % for math fonts
\usepackage{tgpagella} % overrides the text fonts in mathpazo
\usepackage{float}
\usepackage{listings} % for code block

\newcommand{\mnras}{Monthly Notices of the Royal Astronomical Society}
\newcommand{\apjl}{Astrophysical Journal Letters}
\newcommand{\apjs}{Astrophysical Journal Supplement}

\newcommand{\aap}{Astronomy \& Astrophysics}
\newcommand{\jcap}{J. Cosmol. Astropart. Phys.}

\usepackage{booktabs}
\usepackage{array, multirow}

\usepackage{lipsum}

\lstdefinestyle{mystyle}{
    basicstyle=\ttfamily\small,
    breakatwhitespace=true,         
    breaklines=true,   
}

\definecolor{cadetblue}{rgb}{0.37, 0.62, 0.63}

\newcommand{\stonybrook}{Physics and Astronomy Department, Stony Brook University, Stony Brook, NY  11794}

\begin{document}

\title{Cosmological Parameter Forecasts for a CMB-HD Survey}

\author{Amanda MacInnis}
\affiliation{\stonybrook}

\author{Neelima Sehgal}
\affiliation{\stonybrook}

\author{Miriam Rothermel}
\affiliation{\stonybrook}

%%%%%%%%%%%%%%%%%%%%%%%%%%%%%%%%%%%%%%%%%%%%%%%%%%%%%%%
\begin{abstract}

We present forecasts on cosmological parameters for a CMB-HD survey.  For a $\Lambda$CDM+$N_\mathrm{eff}$+$\sum m_\nu$ model, we find $\sigma(n_\mathrm{s}) = 0.0013$ and $\sigma(N_{\mathrm{eff}}) = 0.014$ using CMB and CMB lensing multipoles in the range of $\ell \in [30,20000]$, after adding anticipated residual foregrounds, delensing the acoustic peaks, and adding DESI BAO data.  This is about a factor of two improvement in ability to probe inflation via $n_\mathrm{s}$ compared to precursor CMB surveys.  The  $N_{\mathrm{eff}}$ constraint can rule out light thermal particles back to the end of inflation with 95\% CL; for example, it can rule out the QCD axion in a model-independent way assuming the Universe's reheating temperature was high enough that the axion thermalized. We find that delensing the acoustic peaks and adding DESI BAO tightens parameter constraints.  
We also find that baryonic effects can bias parameters if not marginalized over, and that uncertainties in baryonic effects can increase parameter error bars; however, the latter can be mitigated by including information about baryonic effects from kinetic and thermal Sunyaev-Zel'dovich measurements by CMB-HD. The CMB-HD likelihood and Fisher estimation codes used here are publicly available; the likelihood is integrated with Cobaya to facilitate parameter forecasting.

\end{abstract}
%%%%%%%%%%%%%%%%%%%%%%%%%%%%%%%%%%%%%%%%%%%%%%%%%%

\maketitle

\section{Introduction}
\label{sec:intro}

Measurements of the Cosmic Microwave Background (CMB) have provided precise information on the cosmological parameters of our Universe~\cite{planck18params,Aiola2020,SPT-3G:2022hvq}.  Near future CMB experiments such as the Simons Observatory (SO)~\cite{SOforecast} and CMB-S4~\cite{S4forecast} will improve measurements of the CMB temperature and polarization power spectra below $\ell \sim 5000$ and the CMB lensing power spectrum considerably, yielding improved constraints on $\Lambda$CDM parameters, the sum of the neutrino masses ($\sum m_\nu$), and the effective number of light relativistic species ($N_\mathrm{eff}$), in particular~\cite{SOforecast,S4forecast}. Beyond that, an even lower noise and higher resolution CMB experiment can improve parameter constraints further by measuring smaller angular scales in both temperature and polarization, measuring the lensing power spectrum with more precision and over smaller scales, and enabling more aggressive foreground removal down to lower flux limits.  CMB-HD is a proposed concept for a Stage-V CMB facility~\cite{HDastro2020,HDsnowmass} that would have three times lower instrument noise and about six times higher resolution than CMB-S4.  In this work, we forecast the cosmological parameter constraints that can be achieved by such a facility, folding in anticipated residual foregrounds, delensing, and baryonic effects.  

Delensing of the CMB acoustic peaks has been demonstrated in a number of recent CMB data analyses~\cite{han20delensing,Millea:2020iuw,Carron:2017vfg}. It can tighten cosmological parameter constraints by sharpening the acoustic peaks, removing uncertainty in the lensing realization, and reducing lensing-induced power spectrum covariances~\cite{Green2017, Hotinli2021}. Baryonic effects such as AGN feedback can move around the matter distribution on moderately small scales, which impacts both the lensed CMB power spectra and the CMB lensing power spectrum; for next-generation CMB experiments, neglecting such effects can bias parameters~\cite{Chung2019,mccarthy22}.  The residual foregrounds we include in this work, in both CMB and CMB lensing power spectra, are those that preliminary analyses suggest can be, and need to be, obtained for a CMB-HD facility to achieve its lensing science goals~\cite{han22}.  We emphasize that a full demonstration that such foreground residuals can be achieved using realistic simulations is a subject of ongoing research.  We also include baryon acoustic oscillation (BAO) data expected from the Dark Energy Spectroscopic Instrument (DESI)~\cite{desi}, which serves to break a number of parameter degeneracies.  

Parameters are forecast with both a Fisher estimation code and a likelihood plus Markov chain; we find that the resulting forecasts from both methods are in agreement. We make publicly available the CMB-HD temperature, polarization, and lensing ($TT, TE, EE, BB, \kappa \kappa$) signal and noise spectra, joint covariance matrix, binning matrix, Fisher forecasting software\footnote{\url{https://github.com/CMB-HD/hdfisher}}, and likelihood\footnote{\url{https://github.com/CMB-HD/hdlike}} in order to enable further forecasting.  In addition, the CMB-HD likelihood has been integrated with the latest version of Cobaya\footnote{\url{https://cobaya.readthedocs.io}}~\cite{cobaya} using CAMB\footnote{\url{https://camb.info/}}~\cite{CAMBLewis:1999bs, CAMBHowlett:2012mh}. 

In Section~\ref{sec:spectra}, we discuss the experimental configuration employed, our generation of CMB and lensing noise spectra, and our method of delensing.  In Section~\ref{sec:covmat}, we discuss the calculation of the joint covariance matrix.  We describe the Fisher estimation and likelihood plus Markov chain procedures for forecasting cosmological parameters in Section~\ref{sec:forecastMethods}.  In Section~\ref{sec:parameterResults}, we present parameter forecasts for a CMB-HD survey and show the impact of foregrounds, delensing, baryonic effects, and BAO.  We discuss the implications of these forecasts for discovering new light relic particles and probing inflation in Section~\ref{sec:discussion}, and we conclude in Section~\ref{sec:conclusion}.

\section{Generating Signal and Noise Spectra}
\label{sec:spectra}

Below we discuss our method of obtaining parameter forecasts for an ultra low-noise, ultra-high-resolution CMB survey, such as CMB-HD~\cite{HDastro2020, HDsnowmass}.  We focus in particular on the $\Lambda$CDM model and minimal extensions.  To assess the additional constraining power of a CMB-HD experiment, we also provide comparisons to experiments similar to the Simons Observatory (SO)~\cite{SOforecast} and CMB-S4~\cite{S4forecast}.  We assume that the CMB datasets we forecast for will provide $TT$, $TE$, $EE$, and $BB$ power spectra, as well as the $\kappa \kappa$ power spectra from CMB lensing.  In Section~\ref{sec:expconfig}, we detail the experimental configurations assumed, and in Sections~\ref{sec:noisespectra} and~\ref{sec:fg}, we describe the instrument noise and residual foreground models.  In Section~\ref{sec:lensingnoise}, we describe the computation of the CMB lensing power spectrum noise, and in Section~\ref{sec:delensedspectra}, we explain the generation of the delensed CMB spectra.

\subsection{Experimental Configuration} \label{sec:expconfig}

We create mock signal, noise, and covariance matrices for experiments similar to CMB-HD, CMB-S4, SO, and DESI.  Note that we do not aim to reproduce the characteristics of these experiments exactly, but instead aim to give approximate estimates based on the experimental configurations described below and in Table~\ref{tab:exps}.

For the CMB experiments we assume $60\%$ of the sky is observed.  We also model only the 90 and 150 GHz channels since they contain the bulk of the CMB signal, and we assume the other frequencies will be utilized primarily for foreground cleaning. We obtain the noise levels for HD-like, S4-like, and SO-like surveys from~\cite{HDastro2020,S4forecast,SOforecast}, respectively, stressing again that we do not aim to present official forecasts for each experiment, but rather to assess their comparative constraining power. Throughout this work, ``SO-like'' and ``S4-like'' surveys will refer to experimental configurations similar to the SO goal and CMB-S4 wide-area surveys~\cite{SOforecast, S4forecast}.  We assume SO-like and S4-like experiments will employ 5-meter dishes and a CMB-HD-like survey will use 30-meter dishes, and we determine the resolution of each facility from these specifications.  

Regarding the CMB multipole ranges, we assume  in our forecasts that a CMB-HD-like survey can measure multipoles in the range of $\ell \in [1000, \ 20000]$ for $TT$, $TE$, $EE$.   
We assume CMB-HD will not measure $\ell < 1000$ since these multipoles are a challenge for a 30-meter dish\footnote{While the baseline CMB-HD design does not assume multipoles below $\ell = 1000$ will be measured by CMB-HD, we note that there are interesting science cases that benefit from CMB-HD measuring the $BB$ spectrum down to $\ell = 100$, e.g.~\cite{Mandal:2022tqu}; such low-$\ell$ measurements will be strived for if they can be achieved technologically.}; however, since both SO and CMB-S4 will measure the $TT$, $TE$, and $EE$ spectra to the sample-variance limit for $\ell \in [30, 1000]$ over the same $60\%$ of the sky as CMB-HD, we extend the CMB-HD multipole range down to $\ell=30$ for $TT$, $TE$, and $EE$. For the $BB$ spectrum in the multipole range of $30 \leq \ell < 1000$, we use the anticipated $BB$ noise from an Advanced Simons Observatory (ASO) type survey; for ASO we assume the same resolution and sky area as SO, but 3.5 and 3.8~$\mu$K-arcmin white noise in temperature for 90 and 150~GHz, respectively.

\begin{table}[t]
    \begin{center}
    \begin{tabular}{cccccr}
      \toprule
      \toprule
      \multirow{2}{*}{Exp.} & \multirow{2}{*}{$f_\mathrm{sky}$} &  Freq., & Noise, & FWHM,  &  \multicolumn{1}{c}{Multipole Range} 
      \\
      & &   GHz & $\mu$K-arcmin & arcmin  & \multicolumn{1}{c}{$\ell, L$} 
      \\
      \midrule
      \multirow{3}{*}{HD} & \multirow{3}{*}{0.6} &  90 & 0.7 & 0.42 & TT, TE, EE: [30,\ 20000]
      \\
      &  &  150  & 0.8  & 0.25  & BB:  [1000,\ 20000]
      \\
      &  &   &   &   & $\kappa\kappa$: [30,\ 20000] 
      \\
      \midrule[0.2pt]
      \multirow{3}{*}{S4} & \multirow{3}{*}{0.6} & 90 & 2.0 & 2.2 & TT: [30,\ 3000]
      \\
      &  & 150 & 2.0 & 1.4 & TE, EE, BB: [30,\ 5000]
      \\
      & & & & & $\kappa\kappa$: [30,\ 3000]
      \\
      \midrule[0.1pt]
      \multirow{3}{*}{SO} & \multirow{3}{*}{0.6} & 90 & 5.8 & 2.2 & TT: [30,\ 3000]
      \\
      & & 150 & 6.3 & 1.4 & TE, EE, BB:  [30,\ 5000] 
      \\
      & & & & & $\kappa\kappa$: [30,\ 3000]
      \\
      \bottomrule
    \end{tabular}
    \caption{For each CMB experiment considered, we list the white noise level in temperature and beam full-width at half-maximum (FWHM) for each frequency used in the forecast, the sky fraction ($f_\mathrm{sky}$), and the multipole ranges used for the power spectra. For the SO-like and S4-like forecasts, we only forecast for the large-aperture telescopes, and use noise levels from~\protect{\cite{SOforecast, S4forecast}}, respectively. For the CMB-HD-like forecasts, we use noise levels from~\protect{\cite{HDsnowmass}}, and we assume CMB-HD only measures $\ell \geq 1000$.  Since SO will measure $TT, TE,$ and $EE$ to the sample variance limit for $\ell < 1000$ over the CMB-HD sky area, we extend the CMB-HD multipole range down to $\ell=30$ for these spectra; for the $BB$ spectrum, we use an ASO-like $BB$ spectrum and noise in the multipole range of $30 \leq \ell < 1000$ (see text for details). We emphasize that we do not aim to present official forecasts for each experiment, but rather to assess the comparative constraining power varying a number of parameters.}
    \label{tab:exps}
    \end{center}
\end{table}

Given the resolution of SO and CMB-S4, we limit the $TE$, $EE$, and $BB$ multipole range to $\ell \leq 5000$.  While we do not explicitly include foreground residuals in our SO-like and S4-like forecasts, we only consider multipoles of $\ell \leq 3000$ for $TT$, assuming higher $TT$ multipoles will be foreground dominated.  Similarly, in light of foreground contamination and bias, we only consider $\kappa \kappa$ for $L \leq 3000$ for SO-like and S4-like experiments.  In contrast, we include $L \in [30, 20000]$ for an HD-like experiment, and use the $\kappa \kappa$ covariance matrix from~\cite{han22} for $L \in [4800, 20000]$, as we discuss in more detail below.  We also include in the HD-like CMB power spectra, the foreground residual levels required for CMB-HD to achieve its lensing measurement target, as specified in~\cite{han22}.

To simulate the mock DESI dataset, we assume the data is taken over 14,000 square degrees (approximately 35\% of the sky) for the baseline galaxy survey and the bright galaxy survey, as specified in~\cite{desi}. Our mock DESI data consists of distance ratio measurements, $r_\mathrm{s} / d_V(z)$, obtained from CAMB~\cite{CAMBLewis:1999bs, CAMBHowlett:2012mh}, and a covariance matrix for those measurements, which we calculate.  Here $r_\mathrm{s}$ is the comoving sound horizon and $d_V$ is a combined distance measurement, which we define in Section~\ref{sec:baoCovmat}. We calculate the covariance matrix using the forecasted fractional errors on $H(z) r_\mathrm{s}$ and $r_\mathrm{s} / d_A(z)$ provided in~\cite{desi}, as we discuss in more detail in Section~\ref{sec:baoCovmat}.

\subsection{CMB Instrumental Noise Spectra}
\label{sec:noisespectra}

A first step towards generating theoretical CMB and CMB lensing power spectra is to compute the expected instrumental noise on the CMB power spectra.  We add to these noise spectra for $TT$ residual extragalactic foregrounds in the case of CMB-HD, as described in Section~\ref{sec:fg}.  These mock CMB power spectra are then used to calculate the expected lensing reconstruction noise (Section~\ref{sec:lensingnoise}), which is then used to calculate the theoretical delensed CMB power spectra (Section~\ref{sec:delensedspectra}).  The lensed or delensed CMB spectra and the lensing  spectrum, along with the CMB and lensing noise spectra, are then used in the analytic calculation of the covariance matrix (Section~\ref{sec:covmat}). This covariance matrix is then used to forecast parameter constraints either via a Fisher matrix (Section~\ref{sec:fisher}) or a likelihood and Markov Chain (Section~\ref{sec:mcmc}).

The power spectra of the CMB temperature and polarization fields and the (projected) lensing potential are determined by the cosmological model that describes our Universe.  We use a flat $\Lambda$CDM+$N_\mathrm{eff}$+$\sum m_\nu$ model described by eight cosmological parameters with fiducial values from the \textit{Planck} baseline cosmological parameter constraints~\cite{planck18params}, which we list in Table~\ref{tab:fisher}. We use CAMB to generate the lensed, unlensed\footnote{These are used in the analytic covariance matrix calculation described in Section~\ref{sec:covmat}.}, and delensed CMB power spectra $C_\ell^{XY}$ for $XY \in [TT,\ TE,\ EE,\ BB]$ and the lensing power spectrum $C_L^{\kappa\kappa}$. We use the 2016 version of HMcode~\cite{Mead2015,Mead2016} to calculate the non-linear matter power spectrum for a cold dark matter model, and increase the accuracy of CAMB beyond its default settings as discussed in detail in Appendix~\ref{app:accuracy}. We also use CAMB to calculate the theoretical BAO signal, i.e.~$r_s / d_V(z)$, using the same fiducial cosmology and the default CAMB accuracy settings.

\begin{table}[t]
    \begin{center}
    \begin{tabular}{l@{\hskip 1.5em} c@{\hskip 1.5em} c c}
      \toprule
      \toprule
      \multirow{2}{*}{Parameter}  & \multirow{2}{*}{Fiducial}  & Step & \multirow{2}{*}{Prior} 
      \\
       & & size (\%) &
      \\
      \midrule
      $\Omega_\mathrm{b} h^2$\dotfill & $0.02237$ & 1 & $[0.005,\ 0.1]$
      \\
      $\Omega_\mathrm{c} h^2$\dotfill & $0.1200$ & 1 & $[0.001,\ 0.99]$
      \\
      $\ln(10^{10} A_\mathrm{s})$\dotfill & $3.044$ & 0.3\footnote{The step size of 0.3\% $\ln(10^{10} A_\mathrm{s})$ corresponds to an approximate step size of 1\% on $A_\mathrm{s}$.} & $[2,\ 4]$
      \\
      $n_\mathrm{s}$\dotfill & $0.9649$ & 1 & $[0.8,\ 1.2]$
      \\
      $\tau$\dotfill & $0.0544$ & 5 & $0.054 \pm 0.007$
      \\
      $100 \theta_\mathrm{MC}$\dotfill & $1.04092$ & 1 & $[0.5,\ 10]$
      \\
      $H_0$ [km s$^{-1}$ Mpc$^{-1}$]\dotfill & $67.36$ & 1 & $[20,\ 100]$
      \\
      $N_\mathrm{eff}$\dotfill & $3.046$ & 5 & $[0.05,\ 10]$
      \\
      $\sum m_\nu$ [eV]\dotfill & $0.06$ & 10 & $[0,\ 5]$
      \\
      \bottomrule
    \end{tabular}
    \caption{Listed in the first two columns are the cosmological parameters varied in the Fisher or MCMC forecasts and their fiducial values based on the \textit{Planck} baseline cosmological parameter constraints~\protect{\cite{planck18params}}. The third column lists the step size for each parameter, as a percentage of the fiducial value, used when calculating the numerical derivatives of the power spectra with respect to that parameter. The last column lists the priors used for the MCMC analysis; the prior on the reionization optical depth, $\tau$, is also applied to the Fisher forecasts. Note that we sample over $100\theta_\mathrm{MC}$ and obtain $H_0$ as a derived parameter in the MCMC analysis. All priors are uniform, with the exception of $\tau$, where we use a Gaussian prior from \textit{Planck}~\protect{\cite{planck18params}}.  }
    \label{tab:fisher}
    \end{center}
\end{table}

We generate noise spectra for each CMB experiment listed in Table~\ref{tab:exps} at 90 and 150 GHz. Each frequency $f$ has a root-mean-square white noise level $\Delta^{TT}_f$ in temperature and a beam full-width at half-maximum of $\theta^\mathrm{FWHM}_f$ given in Table~\ref{tab:exps} . We assume that the noise in polarization is given by $\Delta^{EE}_f = \Delta^{BB}_f = \sqrt{2} \Delta^{TT}_f$, and set $\Delta^{TE}_f = 0$ assuming uncorrelated noise in temperature and polarization maps. For a given frequency, we model the beam-deconvolved noise power spectrum as
\begin{equation} \label{eq:noise}
    N_{\ell_f}^{XY} = \left(\Delta^{XY}_f\right)^2 \exp\left[\frac{\ell (\ell + 1) \left(\theta_f^\mathrm{FWHM}\right)^2}{8 \ln 2}\right],
\end{equation}
for $XY \in [TT,\ EE,\ BB]$. 

To model a CMB-HD-type survey, we add the anticipated residual foreground power spectra (see Section~\ref{sec:fg}) to the temperature noise spectrum at each frequency, so that ${N_{\ell_f}^{TT}}^\mathrm{HD} = N_{\ell_f}^{TT} + C_{\ell_f}^{FG}$. Since we assume that CMB-HD will not measure $\ell<1000$ (and does not need to for $TT, TE, EE$ as discussed in Section~\ref{sec:expconfig}), we use an ASO-like $BB$ noise spectrum for $\ell \in [30, 1000)$. We assume polarized foregrounds can be removed to levels below the instrument noise for all CMB experiments considered here by making use of the additional frequency coverage outside of 90 and 150 GHz that each experiment will have.

\begin{figure*}
    \centering
    \includegraphics[width=\textwidth]{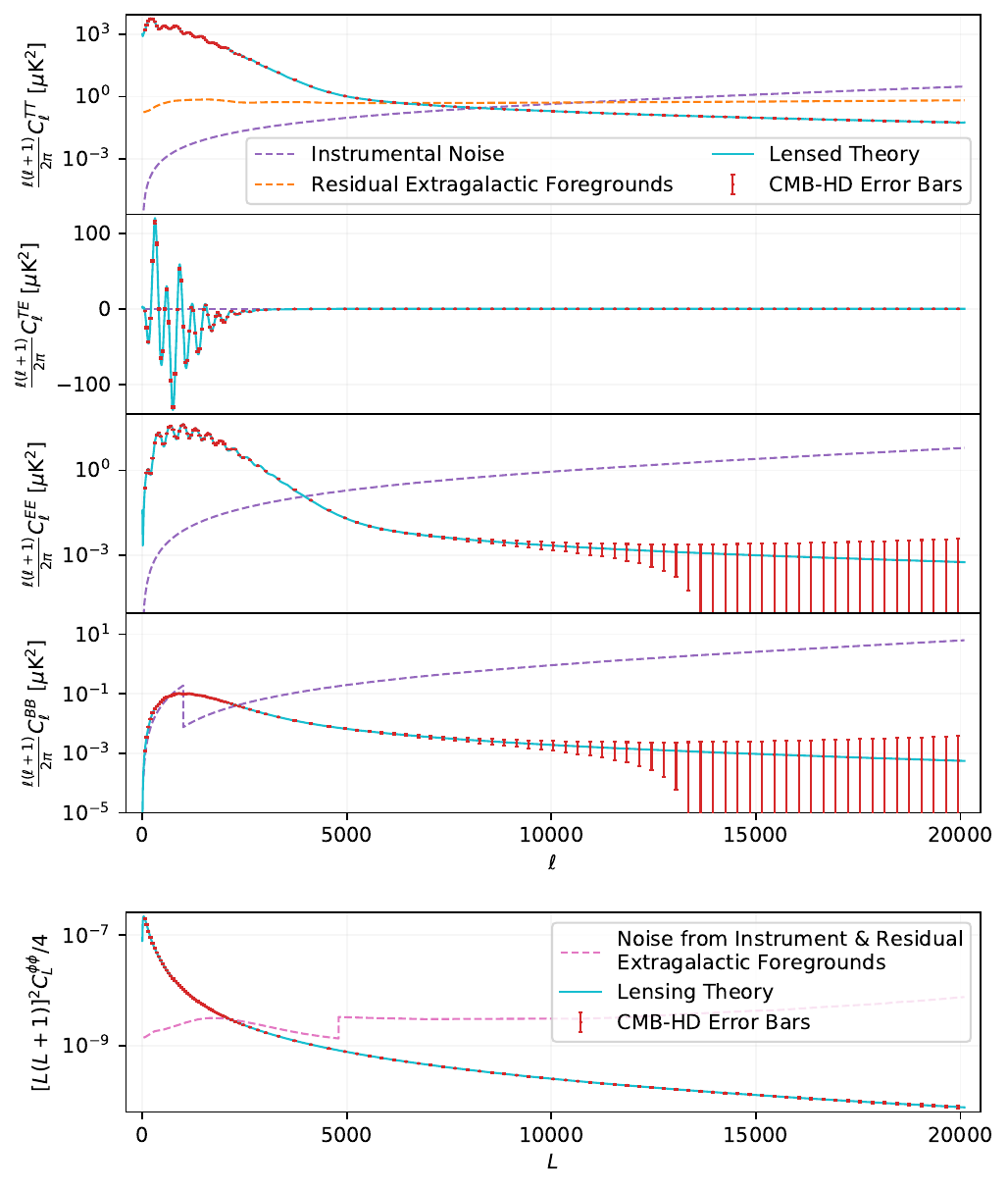}
    \caption{ Expected CMB $TT$, $TE$, $EE$, $BB$, and $\kappa \kappa$ power spectra for a CMB-HD-type survey, with instrument noise and residual extragalactic foregrounds coadded from $90$ and $150$ GHz. We show the theory spectra (blue), the per-mode beam-deconvolved instrument noise (purple), the residual temperature foregrounds (orange), and the expected CMB-HD error bars, plotting the diagonals of the binned covariance matrix (red)  (see Sections~\ref{sec:expconfig},~\ref{sec:noisespectra}, and~\ref{sec:fg} for details). We also show the noise curve for the lensing power spectrum, $\kappa\kappa$, (pink), which is a combination of noise from the primordial CMB, instrument, and residual foregrounds (see Section~\ref{sec:lensingnoise} for details). Note that in many cases the error bars are  smaller than the point indicating the bin center.}
    \label{fig:spectra}
\end{figure*}

We coadd the noise spectra, which include both instrument noise and residual foregrounds, for 90 and 150 GHz for each experiment using inverse-noise weights, $W_{\ell_f}^{XY} = (N_{\ell_f}^{XY})^{-1}$, such that the coadded noise power spectrum is given by
\begin{equation} \label{eq:coaddnoise}
    N_\ell^{XY} = \frac{\sum_f \left(W_{\ell_f}^{XY}\right)^2 N_{\ell_f}^{XY}}{\left(\sum_f W_{\ell_f}^{XY}\right)^2} .
\end{equation}
The total CMB power spectra are obtained by adding the coadded noise spectra to the theoretical CMB power spectra, i.e.~$C_\ell^{XY,\mathrm{tot}} = C_\ell^{XY} + N_\ell^{XY}$.
For parameter forecasts (Sections~\ref{sec:fisher} and~\ref{sec:mcmc}), we use a binned signal spectra given by
\begin{equation} \label{eq:binnedspectra}
        C_{\ell_b}^{XY} = \sum_{\ell} M_{b\ell} C_{\ell}^{XY},
\end{equation}
for $XY \in [TT,\ TE,\ EE,\ BB,\ \kappa\kappa]$, where $M_{b\ell}$ is the binning matrix and $C_{\ell_b}^{XY}$ is the binned spectrum with bin centers $\ell_b$. The binning is chosen to capture the CMB acoustic peaks on scales of $\ell \lesssim 5000$ and uses a uniform bin width of $\Delta \ell = 300$ on smaller scales.
The first four panels of Fig.~\ref{fig:spectra} show the CMB theory, instrumental noise, residual extragalactic foregrounds, and binned forecasted error bars for CMB-HD, plotting the binned diagonals of the covariance matrix (see Section~\ref{sec:covmat}). In the last panel, we show the lensing theory power spectrum, the lensing reconstruction noise (see Section~\ref{sec:lensingnoise}), and the binned forecasted lensing error bars.

\subsection{Residual Extragalactic Foregrounds}
\label{sec:fg}

The observed CMB is contaminated by astrophysical foregrounds, which can be both Galactic and extragalactic in origin.  Galactic foregrounds dominate on relatively large-scales, i.e.~$\ell<500$, and are frequency dependent.  As mentioned above, we assume CMB-HD only measures multipoles above $\ell=1000$, and will use lower multipoles from SO for $TT, TE$, and $EE$, which will already be sample variance limited over that range.  SO will have observations over at least six different frequencies, and we assume in this work that SO can remove the Galactic foreground contribution to below the CMB sample variance limit for $TT, TE$, and $EE$.  We assume CMB-HD will use the $BB$ measurement from ASO below $\ell=1000$, and we assume that ASO will remove Galactic foregrounds to below their noise levels for $BB$ over this range.

For extragalactic foregrounds, the six times higher resolution and more than three times lower noise of CMB-HD compared to precursor wide-area CMB surveys, in addition to seven frequency channels, will allow for much greater foreground subtraction. These extragalactic foregrounds include the thermal and kinetic Sunyaev-Zel'dovich effects (tSZ and kSZ) from hot gas in galaxy clusters, groups, and the intergalactic medium~\cite{SZ1969,SZ1970,SZ1972}, the cosmic infrared background (CIB) from dusty star-forming galaxies, and the radio emission from active galactic nuclei.  For CMB-HD, we assume the foreground levels obtained in~\cite{han22}; for radio and CIB sources these levels assume sources detected at 5$\sigma$ in at least one of the CMB-HD channels can be template-subtracted from all channels by extrapolating the measured flux to the other channels using a spectral template. 
In particular, these residual foreground levels assume a CMB-HD-type experiment will template-subtract CIB sources above 0.03 and 0.008 mJy at 150 and 90 GHz, respectively, in part exploiting CMB-HD observations at 220 and 280 GHz to find sources and extrapolate their fluxes to lower frequencies.\footnote{In the case of the CIB, see~\cite{han22} for discussion of how these flux levels were obtained, accounting for confusion from blended sources and CIB spectral index uncertainties.} These levels also assume radio sources above 0.03 and 0.04 mJy at 150 and 90 GHz, respectively, are template subtracted, as well as all tSZ sources detected at $3\sigma$ significance or higher.  We do not assume the kSZ effect from reionization can be removed, since that is quite challenging given its frequency independence and high redshift; however, we do assume the reionization kSZ only adds Gaussian noise uncorrelated with the lensing signal.  We assume that the late-time kSZ effect can be removed from the CMB-HD maps through innovative ``de-kSZ-ing'' techniques that are currently in development~\cite{Foreman:2022ves}, and do not include it here. We stress again that the residual foreground levels assumed are what preliminary studies suggest is possible and necessary for CMB-HD to achieve an unbiased CMB lensing power spectrum over the lensing multipole range of $L\in [5000, 20000]$~\cite{han22}. However, we leave it to future work to robustly demonstrate with realistic simulations and an end-to-end pipeline that these residual foreground levels can be achieved in practice. 

The total residual extragalactic foreground power from the sources described above is shown as the dashed orange curve in the top panel of Fig.~\ref{fig:spectra} for the coadded 90 and 150 GHz temperature power spectra; this residual foreground power is included in the total noise spectra for the CMB-HD temperature power spectrum.  We include only instrumental noise for the $TE, EE$, and $BB$ spectra since we assume that the combination of high resolution and seven frequency channels for CMB-HD will be sufficient to reduce polarized extragalactic foregrounds to residual levels below the instrument noise. 

As mentioned in Section~\ref{sec:expconfig}, while we do not explicitly include residual extragalactic foreground levels in the temperature spectra for SO and CMB-S4, we only include multipoles below $\ell = 3000$ in those spectra, assuming higher multipoles will be dominated by foreground uncertainty.

\subsection{CMB Lensing Power Spectrum Noise}
\label{sec:lensingnoise}

Lensing breaks the Gaussianity of the primordial CMB, introducing mode-coupling in the CMB on different scales.  A quadratic estimator~\cite{hu2001mappingDM,huokamoto2002,OkamotoHu2003,hdv2007} takes advantage of this lensing-induced mode-coupling of the CMB to reconstruct the lensing potential from pairs of CMB maps that are filtered to isolate this mode-coupling. The lensing reconstruction can then be used to estimate the lensing potential power spectrum, which is the connected (non-Gaussian) part of the CMB four-point function.  The disconnected (Gaussian) part of the four-point function, called the Gaussian $N_L^{(0)}$ bias, arises from the primordial CMB and instrument noise even in the absence of lensing~\cite{Kesden2003}; we assume that this will be subtracted with traditional realization-dependent (RDN0) subtraction techniques~\cite{Namikawa2013}.

We use two types of quadratic estimators in this work.  The traditional estimator from~\cite{hu2001mappingDM,huokamoto2002,OkamotoHu2003}, henceforth called H\&O, which uses the same CMB multipole ranges in both CMB maps that go into the estimator.  The other quadratic estimator we use is from~\cite{hdv2007}, henceforth called HDV, which allows one map to contain only large scales ($\ell<2000$), called a gradient map, and one map to contain only small scales ($\ell \in (5000, 20000)$). The HDV estimator is well suited to returning the small-scale signal, and allows one to minimize bias from non-Gaussian astrophysical foregrounds by utilizing two CMB maps that do not overlap in scales~\cite{han22}.

While we can remove the Gaussian noise {\it{bias}} by RDN0 subtraction techniques, the Gaussian part of the four-point function still contributes {\it{noise}} to the lensing power spectrum which can not be completely removed.\footnote{We note that split-based lensing estimators can remove the contribution to $N_L$ from instrument noise~\cite{Madhavacheril:2020ido}, but not from the primordial CMB.  In this work, we do not assume split-based estimators are used, but their use is a potential avenue for gaining improved signal-to-noise.}  We hereafter call this lensing spectrum noise $N_L^{\kappa\kappa}$ to distinguish it from the CMB noise spectra $N_{\ell}^{XY}$ for $XY \in [TT,\ EE,\ BB]$.

We use the public CLASS delens package~\cite{Hotinli2021}\footnote{\url{https://github.com/selimhotinli/class_delens}} to calculate $N_L^{\kappa\kappa}$ for the H\&O estimator~\cite{OkamotoHu2003}.\footnote{We cross checked that the CLASS delens package returns the same $N_L^{\kappa\kappa}$ as the public tempura package (\url{https://github.com/simonsobs/tempura}).  We use the CLASS delens package because it can iteratively delens all the estimator combinations, not just $EB$, as we discuss in Section~\ref{sec:delensedspectra}.}  
For SO-like and CMB-S4-like experiments, we use the H\&O estimator to generate $N_L^{\kappa\kappa}$ curves using the CMB multipole ranges specified in Table~\ref{tab:exps} and four-point lensing estimators from pairs of $TT, TE, EE, EB$, and $TB$ maps, noting that the other possible four-point combinations contribute less significantly to the overall signal-to-noise.  For all the H\&O $N_L^{\kappa\kappa}$ curves, discussed here and below, we reduce their noise levels further by iterative delensing, which we describe in Section~\ref{sec:delensedspectra}. We coadd these $N_L^{\kappa\kappa}$ spectra to obtain a minimum variance (MV) noise curve, $MV~N_L^{\kappa\kappa}$.   

For the CMB-HD-like experiment, for lensing multipoles below $L=5000$, we generate $N_L^{\kappa\kappa}$ curves using all CMB multipoles within $\ell \in [30, 20000]$ for $T, E$, and $B$\footnote{Note that we use ASO-like noise for the $BB$ spectrum for $\ell<1000$, and since CLASS delens does not allow different noise levels for $E$ and $B$ for the $EB$ estimator, for $EB$ we set both $E$ and $B$ noise levels to that of ASO below $\ell<1000$.}, but note that there is minimal change to $N_L^{\kappa\kappa}$ for $L<5000$ if we limit the $T$ spectra to $\ell \in [30, 10000]$ due to foreground concerns.  We calculate $N_L^{\kappa\kappa}$ with the H\&O estimator for $TT, TE, EE, EB$, and $TB$ maps, reducing these noise levels further by iterative delensing, and coadd these to obtain an $MV~N_L^{\kappa\kappa}$.   We assume the $C_L^{\kappa\kappa}$ covariance matrix for $L \in [30, 5000]$ has only diagonal contributions since the lensing convergence is assumed to be Gaussian on relatively large scales. 

\begin{figure}[t]
    \centering
    \includegraphics[width=\columnwidth]{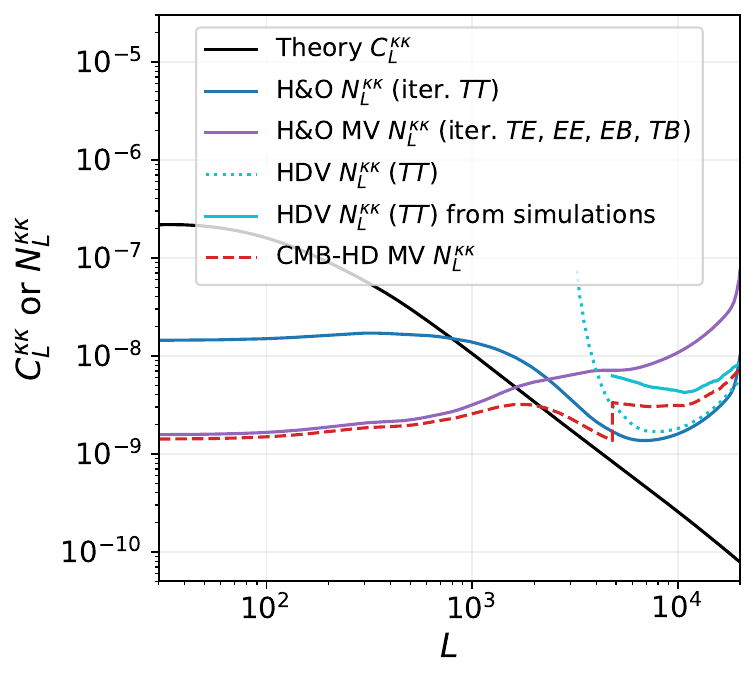}
    \caption{We show the theoretical lensing convergence power spectrum, $\kappa\kappa$, (black solid) along with the CMB-HD minimum-variance (MV) lensing power spectrum noise, $N_L^{\kappa\kappa}$, (red dashed). As discussed in Section~\ref{sec:lensingnoise}, the latter is a combination of the estimators of~\protect{\cite{huokamoto2002}} (H\&O) and~\protect{\cite{hdv2007}} (HDV); in the lensing multipole range $L \in [30, 5000]$, we use the H\&O MV reconstruction noise from the $TT$ (blue) and $TE$, $EE$, $EB$, and $TB$ estimators (the MV of the latter four estimators are shown in purple).  For each of these estimators, we assume iterative delensing will be done, and lower the noise curves accordingly following the method of~\protect{\cite{Hotinli2021}} and summarized in Section~\ref{sec:delensedspectra}. In the range $L \in [5000, 20000]$, to reduce bias from residual extragalactic foregrounds, we replace the $N_L^{\kappa\kappa}$ from the H\&O $TT$ estimator with that from the HDV $TT$ estimator, which we obtain from the diagonal elements of the simulation-based covariance matrix of~\protect{\cite{han22}} (cyan solid); to this we coadd the H\&O MV $N_L^{\kappa\kappa}$ from $TE$, $EE$, $EB$, and $TB$.  For comparison, we also show the analytically-derived HDV $TT$ $N_L^{\kappa\kappa}$ (cyan dotted); the excess variance of the simulation-based HDV $N_L^{\kappa\kappa}$ compared to the analytic one is likely due to
    higher-order lensing corrections and non-Gaussian fluctuations of the matter power spectrum that are significant on these small scales and are captured by the simulations, as discussed in~\protect{\cite{Nguyen2017, han22}}. See Section~\ref{sec:lensingnoise} for full details. }
    \label{fig:LensingNoise}
\end{figure}

In Fig.~\ref{fig:LensingNoise}, we show $N_L^{\kappa\kappa}$ for CMB-HD from the H\&O estimator from $TT$ alone (solid dark blue curve), and from everything other than $TT$ (i.e.~minimum variance coadd of  $TE, EE, EB, TB$) (solid purple curve).  
From the comparison of these curves, we see that the polarization estimators dominate the signal-to-noise below $L=2500$; however, on smaller scales, the $TT$ estimator has significantly lower noise than all the other estimators combined.  For this reason, on small scales, one must be concerned about the impact of astrophysical foregrounds in the temperature maps and how they may bias the lensing power spectrum.  We also show in this figure the $N_L^{\kappa\kappa}$ for the HDV quadratic estimator using only $TT$ and $\ell \in [30,2000]$ in the gradient map and $\ell \in [5000, 20000]$ in the small-scale map (dotted cyan curve).  This noise curve was made using the symlens package\footnote{\url{https://github.com/simonsobs/symlens}}. We see that the HDV curve matches the H\&O curve well on small-scales; however, it is not as useful on large scales.  We note that all the $N_L^{\kappa\kappa}$ curves shown in Fig.~\ref{fig:LensingNoise} were calculated including both the instrument noise and residual foregrounds described in Sections~\ref{sec:noisespectra} and~\ref{sec:fg} for the CMB spectra.  

For CMB-HD, on small lensing scales from $L \in [5000, 20000]$, we use the simulation-based lensing covariance matrix from~\cite{han22} for $TT$, which incorporates the HDV quadratic estimator with the $\ell$ ranges for large-scale and small-scale CMB maps given above, as well as higher-order lensing corrections and non-Gaussian fluctuations of the matter power spectrum that are significant on these small scales and are captured by the simulations. The solid cyan curve in Fig.~\ref{fig:LensingNoise} shows the lensing noise obtained from the diagonals of this covariance matrix.  This $C_L^{\kappa\kappa}$ covariance matrix also includes off-diagonal contributions (which we show in Fig.~\ref{fig:corr}); in generating this covariance matrix, simulation-based calculations of the RDN0 and $N_1$ biases (the latter is a higher-order correction) are subtracted, which reduces the off-diagonal correlations significantly, as discussed in~\cite{han22} and~\cite{Nguyen2017}. Restricting the HDV gradient map to $\ell<2000$ is also important for removing the $N_2$ bias~\cite{Nguyen2017, Hanson2011}.  For the final CMB-HD $N_L^{\kappa\kappa}$ spectra in the range of $L \in [5000, 20000]$, we coadd the simulation-based $TT$ HDV $N_L^{\kappa\kappa}$ with the H\&O $N_L^{\kappa\kappa}$ from $TE, EE, EB,$ and $TB$.  For $L \in [30, 5000]$, we use the H\&O minimum variance $N_L^{\kappa\kappa}$ coadding $TT, TE, EE, EB,$ and $TB$ noise curves.  We show this CMB-HD MV $N_L^{\kappa\kappa}$ curve in Fig.~\ref{fig:spectra} (dashed pink) and in Fig.~\ref{fig:LensingNoise} (dashed red).

For CMB-HD, the $EB$ estimator dominates below $L = 2500$, and foreground biases are a negligible concern.  Above $L=5000$, we attempt to be conservative by adopting the HDV estimator and a simulation-based covariance matrix; we note that there are other more optimal estimators that can be used to reconstruct lensing at small scales that may yield a higher signal-to-noise ratio and which also hold promise for mitigating foreground bias~\cite{Horowitz:2017iql,Hadzhiyska2019,Millea:2020iuw,Chan2023}.  In the regime of $L \in [2500, 5000]$, where the $TT$ estimator dominates, we are optimistic that foreground bias can be mitigated, but emphasize that this still needs to be demonstrated with realistic simulations; as mentioned above, the $TT$ spectrum can be cut to only include $\ell \in [30, 10000]$ as opposed to $\ell \in [30, 20000]$, with no loss of lensing signal-to-noise for $L<5000$.

\subsection{Creation of Delensed CMB Spectra}
\label{sec:delensedspectra}

\begin{figure}[t]
    \centering
    \includegraphics[width=\columnwidth]{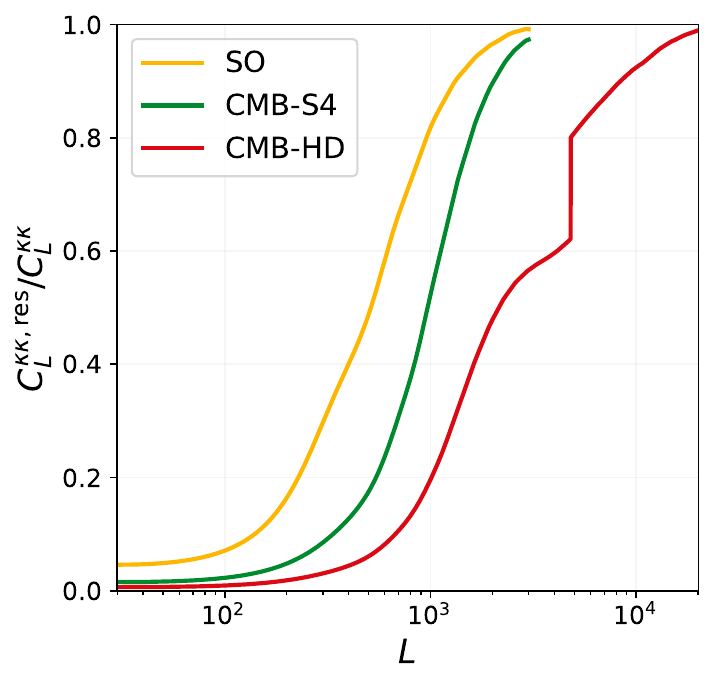}
    \caption{The residual lensing power, $C_L^{\kappa\kappa,\mathrm{res}}$, remaining after delensing (Eq.~\ref{eq:residuallens}), shown as a fraction of the full lensing power spectrum, $C_L^{\kappa\kappa}$, for SO, CMB-S4, and CMB-HD-like experiments with the specifications listed in Table~\ref{tab:exps}. For CMB-HD, the change near $L = 5000$ is due to using different lensing estimators above and below that scale, as discussed in Section~\ref{sec:lensingnoise}. Below $L = 650$, CMB-HD removes over 90\% of the lensing power. }
    \label{fig:ResidualLensing}
\end{figure}

Given the ultra low noise of a CMB-HD-type survey, it will be possible to remove a significant amount of the lensing in the CMB spectra, a procedure called delensing.
Delensing has the effect of tightening cosmological parameter constraints because it sharpens the acoustic peaks, removes uncertainty in the lensing realization that distorted the primordial CMB, and reduces lensing-induced off-diagonal power spectrum covariances~\cite{Green2017, Hotinli2021}.
In order to achieve tighter parameter constraints via delensing, it is important to replace the lost lensing information in the CMB power spectra with that from the reconstructed lensing power spectrum by combining them in the likelihood, properly accounting for the delensed CMB and CMB lensing power spectrum covariances~\cite{Green2017}.  

\begin{figure}[t]
    \centering
    \includegraphics[width=\columnwidth]{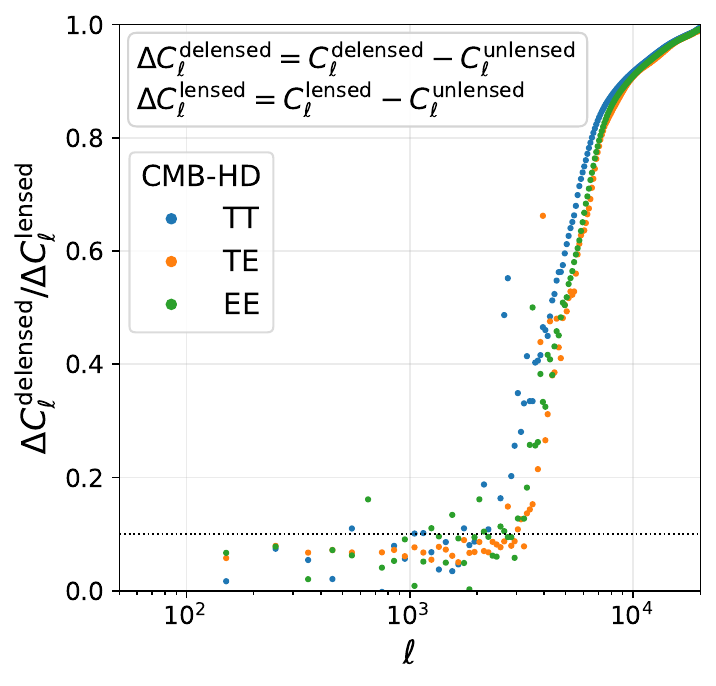}
    \caption{Shown is the amount of lensing power remaining in CMB-HD $TT$, $TE$, and $EE$ power spectra after delensing, $\Delta C_\ell^\mathrm{delensed}$, as a fraction of the total amount of lensing present in the CMB spectra, $\Delta C_\ell^\mathrm{lensed}$. $\Delta C_\ell$ is defined as the difference between the lensed/delensed CMB power spectra and the unlensed power spectra. Since both $\Delta C_\ell$ curves cross zero at the same multipoles, we bin them before taking their ratios, and show the bin centers as points. The lensing in the CMB-HD spectra is reduced on all scales, and about 90\% of the lensing is removed in the CMB spectra for $\ell < 2000$ (indicated by points below the dotted line). See Section~\ref{sec:delensedspectra} for details.}
    \label{fig:CMBDelensing}
\end{figure}

To predict the amount of lensing that can be removed from the CMB power spectra for each experiment discussed in Section~\ref{sec:expconfig}, we first calculate the residual lensing power that will remain in the CMB maps after delensing.  We then lens unlensed CMB spectra by this remaining lensing power using CAMB, following~\cite{han20delensing}.  The residual lensing power, $C_L^{\kappa\kappa,\mathrm{res}}$, is calculated by subtracting the Wiener-filtered lensing power spectrum from the total theoretical $C_L^{\kappa\kappa}$, as follows
\begin{equation} \label{eq:residuallens}
        C_L^{\kappa\kappa,\mathrm{res}} = C_L^{\kappa\kappa} \left(1 - \frac{C_L^{\kappa\kappa}}{C_L^{\kappa\kappa} + N_L^{\kappa\kappa}}\right),
\end{equation}
where $N_L^{\kappa\kappa}$ is the expected lensing reconstruction noise for the MV estimator described in Section~\ref{sec:lensingnoise}. This removes lensing on scales where the lensing power spectrum is detected with high signal-to-noise ratio, but not on scales where the lensing reconstruction is too noisy to estimate and remove the lensing signal. We show the residual lensing power as a fraction of the total power for SO, CMB-S4, and CMB-HD-like experiments in Fig.~\ref{fig:ResidualLensing}.  We see that below $L = 650$, delensing with CMB-HD removes over 90\% of the lensing power.  The abrupt change near $L = 5000$ is due to using different lensing estimators above and below that scale, as discussed in Section~\ref{sec:lensingnoise}.

As mentioned in Section~\ref{sec:lensingnoise}, for all of the H\&O $N_L^{\kappa\kappa}$ curves discussed above, we reduce their noise levels further by iterative delensing.
Iterative delensing of the $EB$ lensing estimator is traditionally assumed in forecasts of future CMB experiments~\cite{S4forecast}.  The reason is that, assuming no primordial $B$-mode power (i.e., assuming the $B$-mode power is entirely due to lensing), there are no \textit{unlensed} $B$-modes to contribute to the variance of the lensing reconstruction.  Thus the $EB$ lensing reconstruction noise can be reduced by delensing, which lowers the $B$-mode power.  Then the delensed $E$ and $B$ maps can be used to obtain a reconstruction of the residual lensing, which can be used to further delens the $E$ and $B$ maps; this procedure can be iterated on until converged~\cite{Seljak:2003pn,Smith2012,Simard:2014aqa}.  On small scales, where most of the CMB power in the temperature and $E$-mode polarization maps is due to lensing, the amount of delensing can be improved by a similar iterative approach~\cite{Hotinli2021}, where all the estimators are iterated instead of only $EB$.

\begin{figure}[t]
    \centering
    \includegraphics[width=\columnwidth]{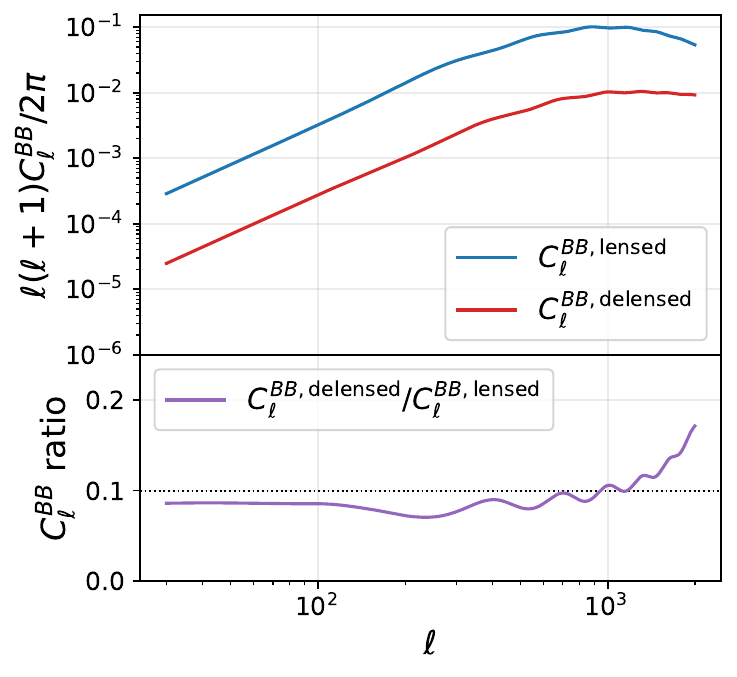}
    \caption{We show the amount of lensing that can be removed from any $BB$ power spectrum via delensing with CMB-HD. In particular, we show the lensed and delensed CMB $BB$ power spectra (top panel), and their ratio (lower panel), in the region where delensing is most effective ($\ell<2000$).  The bottom panel shows that CMB-HD delensing can remove over 90\% of the lensing signal in the $BB$ spectrum for $\ell < 1000$; this is the case even when assuming CMB-HD will not measure $\ell<1000$ and instead using ASO-like $BB$ spectra and instrument noise for $\ell \in [30, 1000)$, as discussed in Section~\ref{sec:noisespectra}. This is because the higher CMB multipoles of CMB-HD are being used to reconstruct low $L$ lensing multipoles used for delensing.}
    \label{fig:BBDelensing}
\end{figure}

As mentioned in Section~\ref{sec:lensingnoise}, we use the CLASS delens code~\cite{Hotinli2021} to calculate the expected lensing reconstruction noise when using the iterative delensing procedure, lowering the final $MV$ lensing noise spectrum, as well as increasing the amount of delensing of the CMB spectra, accordingly.  We show the effect of iterative delensing on the CMB $TT$, $TE$, and $EE$ power spectra in Fig.~\ref{fig:CMBDelensing}, and on the CMB $BB$ power spectrum in Fig.~\ref{fig:BBDelensing}. We see that CMB-HD removes about 90\% of the lensing in the $TT$, $TE$, and $EE$ spectra for $\ell < 2000$, and 90\% of the lensing in the $BB$ spectra for $\ell < 1000$.

In this work we assume that the bias that arises from internally delensing the acoustic peaks~\cite{Sehgal:2016eag, Carron:2017vfg} can either be removed following a method similar to~\cite{han20delensing} or will not arise as is the case for forward modeling methods~\cite{Millea:2020iuw}. We leave for future work the demonstration of the mitigation of this bias. 

\section{Covariance Matrix Calculation}
\label{sec:covmat}

In addition to the signal and noise spectra described in Section~\ref{sec:spectra}, a full covariance matrix of the expected data is required to forecast cosmological parameter constraints.  

\subsection{CMB Covariance Matrix}
\label{sec:cmbCovmat}

To construct the covariance matrix for the CMB experiments discussed above, we assume we will have $TT, TE, EE, BB$, and $\kappa\kappa$ spectra for each experiment for the multipole ranges listed in Table~\ref{tab:exps}.  We analytically calculate the covariance matrix following~\cite{Hotinli2021} as described below.

\begin{figure*}[t]
    \centering
    \includegraphics[width=0.49\textwidth]{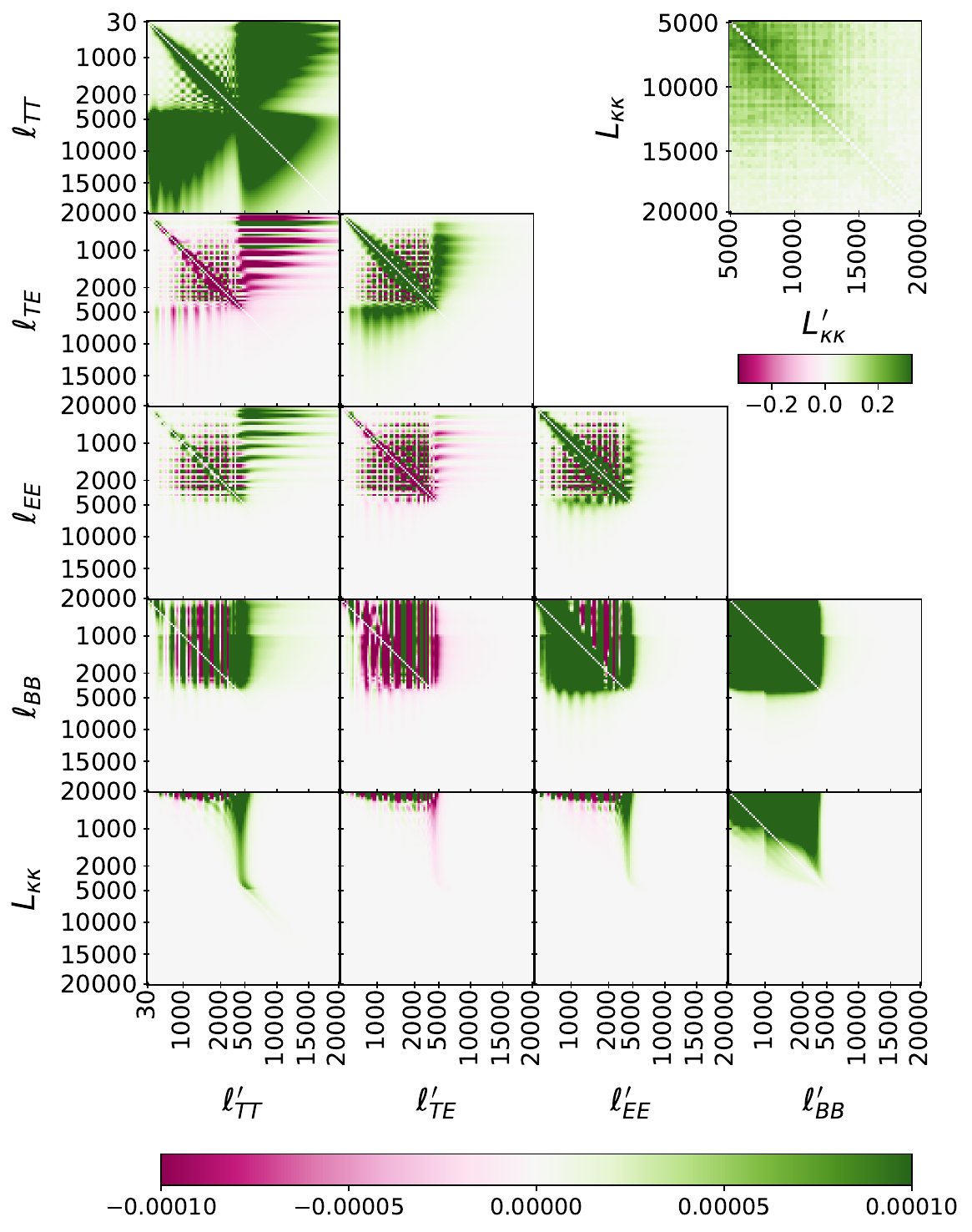}
    \includegraphics[width=0.49\textwidth]{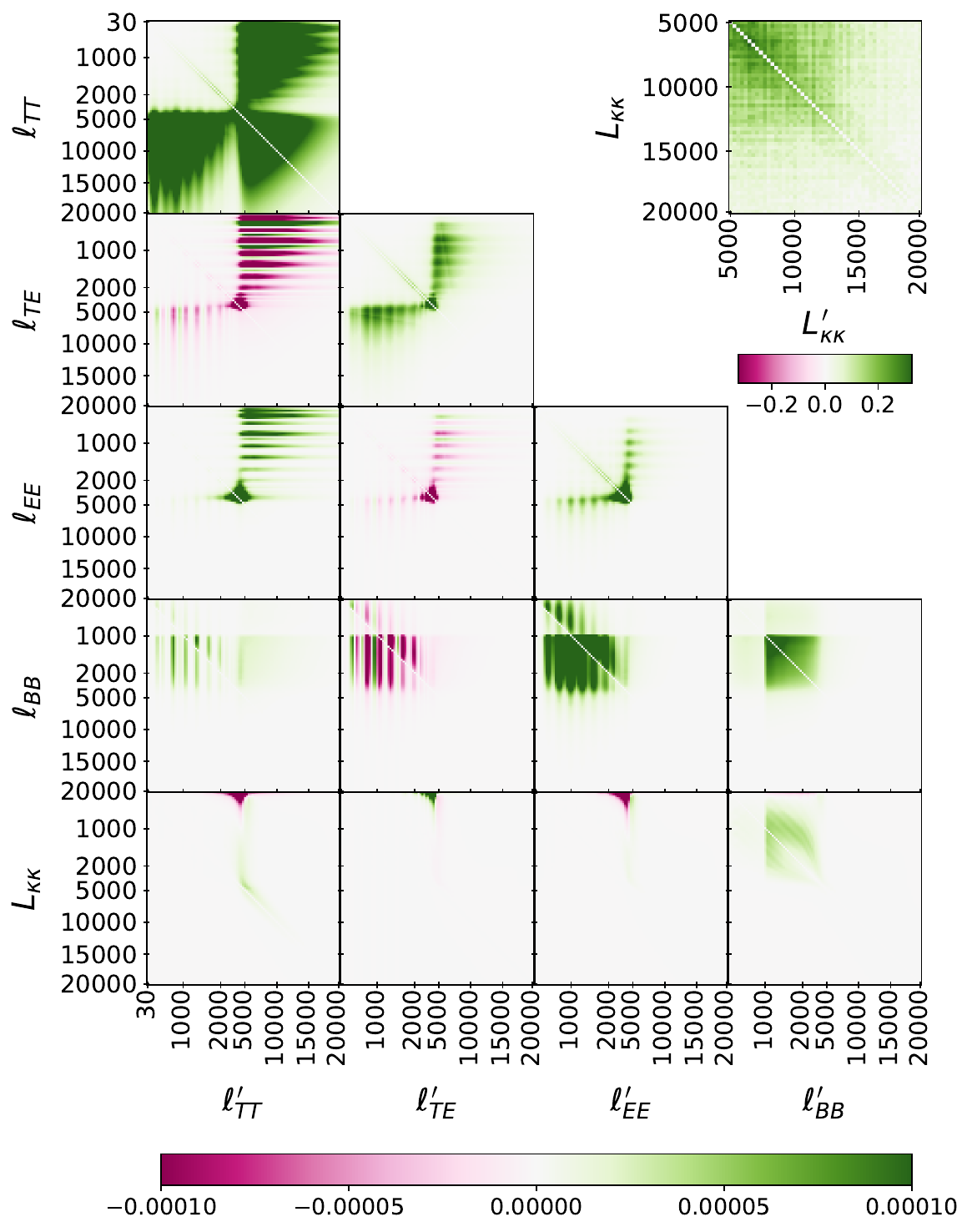}
    \caption{{\it{Left:}} We show the binned analytic correlation matrices for a CMB-HD survey between the lensed CMB $TT$, $TE$, $EE$, and $BB$ power spectra (first four rows), and between the lensed CMB and lensing $\kappa\kappa$ power spectrum (last row). We set the diagonals of each block to zero to highlight the off-diagonal elements, which arise due to the non-Gaussian covariances, as discussed in Section~\ref{sec:covmat}.
    We also show the simulation-based $\kappa\kappa \times \kappa\kappa$ covariance matrix from~\protect{\cite{han22}} which we use on scales $L, L' \gtrsim 5000$; note that this is shown with a different color scale than the analytic correlation matrices. {\it{Right:}} We show the corresponding binned correlation matrices for the delensed CMB-HD spectra. We find that the correlation coefficients for the off-diagonal terms of the binned analytic correlation matrices (not including $\kappa\kappa \times \kappa\kappa$) are less than $\pm 1\%$ everywhere.}
    \label{fig:corr}
\end{figure*}

For the diagonal components of the covariance matrix, for the blocks giving the variance between different CMB spectra (see Fig.~\ref{fig:corr}), we calculate
\begin{equation} 
    \mathbb{C}_{\ell_1 \ell_2,\mathrm{G}}^{XY,WZ} = \frac{\delta_{\ell_1 \ell_2}f_\mathrm{sky}^{-1}}{2\ell_1 + 1} \left[C_{\ell_1}^{XW,\mathrm{tot}}C_{\ell_1}^{YZ,\mathrm{tot}} + C_{\ell_1}^{XZ,\mathrm{tot}}C_{\ell_1}^{YW,\mathrm{tot}}\right], 
\label{eq:cmbxcmb-diags}
\end{equation}
Here $C_{\ell}^{\mathrm{tot}}$ is the total CMB signal plus noise spectrum, including the residual extragalactic foreground levels,  (i.e.~$C_{\ell}^{\mathrm{tot}} = C_\ell + N_\ell)$, and combinations of any two of $X$, $Y$, $Z$, and $W$ can each be $TT, TE, EE$, or $BB$.  The sky fraction observed by each experiment is indicated by $f_\mathrm{sky}$. We label this term given by Eq.~\ref{eq:cmbxcmb-diags} with the subscript G for ``Gaussian'' to indicate that it arises from the Gaussian fluctuations of the lensed or delensed CMB field. 

As discussed in Section~\ref{sec:lensingnoise}, lensing induces mode coupling between different multipoles of the unlensed CMB field, giving the originally Gaussian field a non-Gaussian structure.  The uncertainty in the amount of mode coupling, arising from uncertainty in either the lensing potential or the unlensed CMB field due to sample variance, generates off-diagonal elements in the CMB x CMB blocks~\cite{BenoitLevySmithHu2012}.  Following~\cite{Hotinli2021}, we calculate these off-diagonal terms as
\begin{equation} 
    \begin{split} \mathbb{C}_{\ell_1 \ell_2,\mathrm{NG}}^{XY,WZ} & = \sum_L \frac{\partial C_{\ell_1}^{XY}}{\partial C_L^{\phi\phi}} \frac{2 f_\mathrm{sky}^{-1}}{2 L + 1} \left(C_L^{\phi\phi}\right)^2 \frac{\partial C_{\ell_2}^{WZ}}{\partial C_L^{\phi\phi}} \\ & \quad + \sum_\ell \left[\frac{\partial C_{\ell_1}^{XY}}{\partial C_\ell^{XY,\mathrm{u}}}  \mathbb{C}_{\ell \ell,\mathrm{G}}^{XY,WZ;\mathrm{u}} \frac{\partial C_{\ell_2}^{WZ}}{\partial C_\ell^{WZ,\mathrm{u}}} \right] (1 - \delta_{\ell_1 \ell_2}) , \end{split}
\label{eq:cmbxcmb-offdiags}
\end{equation}
where 
\begin{equation}
    \mathbb{C}_{\ell \ell,\mathrm{G}}^{XY,WZ;\mathrm{u}} = f_\mathrm{sky}^{-1} \frac{\left(C_{\ell}^{XW,\mathrm{u}}C_{\ell}^{YZ,\mathrm{u}} + C_{\ell}^{XZ,\mathrm{u}}C_{\ell}^{YW,\mathrm{u}}\right)}{2\ell + 1} 
\label{eq:unlensedCov}    
\end{equation}
is the sample variance of the unlensed CMB spectra $C_\ell^\mathrm{u}$, and again combinations of any two of $X$, $Y$, $Z$, and $W$ can each be $TT, TE, EE$, or $BB$.\footnote{Note that when a $BB$ combination is involved, we substitute the unlensed $EE$ spectrum in the second term of Eq.~\ref{eq:cmbxcmb-offdiags}  and in Eq.~\ref{eq:unlensedCov}, following~\cite{Hotinli2021,Peloton2017}.}  Similarly $2f_\mathrm{sky}^{-1} (C_L^{\phi\phi})^2/(2 L + 1)$ is the sample variance of the lensing potential spectrum.  The lensing potential power spectrum, $C_L^{\phi\phi}$, is related to the lensing deflection power spectrum, $C_L^{dd}$, and the lensing convergence power spectrum, $C_L^{\kappa\kappa}$, by
\begin{equation}
    C_L^{\kappa\kappa} = \frac{L (L + 1)}{4} C_L^{dd} = \frac{L^2 (L + 1)^2}{4} C_L^{\phi\phi}.
\end{equation}
The subscript NG in Eq.~\ref{eq:cmbxcmb-offdiags} indicates that it is the ``non-Gaussian'' component. The full covariance matrix for the CMB x CMB blocks is given by
\begin{equation}
    \mathbb{C}_{\ell_1 \ell_2}^{XY,WZ} = \mathbb{C}_{\ell_1 \ell_2,\mathrm{G}}^{XY,WZ} + \mathbb{C}_{\ell_1 \ell_2,\mathrm{NG}}^{XY,WZ}.
\end{equation}

For the CMB lensing x CMB lensing block of the covariance matrix, the variance of the lensing potential spectrum, after RDN0 subtraction, is approximately Gaussian~\cite{Peloton2017} and is given by
\begin{equation} 
    \mathbb{C}_{L_1 L_2}^{\phi\phi,\phi\phi} = \delta_{L_1 L_2}  \frac{2 f_\mathrm{sky}^{-1}}{2 L_1 + 1} \left(C_{L_1}^{\phi\phi} + N_{L_1}^{\phi\phi}\right)^2.
\label{eq:phixphi}    
\end{equation}
For CMB-HD, we use Eq.~\ref{eq:phixphi} for $L<5000$, and 
for $L \ge 5000$ we replace the analytic covariance matrix with a simulation-based one from~\cite{han22} when including foregrounds, or from~\cite{Nguyen2017} when not including foregrounds, both of which use the HDV $TT$ estimator as discussed in Section~\ref{sec:lensingnoise}.  Note importantly that the simulation-based covariance matrices have off-diagonal terms, as discussed in more detail in Section~\ref{sec:lensingnoise}, and as shown in Fig.~\ref{fig:corr} for the $\kappa\kappa$ block.\footnote{Note that on large-scales the lensing field is approximately Gaussian, with only diagonal contributions to the covariance matrix. In addition, RDN0 subtraction removes most residual off-diagonal contributions~\cite{Peloton2017}. However, a more accurate treatment would use a simulation-based covariance matrix for all scales.}

We calculate the correlations between the lensing potential spectra and the CMB spectra from Eq.~36 in~\cite{Peloton2017}, which can also be obtained by extending Eq.~5 in~\cite{Hotinli2021} to allow either $XY$ or $WZ$ to be $\phi\phi$, as 
\begin{equation} \label{eq:phixcmb}
    \mathbb{C}_{L_1 \ell_2}^{\phi\phi,XY} = \frac{2}{2 L_1 + 1} \left(C_{L_1}^{\phi\phi}\right)^2 \frac{\partial C_{\ell_2}^{XY}}{\partial C_{L_1}^{\phi\phi}}.
\end{equation}
This covariance arises because the same lenses  that make up the lensing potential are also responsible for lensing the CMB spectra.

The terms given by Eqs.~\ref{eq:cmbxcmb-diags},~\ref{eq:cmbxcmb-offdiags},~\ref{eq:phixphi}, and~\ref{eq:phixcmb} are the most important terms in the analytic covariance matrix calculation, with the diagonal/Gaussian terms having the largest impact on parameters.  However, there are some additional terms that are not included here~\cite{Schmittfull2013,Manzotti2014,Peloton2017}, which we discuss below.  The first term we neglect is the super-sample covariance~\cite{Manzotti2014} arising from lenses on scales larger than the survey area.  Since all three experiments listed in Table~\ref{tab:exps} will survey half the sky, this effect is expected to be negligible as suggested by~\cite{Manzotti2014}.

For the covariance between the lensing potential spectrum and the CMB spectra, there are four terms given in Eq.~33 of~\cite{Peloton2017}.  As discussed in~\cite{Peloton2017}, the first two terms cancel if one does RDN0 subtraction, which we assume will be done for all the CMB experiments considered here, as mentioned in Section~\ref{sec:lensingnoise}.  The third term in Eq.~33 of~\cite{Peloton2017} is Eq.~\ref{eq:phixcmb} above, and we found that the fourth term (Type B Primary Trispectrum) had negligible impact on the parameter forecasts and is thus omitted. 

In this work, we neglect a first-order correction to the lensing power spectrum  known as the ``N1 bias''~\cite{Kesden:2003cc,Hanson2011}.  While this term is called a ``bias'', it is in fact a higher-order correction to the lensing signal that can be calculated from first principles similar to the zeroth-order lensing spectrum.  This N1 signal is of similar or larger amplitude than the standard CMB power spectrum at high multipoles, and its inclusion may yield a higher lensing signal-to-noise ratio than we forecast here.  We leave further exploration of this to future work, and note that ignoring this term is likely conservative.

We use the CLASS delens package~\cite{Hotinli2021} to evaluate the derivatives $\partial C_\ell^{XY} / \partial C_L^{\phi\phi}$ and $\partial C_\ell^{XY} / \partial C_{\ell'}^{XY,\mathrm{u}}$ in Eq.~\ref{eq:cmbxcmb-offdiags} and~\ref{eq:phixcmb}.\footnote{We increase the CLASS precision parameters beyond the default values to the following: \texttt{accurate\_lensing = 1}, \texttt{delta\_l\_max = 2000}, \texttt{perturbations\_sampling\_stepsize = 0.05}, \texttt{k\_max\_tau0\_over\_l\_max = 50}, and \texttt{P\_k\_max\_1/Mpc = 500} to be consistent with the accuracy settings used for CAMB (see Appendix~\ref{app:accuracy}.)}  With these derivatives, we assemble the final covariance matrix combining the terms in  Eqs.~\ref{eq:cmbxcmb-diags},~\ref{eq:cmbxcmb-offdiags},~\ref{eq:phixphi}, and~\ref{eq:phixcmb}.   We cross checked the CLASS delens $\partial C_\ell^{XY} / \partial C_L^{\phi\phi}$ derivatives by comparing to those we obtained using the lenscov\footnote{\url{https://github.com/JulienPeloton/lenscov}} package~\cite{Peloton2017} up to $\ell, L = 10,000$.  Since lenscov does not calculate covariances involving delensed CMB spectra, nor does it compute the derivatives $\partial C_\ell^{XY} / \partial C_{\ell'}^{XY,\mathrm{u}}$, we constructed covariance matrices from the CLASS delens and lenscov packages using lensed CMB spectra and Eqs.~\ref{eq:cmbxcmb-diags},~\ref{eq:cmbxcmb-offdiags},~\ref{eq:phixphi}, and~\ref{eq:phixcmb} (excluding the second term in Eq.~\ref{eq:cmbxcmb-offdiags}).  We confirmed that  the forecasted parameter errors obtained from either covariance matrix were consistent 
(within $< 0.05$\% for each parameter).

We generated covariance matrices for both lensed and delensed CMB spectra, by setting the $C_{\ell_1}$ in Eq.~\ref{eq:cmbxcmb-diags} to either lensed or delensed spectra, with the latter obtained as described in Section~\ref{sec:delensedspectra}. For the derivatives in Eqs.~\ref{eq:cmbxcmb-offdiags} and~\ref{eq:phixcmb}, CLASS delens provides an option to output them for delensed CMB spectra, given the lensing spectrum noise, $N_L^{\kappa\kappa}$.  

We bin our final covariance matrix using the same binning matrix, $M_{b\ell}$, as used for the one-dimensional spectra (see Eq.~\ref{eq:binnedspectra} of Section~\ref{sec:noisespectra}), so that we obtain a binned covariance matrix given by
\begin{equation} \label{eq:binnedcov}
        \mathbb{C}_{\ell_b \ell_{b'}}^{XY,WZ} = \sum_{\ell \ell'} M_{b\ell} \mathbb{C}_{\ell \ell'}^{XY,WZ} M_{b'\ell'}^T,
\end{equation}
for $XY$ and $WZ$ in $[TT,\ TE,\ EE,\ BB, \ \kappa\kappa]$.  We then form a single block covariance matrix from the 25 binned blocks.  We replace the binned CMB-HD $\kappa\kappa \times \kappa\kappa$ covariance matrix on scales $L_1,\ L_2 \gtrsim 5000$,  with the simulation-based one from~\cite{han22} when including foregrounds, or~\cite{Nguyen2017} when not including foregrounds, as mentioned above.

We show in Fig.~\ref{fig:corr} the binned analytic correlation matrices corresponding to the lensed and delensed covariance matrices, calculated using Eq.~\ref{eq:cmbxcmb-diags},~\ref{eq:cmbxcmb-offdiags},~\ref{eq:phixphi}, and~\ref{eq:phixcmb}, for the CMB-HD experiment. We also show the binned simulation-based $\kappa\kappa \times \kappa\kappa$ correlation matrix from~\cite{han22} which is used on scales $L_1,\ L_2 \gtrsim 5000$.
We calculate the unbinned elements of the correlation matrix, $\rho_{\ell_1 \ell_2}^{XY, WZ}$, between $C_{\ell_1}^{XY}$ and $C_{\ell_2}^{WZ}$ by~\cite{Peloton2017,Hotinli2021}
\begin{equation} \label{eq:corr}
    \rho_{\ell_1 \ell_2}^{XY, WZ} = \frac{\mathbb{C}_{\ell_1 \ell_2}^{XY, WZ}}{\sqrt{\mathbb{C}_{\ell_1 \ell_1}^{XY, XY} \mathbb{C}_{\ell_2 \ell_2}^{WZ, WZ}}},
\end{equation}
where $XY$ and $WZ$ are $TT$, $TE$, $EE$, $BB$, or $\kappa\kappa$ (we have used lowercase $\ell$'s for simplicity). We then bin this in the same way as Eq.~\ref{eq:binnedcov} to generate Fig.~\ref{fig:corr}. For this figure, in order to highlight the off-diagonal elements, the diagonal elements of each block were set to zero. In the CMB x CMB blocks for lensed spectra at low multipoles (Fig.~\ref{fig:corr}, left), we see the familiar ``checkerboard'' pattern from the lensing-induced peak-smoothing, as noted by~\cite{BenoitLevySmithHu2012,Schmittfull2013,Peloton2017,Green2017}.  For the delensed correlation matrix (Fig.~\ref{fig:corr}, right), correlations in these blocks are reduced at low multipoles, as also seen in~\cite{Green2017,Hotinli2021}; in fact they are nearly zero on scales where delensing removes much of the lensing from the CMB spectra (as shown in Fig.~\ref{fig:CMBDelensing}). In the CMB x $\kappa\kappa$ blocks, correlations are visible between large-scale (low-$L$) lensing and lensed CMB spectra for a wide range of scales (up to $\ell \sim 5000$), as also discussed in~\cite{Green2017}. The $\kappa\kappa$ x $\kappa\kappa$ blocks show off-diagonal terms from higher-order lensing corrections and non-Gaussian fluctuations of the matter power spectrum that are captured by simulations~\cite{han22}. We find that in both the lensed and delensed binned analytic correlation matrices (not including $\kappa\kappa \times \kappa\kappa$), the correlation coefficients, $\rho_{\ell_1 \ell_2}^{XY, WZ}$, for the off-diagonal terms are less than $\pm 1\%$.

\subsection{BAO Covariance Matrix}
\label{sec:baoCovmat}

We also construct a covariance matrix for BAO data from a DESI-like survey.  The BAO data measures the transverse and radial scales, $r_s/d_A(z)$ and $H(z) r_s$, respectively, where $r_s$ is the comoving sound horizon at the end of the baryon drag epoch (i.e.~$r_s(z_d)$), $d_A(z)$ is the angular diameter distance to redshift $z$, and $H(z)$ is the expansion rate at redshift $z$~\cite{desi}.  These are combined into a distance measurement defined by~\cite{EisensteinSDSS2005}
\begin{equation} \label{eq:BAOdV}
        d_V(z) \equiv \left[(1+z)^2 d_A^2(z)\frac{cz}{H(z)}\right]^{1/3}.
\end{equation}
We assume a Gaussian (diagonal) covariance matrix for the DESI BAO data. We calculate the uncertainty $\sigma_j$ on $r_s/d_V(z_j)$ following the approach of~\cite{Allison2015} as 
\begin{equation} 
    \begin{split}
            \sigma_j & \equiv \sigma\left(\frac{r_s}{d_V(z_j)}\right) \\ & = \frac{1}{3} \frac{r_s}{d_V(z_j)} \sqrt{\left[\frac{\sigma\left(H(z_j)r_s\right)}{H(z_j)r_s}\right]^2 + \left[2\frac{\sigma\left(d_A(z_j)/r_s\right)}{d_A(z_j)/r_s}\right]^2}. 
    \end{split}
\label{eq:BAOerrorbar}    
\end{equation}
We use CAMB to calculate the theoretical BAO signal $r_s / d_V(z_j)$, at each redshift $z_j$, for the {\it{Planck}} fiducial cosmology used in this work. We use the forecasted fractional uncertainties on the quantities $H(z_j) r_s$ and $d_A(z_j)/r_s$, at redshifts $z_j$, given by Table~2.3 (for $z\in[0.65, 1.85]$) and Table~2.5 (for $z\in[0.05, 0.45]$) in~\cite{desi} for the DESI experiment covering 14,000 square degrees of sky (i.e., the terms under the square root in the second line of Eq.~\ref{eq:BAOerrorbar}).  The diagonal elements of the covariance matrix are $\sigma_j^2$, with all off-diagonal elements set to zero.

\section{Methods for Parameter Forecasts}
\label{sec:forecastMethods}

We use both a Fisher matrix method (described in Section~\ref{sec:fisher}) and a likelihood plus Markov chain Monte Carlo (MCMC) method (described in Section~\ref{sec:mcmc}) to forecast cosmological parameter constraints from the combination of CMB and BAO data.  These forecasts are presented in Section~\ref{sec:parameterResults}.  

We forecast uncertainties on the six parameters of the $\Lambda$CDM model: the physical baryon and cold dark matter densities, $\Omega_\mathrm{b} h^2$ and $\Omega_\mathrm{c} h^2$, respectively; the primordial comoving curvature power spectrum amplitude, $\ln (10^{10} A_\mathrm{s})$, and scalar spectral index, $n_\mathrm{s}$,  defined at the pivot scale $k_0 = 0.05$~Mpc$^{-1}$; the reionization optical depth, $\tau$; and the Hubble constant $H_0$. For MCMC runs, we treat $H_0$ as a derived parameter, and instead sample over the parameter $100 \theta_\mathrm{MC}$, where $\theta_\mathrm{MC}$ is the \texttt{cosmoMC}\footnote{\url{https://cosmologist.info/cosmomc/readme.html}} approximation to the angular scale of the sound horizon at last scattering. We additionally vary the effective number of relativistic species $N_\mathrm{eff}$ and the sum of the neutrino masses $\sum m_\nu$. We refer to these eight parameters (listed in Table~\ref{tab:fisher}) as the $\Lambda$CDM+$N_\mathrm{eff}$+$\sum m_\nu$ model.
Since the CMB experiments considered here do not observe very large scales ($\ell < 30$), we apply a prior on the optical depth $\tau$ of $\sigma(\tau) = 0.007$ from {\it{Planck}}~\cite{planck18params}.

Both the Fisher and MCMC methods assume that the CMB and BAO data can be described by a Gaussian likelihood function $\mathcal{L}(\hat{d}|\vec{\theta})$ for the probability distribution of the data $\hat{d}$ given a set of model parameters $\vec{\theta}$.
For the CMB data described in Sections~\ref{sec:spectra} and~\ref{sec:cmbCovmat}, our likelihood function is given by
\begin{equation} 
    -2 \ln \mathcal{L}_\mathrm{CMB}\left(\hat{C}_{\ell_b} | \vec{\theta} \right) = \sum_{\ell_b \ell_{b'}} \Delta C_{\ell_b}(\vec{\theta}) \mathbb{C}^{-1}_{\ell_b \ell_{b'}}  \Delta C_{\ell_{b'}}(\vec{\theta}),
\label{eq:likelihood}        
\end{equation}
where $\Delta C_{\ell_b}(\vec{\theta}) = \hat{C}_{\ell_b} - C_{\ell_b}(\vec{\theta})$, $\hat{C}_{\ell_b}$ is the binned (lensed or delensed) data spectra\footnote{We use theory spectra at our fiducial cosmology, with no scatter, to simulate the data spectra. We verified that this yields the same error bars in our MCMC analysis as we would get if we had added scatter to the spectra drawn from our covariance matrix. \label{MCMCbandpowers}} with bin centers $\ell_b$, $C_{\ell_b}(\vec{\theta})$ is the binned theory spectra evaluated at a given set of cosmological parameters $\vec{\theta}$, and $\mathbb{C}_{\ell_b \ell_{b'}}$ is the binned covariance matrix of the data. $\hat{C}_{\ell_b}$ and $C_{\ell_b}(\vec{\theta})$ hold the binned $TT$, $TE$, $EE$, $BB$, and $\kappa\kappa$ spectra (in the same format and order as the covariance matrix, $\mathbb{C}_{\ell_b \ell_{b'}}$).\footnote{While the $BB$ power spectrum does not add significantly to the parameter constraints, we include it for completeness.}

For the BAO data, given by $f_j \equiv r_s/d_V(z_j)$ and described in Section~\ref{sec:baoCovmat}, we use the likelihood function
\begin{equation} 
 -2 \ln \mathcal{L}_\mathrm{BAO}\left(\hat{f}_j\ | \vec{\theta}\right) = \sum_{j} \frac{\left[\hat{f}_j - f_j(\vec{\theta})\right]^2}{\sigma_j^2},
\label{eq:likelihoodBAO}
\end{equation}
where $\hat{f}_j$ is the BAO data with variance $\sigma_j^2$ at redshift $z_j$, and $f_j(\vec{\theta})$ is the theory evaluated at the set of cosmological parameters $\vec{\theta}$~\cite{Dodelson2003}.  

We make both our Fisher and likelihood code publicly available, as mentioned above, and describe them in more detail below.

\subsection{Fisher Matrix Estimation}
\label{sec:fisher}

Fisher matrices can be used as an alternative to MCMC runs for parameter forecasts. While the Fisher approach is much quicker than a full MCMC run, it assumes that the parameters follow Gaussian posterior distributions; this is a good approximation for the six $\Lambda$CDM parameters, but may not be suitable when allowing other parameters (such as neutrino mass) to vary.  Thus we cross check our Fisher matrix results with a likelihood/MCMC analysis and find good agreement (see Appendix~\ref{app:mcmc_vs_fisher}).

To form a Fisher matrix, we calculate the change in the theoretical signal as each parameter is varied (i.e.~the numerical derivatives of spectra with respect to each parameter); the covariance matrix of the data tells us how well we can detect this change, and therefore how well we can constrain parameters.
We assume that the data can be described by a Gaussian likelihood, such as the ones in Eqs.~\ref{eq:likelihood} and~\ref{eq:likelihoodBAO}.
We also assume that the likelihood is maximized when evaluated at a fiducial set of parameters $\vec{\theta}_0$, and Taylor expand about this point. The elements $F_{\alpha\beta}$ of the Fisher matrix $F$ are given by the coefficients of the quadratic terms, i.e.
\begin{equation*}
    F_{\alpha\beta} \equiv \left.-\left\langle \frac{\partial^2 \ln\mathcal{L}}{\partial \theta_\alpha \partial \theta_\beta} \right\rangle \right|_{\vec{\theta}_0},
\end{equation*}
where the indices $\alpha$ and $\beta$ correspond to specific parameters. This characterizes how rapidly the likelihood changes as the parameters move away from their fiducial values.
The covariance matrix {\it{for the parameters}} is found by taking the inverse of the Fisher matrix, i.e.~$\mathrm{cov}\left(\theta_\alpha,~\theta_\beta\right) = \left(F^{-1}\right)_{\alpha\beta}$. The forecasted marginalized error $\sigma_\alpha$ on parameter $\theta_\alpha$ is then found from, $\sigma_\alpha = \sqrt{\left(F^{-1}\right)_{\alpha\alpha}}$.
    
For the CMB likelihood function given by Eq.~\ref{eq:likelihood}, the elements of the Fisher matrix are
\begin{equation} 
    F_{\alpha\beta}^\mathrm{CMB} = \sum_{\ell_b, \ell_b'} \left. \left(\frac{\partial C_{\ell_b}}{\partial \theta_\alpha} \mathbb{C}^{-1}_{\ell_b \ell_b'} \frac{\partial C_{\ell_b'}}{\partial \theta_\beta} \right) \right|_{\vec{\theta}_0}.
\label{eq:fisherCMB}    
\end{equation}
We use Eq.~\ref{eq:fisherCMB} to compute the CMB Fisher matrix for a given set of cosmological parameters, numerically evaluating the derivatives with respect to parameters using a finite difference method with the parameter step sizes listed in Table~\ref{tab:fisher}.  We apply the Gaussian prior on $\tau$  by adding its inverse variance $1 / \sigma^2(\tau)$ to the corresponding element $F_{\tau\tau}$ of the Fisher matrix~\cite{DarkEnergyTaskForce}. 

For the BAO likelihood function given by Eq.~\ref{eq:likelihoodBAO}, the elements of the Fisher matrix are given by
\begin{equation} 
    F_{\alpha\beta}^\mathrm{BAO} = \sum_j \left. \left(\frac{\partial f_j}{\partial \theta_\alpha} \frac{1}{\sigma^2_j} \frac{\partial f_j}{\partial \theta_\beta} \right) \right|_{\vec{\theta}_0},
\label{eq:fisherBAO}    
\end{equation}
where the sum is taken over the redshifts $z_j$, and the derivatives $\partial f_j / \partial \theta_\alpha$ are calculated in the same way as in Eq.~\ref{eq:fisherCMB}. 
To forecast parameter errors from the combination of CMB and BAO data, we take the sum of the two Fisher matrices (since they are from independent data sets), and then invert this sum to obtain the parameter covariance matrix and marginalized error bars~\cite{DarkEnergyTaskForce}.

The Fisher matrix method described above gives the statistical uncertainties on a set of parameters, assuming that the fiducial model used in the calculation of the numerical derivatives accurately describes the true universe.
We can extend this method to calculate the expected {\it{bias}} to the estimated parameter values due to an incorrect fiducial model, following the approach described in~\cite{Huterer2004,Amara2007,Bernal2020} and adopted by~\cite{mccarthy22,Hotinli2021}.  
If we assume a fiducial model with power spectra $C_\ell^\mathrm{fid}$, while the true model that describes our Universe is $C_\ell^\mathrm{true}$, the bias $\Delta \theta_\alpha$ to the parameter $\theta_\alpha$ is given by
\begin{equation}
        \Delta \theta_\alpha = \sum_\beta F^{-1}_{\alpha\beta} \sum_{\ell_b, \ell_b'} \frac{\partial C_{\ell_b}^\mathrm{fid}}{\partial \theta_\beta} \mathbb{C}^{-1}_{\ell_b \ell_b'} \left(C_{\ell_b'}^\mathrm{true} - C_{\ell_b'}^\mathrm{fid}\right).
    \label{eq:bias}
    \end{equation}
Here the Fisher matrix $F_{\alpha\beta}$ is calculated with the fiducial model.
We note that the parameter biases obtained using Eq.~\ref{eq:bias} are technically conservative for the case where one has a prior on the value of a parameter (see Appendix~\ref{app:accuracy}). This is because, while Eq.~\ref{eq:bias} allows one to specify a parameter uncertainty from prior data via the Fisher matrix $F_{\alpha\beta}$, it does not allow one to specify the mean value of this prior, potentially resulting in artificially larger bias estimates than is the actual case.  The bias estimates from Eq.~\ref{eq:bias} can be cross checked with an MCMC method to verify consistency, as done in Appendix~\ref{app:accuracy}.  

\subsection{Likelihood and Markov Chain}
\label{sec:mcmc}

We use an MCMC method to verify our Fisher parameter forecasts. With an MCMC analysis, we want to recover the posterior probability distribution $P(\vec{\theta}\ | \hat{d})$ for a set of model parameters $\vec{\theta}$ that are most likely to describe the observed data $\hat{d}$. The posterior distribution is proportional to the product of the likelihood function, $\mathcal{L}(\hat{d}\ | \vec{\theta})$, and the prior probability distribution, $P(\vec{\theta})$, for the parameter values $\vec{\theta}$. 
The posterior distribution is estimated by drawing random samples of sets of parameters, $\vec{\theta}$, via an MCMC sampler, and evaluating the likelihood and prior probability for each~\cite{Hogg2017MCMC}.

For the results presented in this work, we use the \texttt{mcmc} sampler~\cite{cobayaMCMClewisbrindle, cobayaMCMClewis} from the Cobaya~\cite{cobaya} package to sample from the likelihood functions defined in Eqs.~\ref{eq:likelihood} and~\ref{eq:likelihoodBAO}, and we use CAMB to calculate the theory $C_{\ell}(\vec{\theta})$ or $f_j(\vec{\theta})$ at each step in parameter space.{\footnote{{Note that Cobaya calls CAMB itself, but we added an interface within our likelihood to obtain the delensed spectra from CAMB via Cobaya.}}}  The priors adopted for each MCMC analysis are listed in Table~\ref{tab:fisher}; all priors are uniform except for the Gaussian prior on the reionization optical depth from \textit{Planck} of $\tau = 0.054 \pm 0.007$~\cite{planck18params}.

MCMC runs in general converge faster if given an initial proposal matrix consisting of the anticipated covariance between the cosmological parameters.  We generate such a proposal matrix by first using a Fisher matrix to obtain the conditional posterior width for each parameter, given by $1 / \sqrt{F_{\alpha\alpha}}$ for parameter $\theta_\alpha$. We use this to determine the step size of each parameter in the sampler. We then generate mock data and run a quick preliminary MCMC using the default CAMB accuracy, except for setting \texttt{lens\_potential\_accuracy=1} to turn on the non-linear matter power spectrum (see Appendix~\ref{app:accuracy}).\footnote{Increasing the CAMB accuracy settings higher than this does not change the parameter error bars as long as the mock data is generated at the same accuracy; however, we show in Appendix~\ref{app:accuracy} that low CAMB accuracy settings in general may result in a bias when running MCMC chains on real data.} We use the resulting MCMC chains to calculate a parameter covariance matrix, which we use as a proposal matrix for subsequent MCMC runs with our baseline CAMB accuracy.

\begin{figure*}[t]
    \centering
    \includegraphics[width=\textwidth]{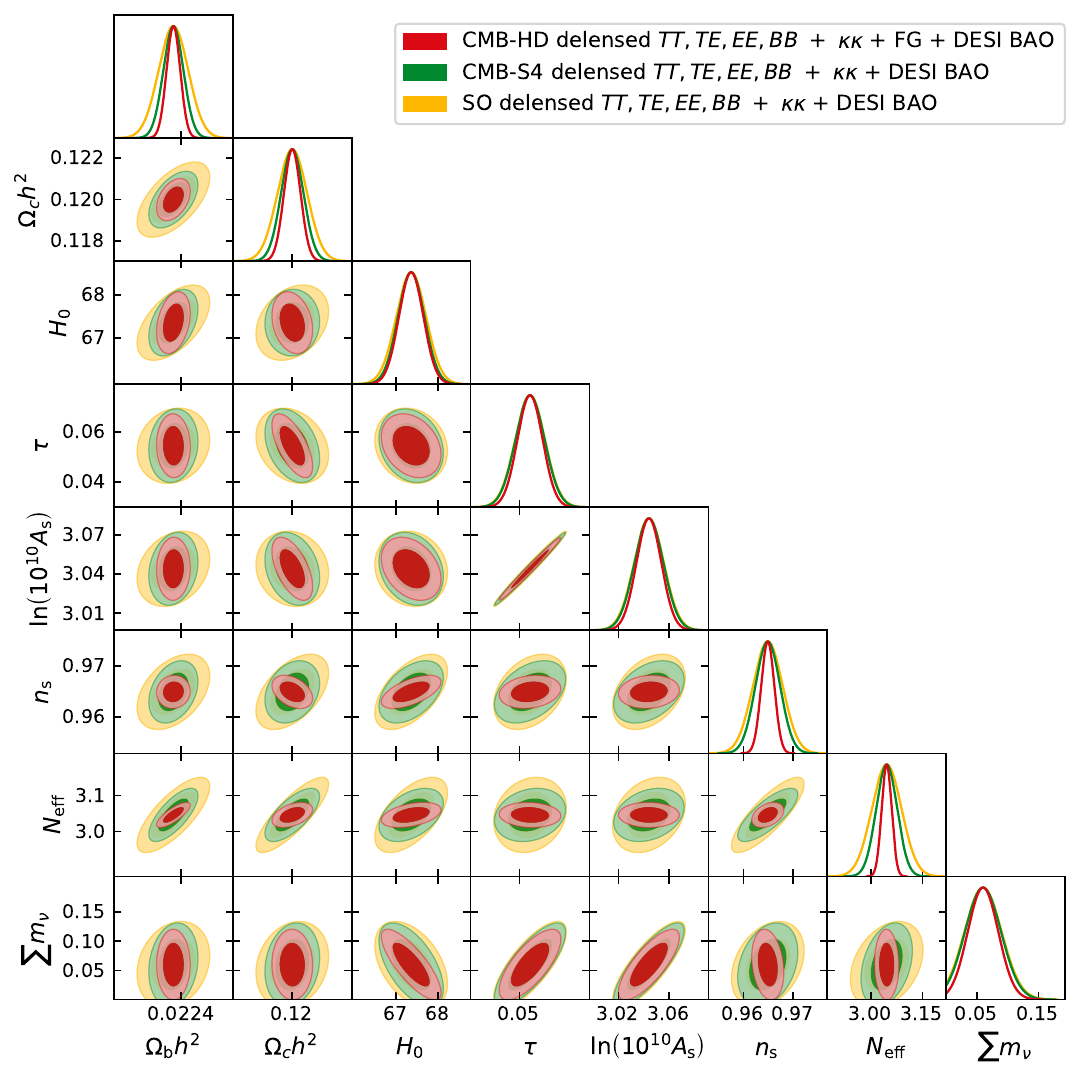}
    \caption{Forecasted cosmological parameter constraints for a $\Lambda$CDM+$N_\mathrm{eff}$+$\sum m_\nu$ model from delensed CMB $TT$, $TE$, $EE$, and $BB$ power spectra and the lensing power spectrum, $\kappa\kappa$, for SO-like, CMB-S4-like, and CMB-HD-like surveys. The experimental configurations for each survey are listed in Table~\ref{tab:exps}. We also include expected DESI BAO data~\protect{\cite{desi}} to these forecasts.  The parameters that see the most improvement from a CMB-HD type survey compared to precursor experiments are $n_\mathrm{s}$ and $N_\mathrm{eff}$.  We give the one-dimensional marginalized parameter errors in Table~\ref{tab:sos4hd}.}
    \label{fig:sos4hd}
\end{figure*}

We consider the MCMC parameter chains to be converged when they reach a value of about $R-1 \leq 0.01$ after removing the first 30\% of each chain as ``burn-in'', where $R-1$ is the Gelman-Rubin convergence parameter~\cite{GelmanRubin1992}.  We find that the likelihood for delensed CMB-HD plus DESI BAO for a $\Lambda$CDM~+~$N_\mathrm{eff}$~+~$\sum m_\nu$ model takes about 12 hours on NERSC\footnote{\url{https://www.nersc.gov/}} using our baseline accuracy to reach $R-1 = 0.03$, and about 16 hours to reach $R-1 = 0.015$. These times are obtained after including a proposal matrix generated as described above.  We confirm that the marginalized mean parameter values that we obtain match the fiducial values used to generate the mock data, listed in Table~\ref{tab:fisher}. We also confirm that we recover parameter errors that match those obtained from the Fisher analysis described above (see Appendix~\ref{app:mcmc_vs_fisher} and Fig.~\ref{fig:mcmc_vs_fisher} for details). Since the MCMC run takes about a day, after we verify a few base cases, we use the Fisher method to explore variations of those cases (e.g.~changes in $\ell_{\rm{max}}$, inclusion of foregrounds or delensing, etc.).

\section{Parameter Forecasts}
\label{sec:parameterResults}

%%%%% SO, S4, HD DELENSED + BAO -- Table III
\begin{table*}[h]
    \centering
    \begin{tabular}{l@{\hskip 2em} c@{\hskip 3em} l@{\hskip 2.5em} l@{\hskip 2.5em} l@{\hskip 4em} c@{\hskip 1.5em} c@{\hskip 0.5em}}
      \toprule
      \toprule
      \multicolumn{1}{l}{} & \multicolumn{1}{l}{} & \multicolumn{3}{c}{$1\sigma$ Error} &  \multicolumn{2}{c}{Ratio of $1\sigma$ Errors}
      \\
      \cmidrule(r){3-5} \cmidrule(r){6-7}
      Parameter & Fiducial &  \multicolumn{1}{l}{SO +} & \multicolumn{1}{l}{CMB-S4 +} & \multicolumn{1}{l}{CMB-HD +} & CMB-HD / SO & CMB-HD / CMB-S4 
      \\
       \multicolumn{1}{l}{$\Lambda$CDM+$N_\mathrm{eff}$+$\sum m_\nu$} & & DESI BAO & DESI BAO  & DESI BAO & 
      \\
     \midrule
      $\Omega_\mathrm{b} h^2$\dotfill & $0.022370$ & $0.000057$ & $0.000039$ & $0.000026$ & $0.46$ & $0.67$  
      \\
      $\Omega_\mathrm{c} h^2$\dotfill & $0.12000$ & $0.00074$ & $0.00056$ & $0.00041$ & $0.55$ & $0.73$ 
      \\
      $\ln(10^{10} A_\mathrm{s})$\dotfill & $3.044$ & $0.012$ & $0.011$ & $0.0098$ & $0.84$ & $0.86$
      \\
      $n_\mathrm{s}$\dotfill & $0.9649$ & $0.0030$ & $0.0025$ & $0.0013$  & $0.43$ & $0.52$ 
      \\
      $\tau$\dotfill & $0.0544$ & $0.0061$ & $0.0060$ & $0.0052$ & $0.85$ & $0.87$
      \\
      $H_0$ [km s$^{-1}$ Mpc$^{-1}$]\dotfill & $67.36$ & $0.36$ & $0.32$ & $0.29$ & $0.81$ & $0.91$ 
      \\
      $N_\mathrm{eff}$\dotfill & $3.046$ & $0.043$ & $0.030$ & $0.014$ & $0.33$ & $0.47$ 
      \\
      $\sum m_\nu$ [eV]\dotfill & $0.06$ & $0.030$ & $0.029$ & $0.025$  & $0.83$ & $0.86$
      \\
      \bottomrule
    \end{tabular}
    \caption{Forecasted cosmological parameter constraints for a $\Lambda$CDM + $N_{\rm{eff}} + \Sigma m_\nu$ model from SO-like, CMB-S4-like, and CMB-HD-like delensed $TT$, $TE$, $EE$, $BB$ and $\kappa\kappa$ power spectra when combined with DESI BAO. The first two columns list the parameters and their fiducial values. The following three columns list the forecasted marginalized $1\sigma$ uncertainties on each parameter for the three experiments, and the last two columns list the ratios of these values. The forecasts for CMB-HD include expected residual extragalactic foregrounds in the temperature power spectrum.  All forecasts include a $\tau$ prior of $\tau = 0.054 \pm 0.007$ from \textit{Planck}~\protect{\cite{planck18params}}. We find improvement in all parameters considered for a CMB-HD-like experiment compared to precursor surveys; we also find the most significant improvement for $n_\mathrm{s}$ and $N_\mathrm{eff}$, of about a factor of two or more.  These forecasts are also depicted in Fig.~\ref{fig:sos4hd}. }     
    \label{tab:sos4hd}
\end{table*}

%%%%% HD only
\begin{table*}[h]
    \begin{center}
    \begin{tabular}{l@{\hskip 2em} c@{\hskip 2em} l@{\hskip 1.5em} l@{\hskip 1.5em} l@{\hskip 1.5em} l@{\hskip 2em} l}
      \toprule
      \toprule
        &  & \multicolumn{4}{c}{$\ell_\mathrm{max}, ~L_\mathrm{max} = 20,000$}  &  $\ell_\mathrm{max}, L_\mathrm{max} = 10,000$
      \\
      \cmidrule(r){3-6} \cmidrule(r){7-7}
      Parameter & Fiducial & HD Lensed & HD Lensed &  HD Delensed & \textbf{HD Delensed}  & HD Delensed 
      \\
       \multicolumn{1}{l}{$\Lambda$CDM+$N_\mathrm{eff}$+$\sum m_\nu$} & & &  + FG & + FG  & \textbf{+ FG + DESI BAO} & + FG + DESI BAO 
      \\
      \midrule
      $\Omega_\mathrm{b} h^2$\dotfill & $0.022370$ & $0.000032$ & $0.000033$  & $0.000027$ & $0.000026$ & $0.000026$
      \\
      $\Omega_\mathrm{c} h^2$\dotfill & $0.12000$ & $0.00064$ & $0.00065$  & $0.00058$ & $0.00041$ & $0.00041$
      \\
      $\ln(10^{10} A_\mathrm{s})$\dotfill & $3.044$ & $0.011$ & $0.011$  & $0.011$ & $0.0098$ & $0.010$
      \\
      $n_\mathrm{s}$\dotfill & $0.9649$ & $0.0023$ & $0.0024$  & $0.0021$ & $0.0013$ & $0.0014$
      \\
      $\tau$\dotfill & $0.0544$ & $0.0056$ & $0.0057$  & $0.0054$ & $0.0052$ & $0.0054$
      \\
      $H_0$ [km s$^{-1}$ Mpc$^{-1}$]\dotfill & $67.36$ & $0.72$ & $0.74$  & $0.65$ & $0.29$ & $0.29$
      \\
      $N_\mathrm{eff}$\dotfill & $3.046$ & $0.017$ & $0.018$  & $0.015$ & $0.014$ & $0.015$
      \\
      $\sum m_\nu$ [eV]\dotfill & $0.06$ & $0.050$ & $0.051$  & $0.047$ & $0.025$ & $0.026$
      \\
      \bottomrule
    \end{tabular}
    \caption{Forecasted cosmological parameter constraints from CMB-HD $TT$, $TE$, $EE$, $BB$ and $\kappa\kappa$ power spectra for a $\Lambda$CDM + $N_{\rm{eff}} + \Sigma m_\nu$ model. All forecasts include a $\tau$ prior of $\tau = 0.054 \pm 0.007$ from \textit{Planck}~\protect{\cite{planck18params}}. The first two columns list the parameters and their fiducial values. The following two columns list their forecasted marginalized $1\sigma$ uncertainties when using lensed spectra with or without foregrounds in the temperature maps. The fifth column shows the forecast when delensing the CMB spectra, and the sixth column shows the change when including DESI BAO data~\protect{\cite{desi}}. In each of these cases we use CMB and CMB lensing multipoles out to $\ell_\mathrm{max}, L_\mathrm{max} =20,000$ for CMB-HD. The last column lists the same information as the sixth column, but instead using a maximum multipole of $\ell_\mathrm{max}, L_\mathrm{max} = 10,000$ for both CMB and CMB lensing power spectra. We see in particular that $n_\mathrm{s}$ and $N_{\rm{eff}}$ constraints are tightened when including multipoles beyond 10,000 for a CMB-HD type survey.  We show corresponding forecasts for $\Lambda$CDM and $\Lambda$CDM + $N_{\rm{eff}}$ models in Appendix~\ref{app:hdparams}.}
    \label{tab:hd}
    \end{center}
\end{table*}

For the discussion below, we show parameter constraints using Fisher matrix estimation since it provides a faster way to explore the impact of several effects.  However, as mentioned above, we confirm that our parameter forecasts are consistent when using either a likelihood and MCMC or Fisher matrix estimation for a subset of cases (see Appendix~\ref{app:mcmc_vs_fisher} for details).  For the SO-like and S4-like experiments, we also confirm that our parameter constraints are consistent with those forecasted in~\cite{SOforecast} and~\cite{S4forecast}, respectively.

In Fig.~\ref{fig:sos4hd}, we show the cosmological parameter constraints for the $\Lambda$CDM+$N_\mathrm{eff}$+$\sum m_\nu$ model from delensed CMB $TT$, $TE$, $EE$, $BB$ spectra and lensing $\kappa\kappa$ spectra when combined with DESI BAO data.  For the CMB-HD-like experiment (red), we have included residual extragalactic foregrounds in the temperature data.  For comparison, we forecast parameter constraints for S4-like (green) and SO-like (yellow) experiments.  We show the $1\sigma$ and $2\sigma$ parameter contours as the dark and lighter colors, respectively, and we list the $1\sigma$ marginalized parameter constraints for all three experiments in Table~\ref{tab:sos4hd}. We find improvement in all parameters considered for a CMB-HD-like experiment compared to precursor surveys; we also find the most significant improvement for $n_\mathrm{s}$ and $N_\mathrm{eff}$, of about a factor of two or more (see last two columns of Table~\ref{tab:sos4hd}).  In particular, for CMB-HD, we find $\sigma(n_\mathrm{s}) = 0.0013$ and $\sigma(N_\mathrm{eff}) = 0.014$.

In the following sections, we focus on a CMB-HD-like experiment and discuss the effects of residual extragalactic foregrounds, delensing of the acoustic peaks, and the combination with DESI BAO data on cosmological parameter constraints. The resulting parameter uncertainties in each case for a $\Lambda$CDM+$N_\mathrm{eff}$+$\sum m_\nu$ model are listed in Table~\ref{tab:hd}; we present results for a $\Lambda$CDM and a $\Lambda$CDM+$N_\mathrm{eff}$ model in Appendix~\ref{app:hdparams}. In the last column of Tables~\ref{tab:hd},~\ref{tab:hdlcdm}, and~\ref{tab:hdlcdmnnu}, we examine the effect of using a lower maximum multipole of $\ell_\mathrm{max}, L_\mathrm{max} = 10,000$ and find, in particular, that $n_\mathrm{s}$ and $N_{\rm{eff}}$ constraints are tightened when including multipoles beyond 10,000. 

\begin{figure}[t]
    \centering
    \includegraphics[width=\columnwidth]{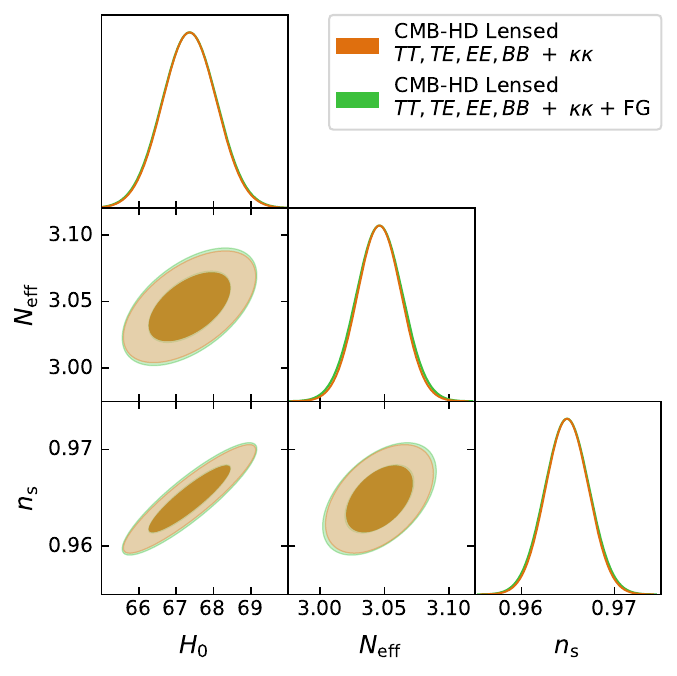}
    \caption{Shown is the impact of including expected residual foregrounds (FG) on the forecasted constraints for a $\Lambda$CDM+$N_\mathrm{eff}$+$\sum m_\nu$ model from CMB-HD lensed $TT$, $TE$, $EE$, $BB$ and $\kappa\kappa$ power spectra. Here we show a subset of parameter constraints for the Hubble constant $H_0$, the effective number of relativistic species $N_\mathrm{eff}$, and the scalar spectral index $n_\mathrm{s}$.  We see that the expected foregrounds do not significantly increase parameter uncertainties; we emphasize that while these are expected foreground levels from preliminary analyses~\protect{\cite{han22}}, it remains to be demonstrated with realistic simulations that such foreground levels can be achieved in practice.}
    \label{fig:HDfg}
\end{figure}

\subsection{Impact of Residual Foregrounds}
\label{sec:paramsFG}

We examine the effect of including residual extragalactic foregrounds in temperature maps, as described in Section~\ref{sec:fg}, on the parameter constraints for CMB-HD from lensed CMB $TT$, $TE$, $EE$, and $\kappa\kappa$ power spectra.  In Fig.~\ref{fig:HDfg}, we show the parameter constraints without (orange) and with (green) foregrounds for a subset of parameters from a $\Lambda$CDM+$N_\mathrm{eff}$+$\sum m_\nu$ model; the full set of forecasted parameter errors are listed in the third and fourth columns of Table~\ref{tab:hd}. We see that the residual foregrounds do not significantly increase the errors on any of the parameters considered here. The ultra-high resolution and low noise of CMB-HD is critical to reducing the foreground levels to those shown in the top panel of Fig.~\ref{fig:spectra}.

\subsection{Impact of Delensing}
\label{sec:paramsDelens}

\begin{figure}[t]
    \centering
    \includegraphics[width=\columnwidth]{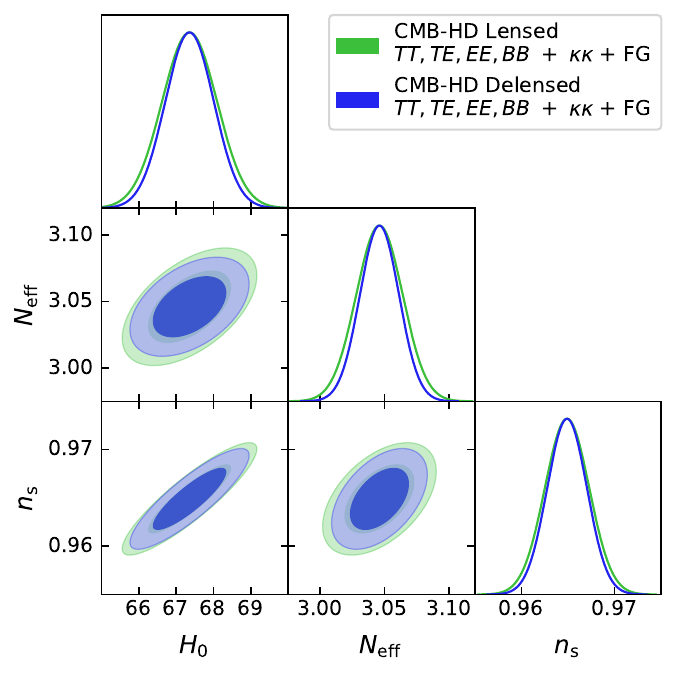}
    \caption{Shown are forecasted constraints for a $\Lambda$CDM+$N_\mathrm{eff}$+$\sum m_\nu$ model from lensed (green) and delensed (blue) CMB-HD $TT$, $TE$, $EE$, $BB$ and $\kappa\kappa$ power spectra. Here we show a subset of parameter constraints for $H_0$, $N_\mathrm{eff}$, and $n_\mathrm{s}$, and show the full parameter constraints in Fig.~\ref{fig:HDdelens-BAO}. We see that delensing the CMB spectra reduces the parameter uncertainties.}
    \label{fig:HDdelens}
\end{figure}

In Fig.~\ref{fig:HDdelens}, we show the parameter constrains from CMB-HD lensed (green) and delensed (blue) CMB spectra and the lensing spectrum for a subset of parameters from a $\Lambda$CDM+$N_\mathrm{eff}$+$\sum m_\nu$ model. In both cases, we include residual extragalactic foregrounds. We see smaller uncertainties in both $n_\mathrm{s}$ and $N_\mathrm{eff}$ after delensing.  We show the comparison between lensed and delensed forecasts for the full parameter set in Fig.~\ref{fig:HDdelens-BAO}.  The fifth column of Table~\ref{tab:hd} gives the marginalized $1\sigma$ parameter errors when including delensing, and shows considerable improvement for all the parameters considered, even when also including residual foregrounds.  We attribute this to the removal of much of the off-diagonal covariance on scales where CMB-HD can delens efficiently, as shown in the right panel of Fig.~\ref{fig:corr}.  Tables~\ref{tab:hdlcdm} and~\ref{tab:hdlcdmnnu} show the impact of delensing for $\Lambda$CDM and $\Lambda$CDM+$N_\mathrm{eff}$ models, respectively.  In Table~\ref{tab:lens_vs_delens}, we show that even when including DESI BAO, delensing still improves parameter errors, especially for $N_\mathrm{eff}$.

\subsection{Impact of DESI BAO}
\label{sec:paramsBAO}

\begin{figure}[t]
    \centering
    \includegraphics[width=\columnwidth]{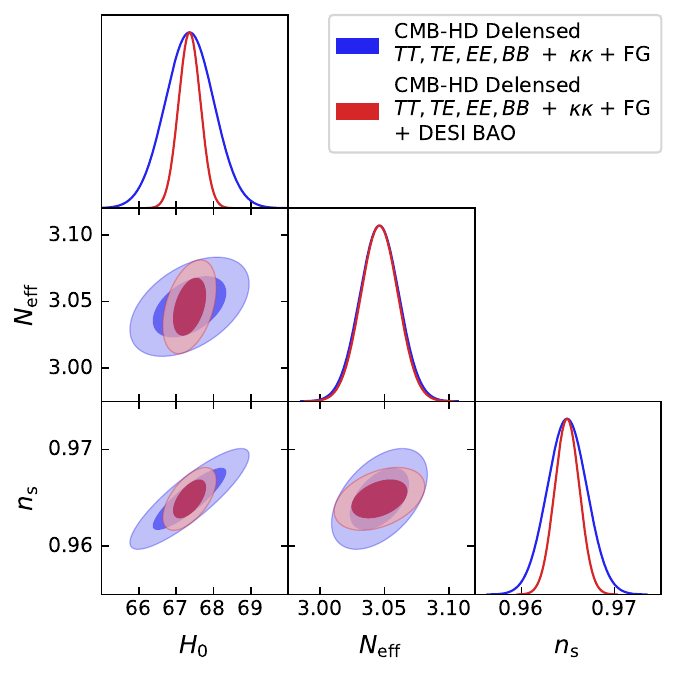}
    \caption{Shown are the forecasted constraints for a $\Lambda$CDM+$N_\mathrm{eff}$+$\sum m_\nu$ model from CMB-HD delensed $TT$, $TE$, $EE$, $BB$ and $\kappa\kappa$ power spectra, with and without including expected DESI BAO data~\protect{\cite{desi}} (blue and red, respectively).  Here we show a subset of parameter constraints for $H_0$, $N_\mathrm{eff}$, and $n_\mathrm{s}$, and show the full parameter constraints in Fig.~\ref{fig:HDdelens-BAO}. We see that the inclusion of DESI BAO data reduces uncertainties on $H_0$ and $n_\mathrm{s}$, but has less impact on $N_\mathrm{eff}$. }
    \label{fig:HDbao}
\end{figure}

\begin{figure*}
    \centering
    \includegraphics[width=\textwidth]{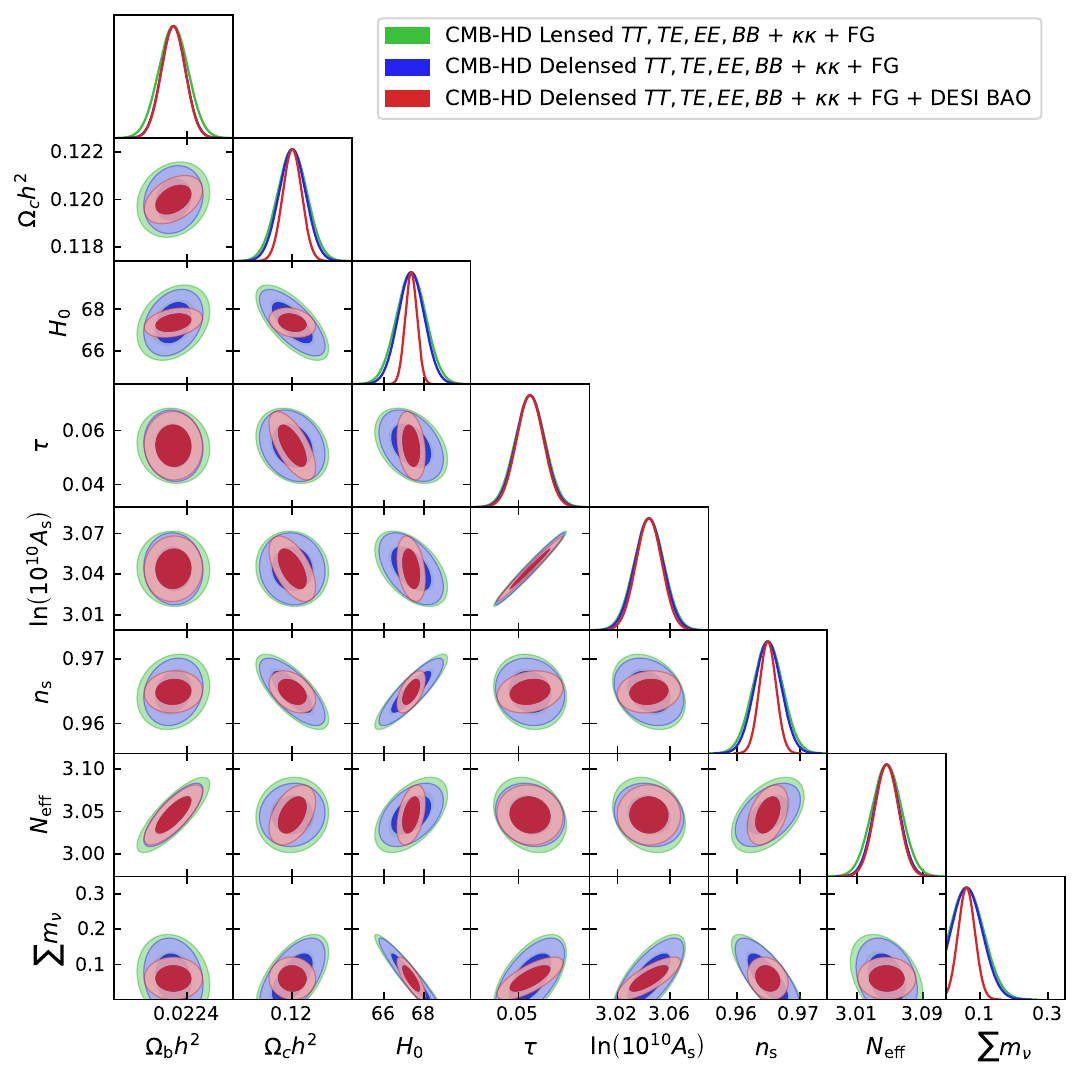}
    \caption{Shown are the forecasted  
    constraints for a $\Lambda$CDM+$N_\mathrm{eff}$+$\sum m_\nu$ model for CMB-HD.  Here we show constraints from lensed CMB $TT$, $TE$, $EE$, $BB$ and $\kappa\kappa$ power spectra (green), delensed CMB $TT$, $TE$, $EE$, $BB$ and $\kappa\kappa$ power spectra (blue), and delensed CMB $TT$, $TE$, $EE$, $BB$ and $\kappa\kappa$ power spectra plus DESI BAO~\protect{\cite{desi}} (red). We see that delensing the CMB spectra and adding DESI BAO data reduce uncertainties on cosmological parameters.} 
    \label{fig:HDdelens-BAO}
\end{figure*}

We show in Figs.~\ref{fig:HDbao} and~\ref{fig:HDdelens-BAO} the parameter constraints for CMB-HD delensed $TT, TE, EE, BB$ and $\kappa\kappa$ spectra, without (blue) and with (red) DESI BAO data.  We see that adding BAO data tightens all the parameter uncertainties, and most significantly impacts $\Omega_\mathrm{c} h^2$, $H_0$, $n_\mathrm{s}$, and $\sum m_\nu$.  The sixth column of Tables~\ref{tab:hd},\ref{tab:hdlcdm} and~\ref{tab:hdlcdmnnu} give the marginalized $1\sigma$ parameter errors when including DESI BAO for $\Lambda$CDM+$N_\mathrm{eff}$+$\sum m_\nu$, $\Lambda$CDM, and $\Lambda$CDM+$N_\mathrm{eff}$ models, respectively.

\subsection{Impact of Baryonic Physics}
\label{sec:paramsBaryons}

The low noise and high resolution of CMB-HD allows it to be sensitive to small scales ($k > 0.5~h$Mpc$^{-1}$) where the effects of baryonic feedback on the matter distribution become important.  On these scales, feedback from active galactic nuclei (AGN) emissions, for example, can push mass out from the centers of halos, while baryonic cores, formed from star-formation and cooling, concentrate mass in the center of halos.  These effects shift the mass distribution, which can be measured with CMB lensing. 

In addition, the hot gas traced by the thermal and kinetic Sunyaev-Zel'dovich effects (kSZ and tSZ, respectively) is also redistributed by baryonic feedback. This makes the kSZ and tSZ effects important external measurements of the amount of feedback that has occurred and the impact of that feedback.  In particular, cross-correlations between CMB lensing and the kSZ/tSZ effects can provide powerful additional constraints on the behavior of baryonic effects~\cite{Troster2021}. 

To model the impact of baryonic feedback, we use the updated \texttt{HMcode-2020} model from~\cite{Mead2020}, and in particular adopt their single-parameter baryonic feedback model characterized by the parameter $\mathrm{log}_{10}({T_\mathrm{AGN}}/{\mathrm{K}})$\footnote{Previous versions of HMCode did not include star-formation in their baryonic model, and only predicted power suppression~\cite{Mead2020}. The \texttt{HMcode-2020} model has six free parameters, but~\cite{Mead2020} find a single-parameter variant is only slightly less accurate (e.g.~see Fig~5 of~\cite{Mead2020}).}; we use a fiducial value of $\mathrm{log}_{10}({T_\mathrm{AGN}}/{\mathrm{K}}) = 7.8$ and a step size of $\pm 0.05$ in the Fisher analysis. This model was used in a recent analysis combining KiDS-1000 optical lensing data with tSZ measurements from {\it{Planck}} and ACT.  This analysis measured the cross-correlation between the lensing shear and tSZ maps, and measured $\mathrm{log}_{10}({T_\mathrm{AGN}}/{\mathrm{K}})$ with a $1\sigma$ uncertainty better than $6\%$~\cite{Troster2021}.

\begin{figure}[t]
    \centering
    \includegraphics[width=\columnwidth]{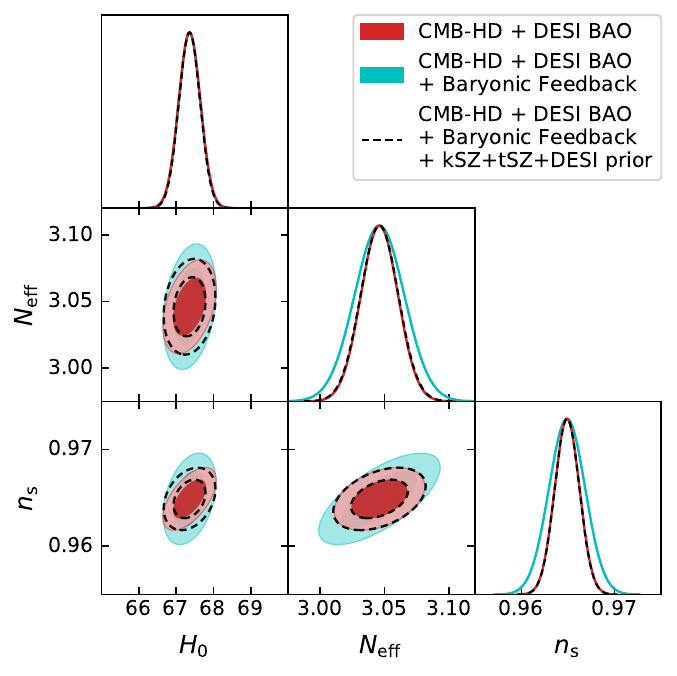}
    \caption{Shown is the impact of freeing the baryonic feedback parameter, $\mathrm{log}_{10}({T_\mathrm{AGN}}/{\mathrm{K}})$, discussed in Section~\ref{sec:paramsBaryons}, on the forecasted constraints for a $\Lambda$CDM+$N_\mathrm{eff}$+$\sum m_\nu$+baryonic feedback model (cyan) from CMB-HD delensed $TT$, $TE$, $EE$, $BB$ and $\kappa\kappa$ power spectra plus DESI BAO. We also show the parameter constraints when fixing this parameter (red) or adding a prior on the baryonic feedback parameter (dashed black); we anticipate such a prior can be obtained from a joint analysis of CMB-HD kSZ, tSZ, and lensing data (see Section~\ref{sec:paramsBaryons} for details). Here we show a subset of parameter constraints for $H_0$, $N_\mathrm{eff}$, and $n_\mathrm{s}$ from a Fisher analysis. }
    \label{fig:HDbaryons}
\end{figure}

We show in Fig.~\ref{fig:HDbaryons} the impact of freeing this baryonic feedback parameter on a subset of parameters for a $\Lambda$CDM+$N_\mathrm{eff}$+$\sum m_\nu$+baryonic feedback model from CMB-HD delensed $TT$, $TE$, $EE$, $BB$ and $\kappa\kappa$ power spectra plus DESI BAO (light blue).  We also show the marginalized $1\sigma$ parameter errors in the second column of Table~\ref{tab:baryonicfeedback}.  We see that the parameter errors do not increase substantially when freeing this baryonic parameter. We also see that CMB-HD can constrain this baryonic feedback parameter to an accuracy of 0.45\% without including any information from kSZ or tSZ.  This tight constraint on baryonic feedback is likely due to the precision CMB-HD will have in measuring the small-scale lensing power spectrum. 

We also add a prior on the baryonic feedback parameter $\mathrm{log}_{10}({T_\mathrm{AGN}}/{\mathrm{K}})$ that we can expect from CMB-HD measurements of the tSZ and kSZ effects in cross-correlation with CMB lensing.  We estimate this prior by extrapolating from the 6\% constraint measured by~\cite{Troster2021} using KiDS-1000 lensing maps cross-correlated with {\it{Planck}} plus ACT tSZ maps over 1000 square degrees of sky.  We note that CMB-HD will survey about 24,000 square degrees of sky, gaining about a factor of 5 improvement over the current feedback constraint.  In addition, the CMB-HD temperature map noise level will be over 20 times deeper than that of {\it{Planck}} plus ACT.  Moreover, the CMB-HD lensing map will be a factor of a few higher in signal-to-noise ratio than that of KiDS-1000, and we can anticipate also folding in information from the cross-correlation of kSZ maps with CMB lensing (not included in the KiDS-1000 analysis).  To be conservative, we assume an overall improvement of two orders of magnitude in ability to constrain $\mathrm{log}_{10}({T_\mathrm{AGN}}/{\mathrm{K}})$ from CMB-HD tSZ and kSZ cross-correlations with CMB lensing compared to the current constraint. (As mentioned above, CMB-HD already can improve the constraint on $\mathrm{log}_{10}({T_\mathrm{AGN}}/{\mathrm{K}})$ by an order of magnitude without any SZ data.) Thus we apply a 0.06\% prior on this baryonic feedback parameter, and show the results in Fig.~\ref{fig:HDbaryons} (dashed black) and Table~\ref{tab:baryonicfeedback} (third column). We see that the inclusion of this additional prior from SZ data returns the parameter errors to what they were before freeing the feedback parameter (shown in the first column of Table~\ref{tab:baryonicfeedback}).

\begin{figure*}
    \centering
    \includegraphics[width=\textwidth]{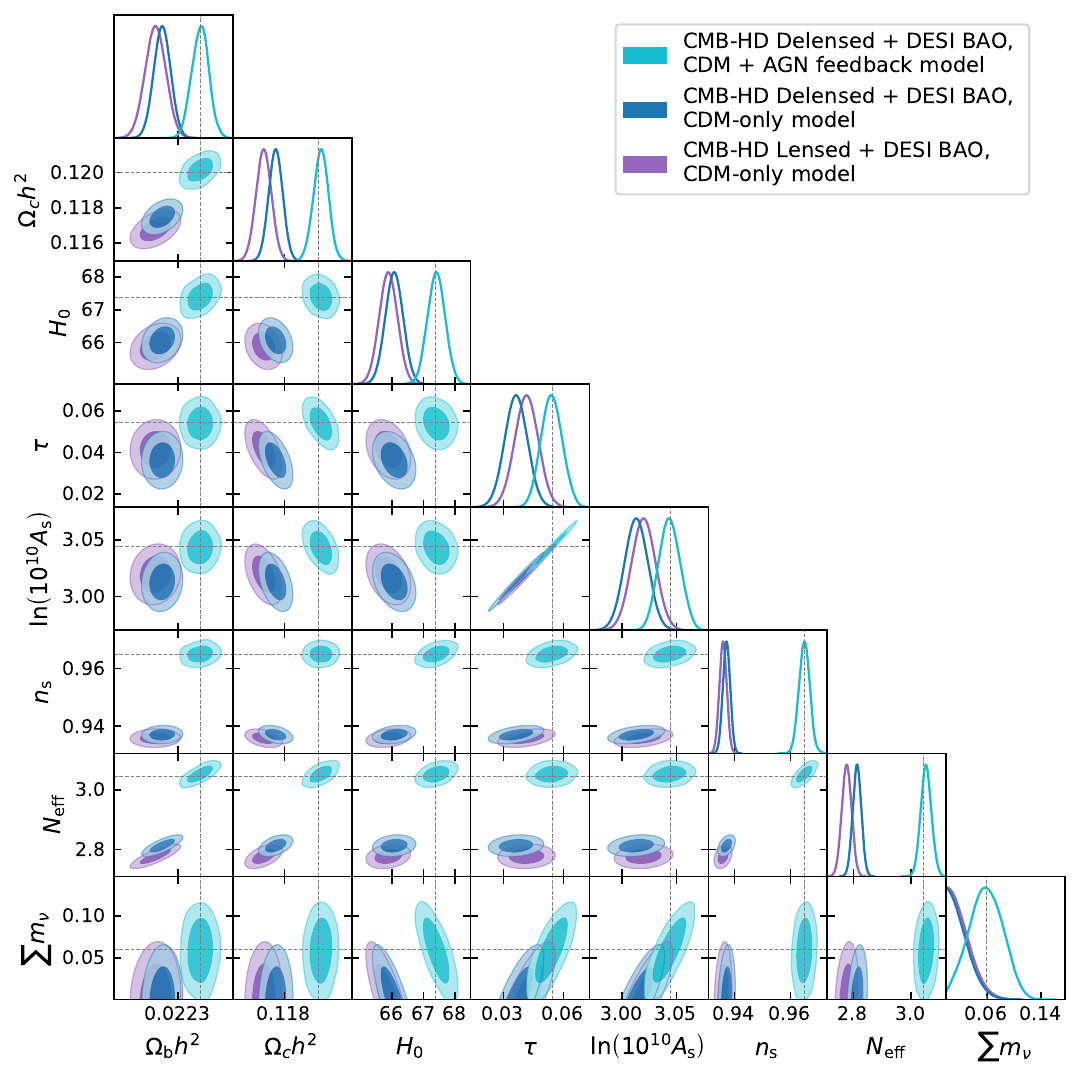}
    \caption{Here we show the expected bias to the parameter constraints from a combination of lensed or delensed (purple and blue, respectively) CMB-HD $TT$, $TE$, $EE$, $BB$ and $\kappa\kappa$ power spectra plus DESI BAO, when baryonic feedback effects are neglected. These biases were computed from Eq.~\ref{eq:bias}, using a CDM-only model as the fiducial model, and assuming the true model includes the effects of baryonic feedback. We center the CDM-only contours at the estimated biased parameter values, and take the error ellipses from the Fisher estimates.    
    In cyan we show the parameter constraints for delensed CMB-HD plus DESI BAO data from an MCMC run where we marginalize over the feedback parameter $\log_{10}\left(T_\mathrm{AGN}/\mathrm{K}\right)$ discussed in Section~\ref{sec:paramsBaryons}.  We see that marginalization over feedback models removes the parameter biases. }
    \label{fig:feedback_bias}
\end{figure*}

We use Eq.~\ref{eq:bias} in Section~\ref{sec:fisher} to predict the bias on the estimated cosmological parameters when AGN feedback is neglected in the model. In this case, we allow $C_\ell^\mathrm{fid}$ to be the CDM-only model whereas the true model $C_\ell^\mathrm{true}$ is CDM+AGN feedback with $\mathrm{log}_{10}({T_\mathrm{AGN}}/{\mathrm{K}}) = 7.8$.  We also use the combination of CMB-HD and DESI BAO data to calculate the Fisher matrix, $F_{\alpha\beta}$, in Eq.~\ref{eq:bias}.
In Fig.~\ref{fig:feedback_bias}, we show the forecasted parameter biases from the combination of lensed (purple) or delensed (blue) CMB-HD and DESI BAO data when assuming a fiducial CDM-only model, as opposed to the true CDM+feedback model.  We can see in Fig.~\ref{fig:feedback_bias} that, while delensing reduces the bias due to an incorrect baryonic feedback model (as discussed in~\cite{mccarthy22, McCarthy2020, Hotinli2021}), for many parameters the remaining bias is much larger than the statistical $1\sigma$ error.  Given that the biases appear quite large, Eq.~\ref{eq:bias}, which assumes small deviations from the true model, may not accurately quantify the bias; however, Fig.~\ref{fig:feedback_bias} gives an indication that qualitatively the biases can be significant.

We also show in Fig.~\ref{fig:feedback_bias} the result of an MCMC run when marginalizing over the baryonic feedback parameter $\log_{10}\left(T_\mathrm{AGN}/\mathrm{K}\right)$ discussed above, applying only a uniform prior of $\log_{10}\left(T_\mathrm{AGN}/\mathrm{K}\right) \in [7.6,~8.0]$ as suggested by~\cite{Mead2020}.  We see that this marginalization over feedback models removes the parameter biases at the expense of some constraining power.  However, as we show in Fig.~\ref{fig:HDbaryons}, adding a prior on the baryonic feedback parameter anticipated from CMB-HD SZ measurements can mitigate this increase in parameter error.  We note that the MCMC run with the baryonic feedback parameter free takes about the same time to converge on NERSC as a run with the feedback parameter fixed.

\begin{table}[t]
    \begin{center}
    \begin{tabular}{l@{\hskip 1.5em} l@{\hskip 1.5em}  l@{\hskip 1.5em} l}
      \toprule
      \toprule
      Parameter & CDM & + feedback & + SZ prior 
      \\
      \midrule
      $\Omega_\mathrm{b} h^2$\dotfill & $0.000026$ & $0.000028$ & $0.000027$
      \\
      $\Omega_\mathrm{c} h^2$\dotfill & $0.00041$ & $0.00046$ & $0.00041$
      \\
      $\ln(10^{10} A_\mathrm{s})$\dotfill & $0.0095$ & $0.010$ & $0.010$
      \\
      $n_\mathrm{s}$\dotfill & $0.0013$ & $0.0021$ & $0.0013$
      \\
      $\tau$\dotfill & $0.0051$ & $0.0056$ & $0.0056$
      \\
      $100\theta_\mathrm{MC}$ & $0.000058$ & $0.000064$ & $0.000060$
      \\
      $N_\mathrm{eff}$\dotfill & $0.014$ & $0.020$  & $0.014$
      \\
      $\sum m_\nu$ [eV]\dotfill & $0.024$ & $0.026$ & $0.027$
      \\
      $\log_{10}\left(T_\mathrm{AGN} / \mathrm{K}\right)$ \dotfill & --- & $0.040$ & $0.0047$
      \\
      \midrule
      $H_0$ [km s$^{-1}$ Mpc$^{-1}$]\dotfill & $0.28$ & $0.29$  & $0.27$
      \\
      $\sigma_8$\dotfill & $0.0033$ & $0.0033$  & $0.0033$
      \\
      \bottomrule
    \end{tabular}
    \caption{Shown are the cosmological constraints from the combination of CMB-HD delensed $TT$, $TE$, $EE$, $BB$, and $\kappa\kappa$ power spectra with DESI BAO for the baseline $\Lambda$CDM + $N_\mathrm{eff}$ + $\sum m_\nu$ model, a $\Lambda$CDM + $N_\mathrm{eff}$ + $\sum m_\nu$ + baryonic feedback model from \texttt{HMCode-2020}~\protect{\cite{Mead2020}}, and a $\Lambda$CDM + $N_\mathrm{eff}$ + $\sum m_\nu$ + baryonic feedback model including a 0.06\% prior on the feedback parameter $\mathrm{log}_{10}({T_\mathrm{AGN}}/{\mathrm{K}})$ expected from from a joint analysis of CMB-HD kSZ, tSZ, and lensing data (see Section~\ref{sec:paramsBaryons} for details). All results shown here are from the likelihood and MCMC chains as opposed to Fisher forecasts. We also include $100 \theta_\mathrm{MC}$, and separate the two derived parameters, $H_0$ and $\sigma_8$, which is the linear root-mean-square matter fluctuations today.}
    %\vspace{-2mm}
    \label{tab:baryonicfeedback}
    \end{center}
\end{table}
%\vspace{-3mm}

\section{Discussion}
\label{sec:discussion}

In this work, we present the parameter forecasts for a CMB-HD survey.  We contrast these forecasts with precursor CMB experiments, showing that the lower noise and higher multipoles of a CMB-HD survey can lead of significant improvement in many of the cosmological parameters, most notably in the scalar spectral index, $n_\mathrm{s}$, and in the effective number of light relativistic species, $N_\mathrm{eff}$.  Specifically we find for delensed CMB-HD $TT,TE,EE,BB$ and $\kappa\kappa$ spectra plus DESI BAO: 
\begin{equation*}
     \sigma\left(n_\mathrm{s}\right) = 0.0013  \quad  (68\%,~TT,TE,EE,BB,\kappa\kappa+\mathrm{BAO}),
\end{equation*}
\begin{equation*}
    \sigma\left(N_\mathrm{eff}\right) = 0.014  \quad (68\%,~TT,TE,EE,BB,\kappa\kappa+\mathrm{BAO}).
\end{equation*}
We find that delensing all the CMB spectra as well as including DESI BAO data are both necessary to achieve these tight constraints.  
We also find that including multipoles out to $\ell_\mathrm{max}, L_\mathrm{max} = 20,000$ is also required.  Since these multipoles are well into the non-linear regime, we must include baryonic effects in modelling the lensing in both the CMB power spectra and CMB lensing spectra.  We find that marginalizing over baryonic effects can mitigate potential bias in parameters at the expense of some constraining power. However, tSZ and kSZ measurements by CMB-HD offer an independent handle on baryonic feedback effects, and folding in that additional information can effectively eliminate the increase in parameter errors due to uncertainty in the baryonic physics. 

The $N_\mathrm{eff}$ parameter uncertainty achieved by CMB-HD+DESI is particularly interesting since any new light particle that was in thermal equilibrium 
at any time after the Universe reheated must change $N_\mathrm{eff}$ by at least $\Delta N_\mathrm{eff} \geq 0.027$~\cite{Wallisch2018, Green2019}.  Here 
reheating refers to the end of inflation and the beginning of the ``Big Bang''.  With $\sigma\left(N_\mathrm{eff}\right)=0.014$, CMB-HD+DESI can either rule out or detect any new light particle species with at least 95\% confidence.  

As a specific example of why this $N_\mathrm{eff}$ constraint is valuable, we consider the QCD axion, which is a well-motivated candidate for being the dark matter  and solving the Strong CP problem~\cite{Peccei2006,Marsh2017,DiLuzio2020}.  We note that if the reheating temperature of the Universe is high enough that the QCD axion thermalized, the $N_\mathrm{eff}$ constraint above can potentially rule out the QCD axion in a model-independent way, or lead to a detection.  

\begin{figure}[t]
    \centering
    \includegraphics[width=\columnwidth]{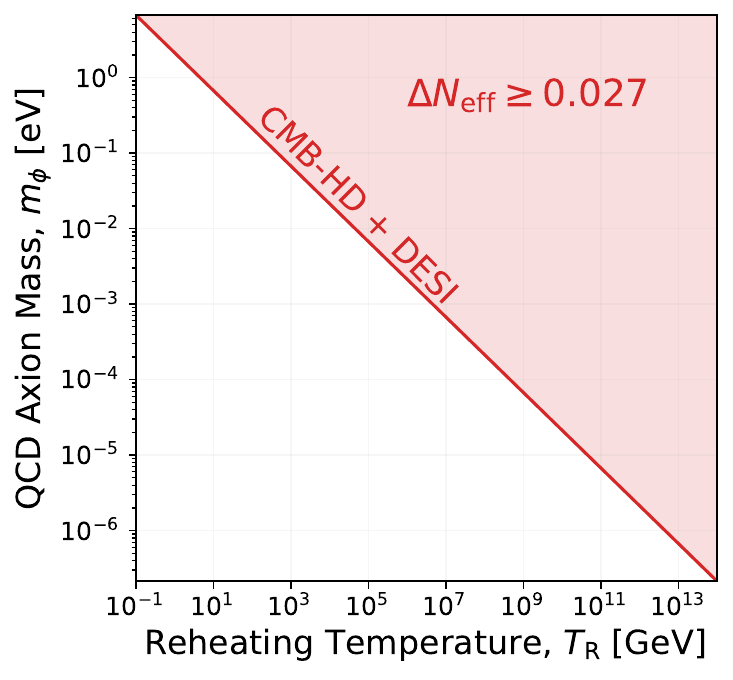}
    \caption{Here we show constraints on the QCD axion mass from delensed CMB-HD plus DESI BAO data. The shaded region shows the region of axion mass, $m_\phi$, and reheating temperature, $T_\mathrm{R}$, that would be excluded if any measurement rules out $\Delta N_\mathrm{eff} \geq 0.027$, which we show in this work CMB-HD+DESI BAO does at the 95\% confidence level.}
    \label{fig:QCDaxion}
\end{figure}

We re-plot Fig.~3 of~\cite{Baumann2016} in Fig.~\ref{fig:QCDaxion} to highlight the QCD axion masses, $m_\phi$, that would be ruled out as a function of reheating temperature, $T_\mathrm{R}$, if CMB-HD+DESI finds no increase in $N_\mathrm{eff}$ at the 95\% confidence level.  To forecast these constraints, we use Eq.~2.11 of~\cite{Baumann2016} to calculate the upper limit of the QCD axion coupling $g_d$ for a given reheating temperature,
\begin{equation}
        g_d < 1.3 \times 10^{-14}~\mathrm{GeV}^{-2} \left(\frac{T_\mathrm{R}}{10^{10}~\mathrm{GeV}}\right)^{-1/2},
\end{equation}
and use Eq.~16 of~\cite{Blum2014} to relate this to the coupling constant $f_a$ via
\begin{equation}
        g_d \approx \frac{2.4 \times 10^{-16}~\mathrm{e}~\mathrm{cm}}{f_a},
\end{equation}
where $1~\mathrm{cm}^{-1} = 1.97 \times 10^{-14}~\mathrm{GeV}$ and $\mathrm{e} = 0.3$.
We then relate the coupling constant to the axion mass using 
\begin{equation}
    m_\phi = 0.6~\mathrm{eV} \left(\frac{10^7~\mathrm{GeV}}{f_a}\right)
\end{equation}
from~\cite{Peccei2006,Hannestad2007,DiValentino2015}.
Given the well-motivated nature of the QCD axion and the many efforts to detect it underway, it is worth highlighting this model-independent additional approach to probing this new physics.

We also note that the $n_\mathrm{s}$ constraint above is about a factor of two tighter than from precursor CMB surveys.  While considerable attention has been devoted to improving constraints on primordial gravitational waves via the tensor-to-scalar ratio, $r$, constraining the scalar spectral index can also rule out interesting inflationary scenarios~\cite{S4forecast}.

\section{Conclusion}
\label{sec:conclusion}

While we present parameter forecasts for CMB-HD above, we note that there are several ways that these forecasts can be improved and made more robust.  As we have stressed a few times, the most challenging aspect of a CMB-HD survey will be removing and mitigating the impact of extragalactic foregrounds.  Demonstrating that this can be achieved with realistic end-to-end simulations is an area of ongoing research, and will be critical for achieving the science presented here.  In addition, more optimal lensing estimators have been proposed and are being explored~\cite{Horowitz:2017iql,Schaan:2018tup,Hadzhiyska2019,Millea:2020iuw,Madhavacheril:2020ido,Legrand:2021qdu,Sailer:2022jwt,Chan2023,Legrand:2023jne}, with the potential to yield higher lensing signal-to-noise ratios than assumed in this work, as well as to provide better paths towards foreground immunity.  We also leave to future work exploring the potential gain in lensing signal-to-noise ratio from exploiting the higher-order lensing corrections contained in the N1 signal.  Furthermore, we have only explored cold dark matter models in this work, with and without baryonic feedback, but note that CMB-HD has sensitivity to alternate dark matter models as well, which will be investigated in 
subsequent work. 

We make the Fisher estimation code used here public along with a Jupyter notebook detailing how to generate new Fisher derivatives for different models.  We also make the likelihood code public and integrate it with Cobaya using CAMB. We hope that this will facilitate additional cosmological parameter forecasts for a CMB-HD survey.

\begin{acknowledgments}
The authors thank Dongwon Han, Gil Holder, Selim Hotinli, Mathew Madhavacheril, Joel Meyers, Vivian Miranda, Rugved Pund, Cynthia Trendafilova, and Alexander van Engelen for useful discussions.  NS also thanks Itay Bloch, Rouven Essig, Peter Graham, Maxim Pospelov, Mauro Valli, and the SCGP Lighting New Lampposts Workshop for useful discussions about the QCD axion.  AM, NS, and MR acknowledge support from DOE award number DE-SC0020441. AM and NS also acknowledge support from NSF award number 1907657.  This research used resources of the National Energy Research Scientific Computing Center (NERSC), a U.S. Department of Energy Office of Science User Facility located at Lawrence Berkeley National Laboratory, operated under Contract No. DE-AC02-05CH11231 using NERSC award HEP-ERCAPmp107. NS also acknowledges the Aspen Center for Physics, which is supported by National Science Foundation grant PHY-1607611, for supporting a workshop on new discoveries in the era of high-resolution, low-noise CMB experiments, which stimulated useful discussions. 
\end{acknowledgments}

%\clearpage

\appendix

\section{Einstein-Boltzmann Code Accuracy Requirements}
\label{app:accuracy}

As discussed in~\cite{mccarthy22}, future high-resolution and low noise CMB experiments will require theory calculations for power spectra beyond the standard default CAMB accuracy settings for the Einstein-Boltzmann solver.  To determine the accuracy settings required for CMB-HD, we focus on the parameter bias induced by modelling the expected data with accuracy settings that are lower than necessary.  For this we use Eq.~\ref{eq:bias} discussed in Section~\ref{sec:fisher}, and we explore four accuracy parameters provided by CAMB (i.e.~\texttt{lens\_potential\_accuracy}, \texttt{AccuracyBoost}, \texttt{lAccuracyBoost}, and \texttt{lSampleBoost}).  We also by default set \texttt{NonLinear=model.NonLinear\_both} to enable the non-linear calculation of both the matter power spectrum and the lensing spectrum.  We note that as long as non-linear corrections are included, changing the four CAMB accuracy settings above does not change the error bars obtained from either Fisher or MCMC analyses, as long as the mock data is generated with the same accuracy.

\begin{figure}[t]
    \centering
    \includegraphics[width=0.5\textwidth]{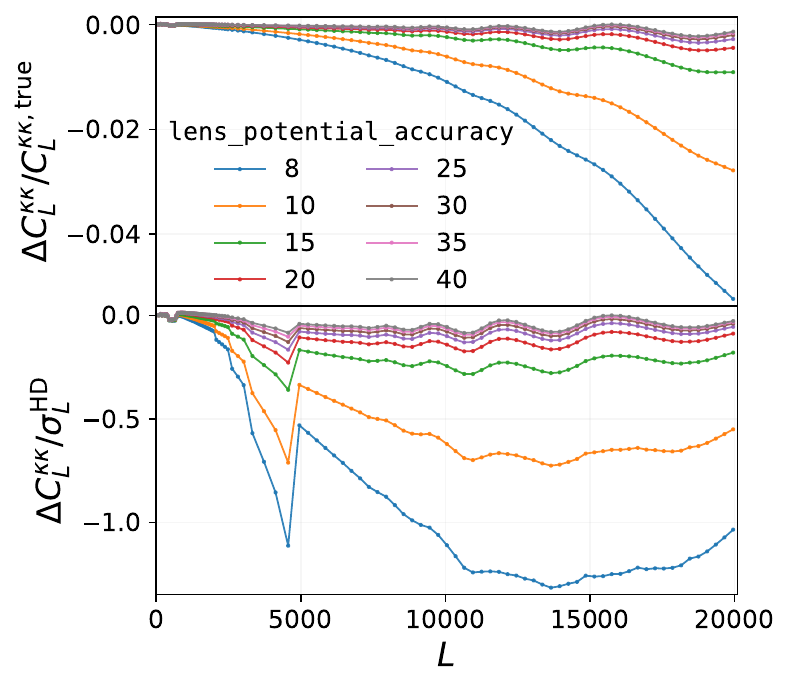}
    \caption{Here we compare the lensing power spectrum for different values of the CAMB accuracy parameter \texttt{lens\_potential\_accuracy}.  The top panel shows the fractional difference between $C_L^{\kappa\kappa}$ and $C_L^{\kappa\kappa,\mathrm{true}}$, 
    while the bottom panel shows the difference as a fraction of the CMB-HD error bar on the lensing power at each multipole, $\sigma_L^\mathrm{HD}$. $C_L^{\kappa\kappa,\mathrm{true}}$ uses the accuracy settings we describe in the text for our approximation of the ``true Universe''.  We see that the lensing power spectrum converges to the ``true Universe'' model when \texttt{lens\_potential\_accuracy} is increased, and for \texttt{lens\_potential\_accuracy > 30} the difference between the spectra is less than 10\% of the CMB-HD error bar for a given $L$.  Here we fix the other parameters to \texttt{AccuracyBoost = 1.1}, \texttt{lAccuracyBoost = 3.0}, and \texttt{lSampleBoost = 3.0}. Since the ``true Universe'' model has \texttt{AccuracyBoost = 3.0}, \texttt{lAccuracyBoost = 5.0}, and \texttt{lSampleBoost = 5.0}, we see that varying these other parameters within these ranges has minimal impact given the CMB-HD error bars.}
    \label{fig:lens_acc}
\end{figure}

We find that the \texttt{lens\_potential\_accuracy} setting has the largest impact on parameter biases.  We show in the top panel of Fig.~\ref{fig:lens_acc}, the change in the lensing power spectrum for different values of \texttt{lens\_potential\_accuracy}, noting that for higher accuracy settings most of the change is at high lensing multipoles.  We show in the bottom panel of Fig.~\ref{fig:lens_acc}, the change in the lensing power spectrum divided by the expected CMB-HD uncertainty per lensing multipole.  We find that above \texttt{lens\_potential\_accuracy~=~30}, the lensing power spectrum theory curves converge with a difference smaller than a tenth of the CMB-HD error bar per multipole.  Thus we take \texttt{lens\_potential\_accuracy~=~40} to be the ``true Universe'' model for this work, and compare our results to this model.

We find that after setting \texttt{lens\_potential\_accuracy = 40}, varying the \texttt{lAccuracyBoost} and \texttt{lSampleBoost} parameters results in parameter biases that are well below $1\sigma$ of the expected CMB-HD+BAO parameter  errors for a $\Lambda$CDM + $N_\mathrm{eff}$ + $\sum m_\nu$ model.  The default CAMB settings for these parameters are 1.0, and we increase the settings to 5.0 for each for our ``true Universe'' model to be conservative.

\begin{figure}[t]
    \centering
    \includegraphics[width=0.5\textwidth]{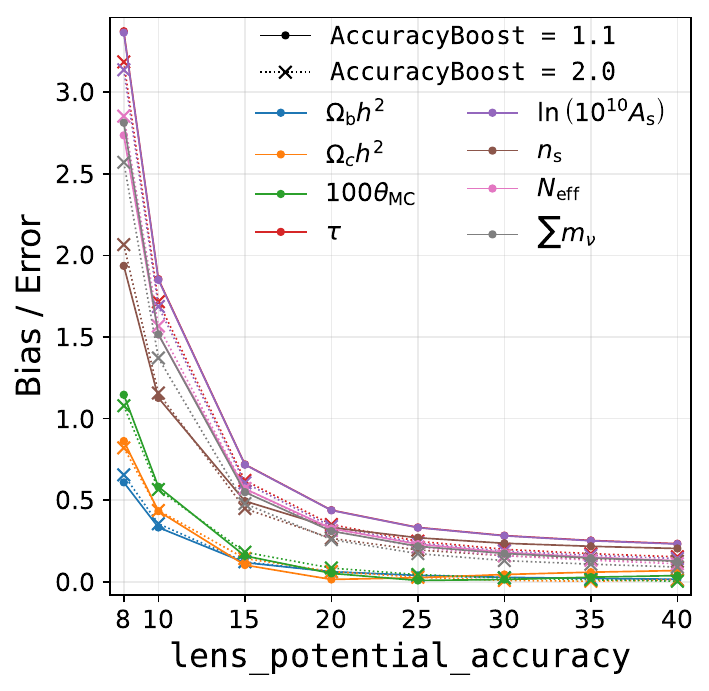}
    \caption{Here we show the expected bias on each cosmological parameter as a fraction of its $1\sigma$ error for CMB-HD+DESI when we calculate the delensed CMB and lensing power spectra with different values of the CAMB accuracy parameters \texttt{lens\_potential\_accuracy} and \texttt{AccuracyBoost}.
    The other parameters are fixed to values of  \texttt{lAccuracyBoost = 3.0} and \texttt{lSampleBoost = 3.0}, with the ``true'' model in Eq.~\ref{eq:bias} calculated with the accuracy settings \texttt{AccuracyBoost = 3.0}, \texttt{lAccuracyBoost = 5.0}, and \texttt{lSampleBoost = 5.0} (see text). We see that the bias converges to below $0.3\sigma$ at \texttt{lens\_potential\_accuracy = 30}, and that increasing the \texttt{AccuracyBoost} does not significantly decrease the bias.}
    \label{fig:biases}
\end{figure}  

In addition, we find that varying the \texttt{AccuracyBoost} parameter, also minimally impacts parameter biases, and also results in biases well below $1\sigma$ of the expected CMB-HD+BAO parameter errors.  The CAMB default setting is \texttt{AccuracyBoost~=~1.0}, and we increase this to \texttt{AccuracyBoost~=~3.0} for our ``true Universe'' model.

We show in Fig.~\ref{fig:biases}, the bias on each parameter divided by the expected CMB-HD+BAO $1\sigma$ parameter error for a $\Lambda$CDM + $N_\mathrm{eff}$ + $\sum m_\nu$ model.  We use Eq.~\ref{eq:bias} to calculate each parameter bias, using the ``true Universe'' model described above as $C_\ell^\mathrm{true}$.  We see that \texttt{lens\_potential\_accuracy~=~30} yields parameter biases that are less than  0.3$\sigma$.  We also find that varying the \texttt{AccuracyBoost} has minimal impact.

\begin{figure}[t]
    \centering
    \includegraphics[width=0.5\textwidth]{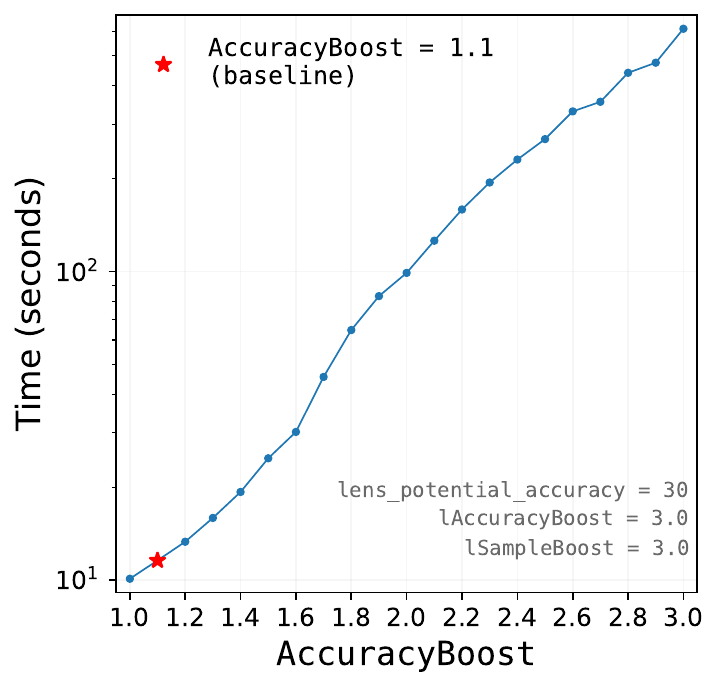}
    \caption{Here we show the increase in the total CAMB computation time to compute the lensed and delensed CMB spectra plus the CMB lensing power spectrum for a maximum $\ell/L$ of 20,000 when the \texttt{AccuracyBoost} parameter is increased. Here the other accuracy parameters are fixed to \texttt{lAccuracyBoost = 3.0}, \texttt{lSampleBoost = 3.0}, and \texttt{lens\_potential\_accuracy = 30}. The computation times were measured on a NERSC \textit{Perlmutter} login node. The red star indicates the baseline accuracy settings used throughout this work, which shows that all the spectra mentioned above can be calculated in a total time of 11 seconds.}
    \label{fig:times}
\end{figure}

In addition, we find that varying the \texttt{AccuracyBoost} has the most significant impact on the computation time. We show in Fig.~\ref{fig:times} the rapid increase in computation time with increasing \texttt{AccuracyBoost}.  Since we see from Fig.~\ref{fig:biases} that an \texttt{AccuracyBoost~=~1.1} yields similar biases to a value double that, we choose as our baseline setting \texttt{AccuracyBoost~=~1.1}; we note that \texttt{AccuracyBoost~=~1.0} is the CAMB default, and we find a significant reduction in bias for a small increase above that.  Similarly, we find little reduction in bias for values of \texttt{lAccuracyBoost} and \texttt{lSampleBoost}  above 3.0.

Thus, we use as our baseline CAMB settings for CMB-HD in this work:
\begin{lstlisting}[language=Python]
import camb
import numpy as np
lmax = 20100 
pars = camb.CAMBparams()
pars.set_cosmology(H0=67.36, ombh2=0.02237, omch2=0.1200, tau=0.0544, num_massive_neutrinos=1, mnu=0.06, nnu=3.046)
pars.InitPower.set_params(As=np.exp(3.044) * 1e-10, ns=0.9649)
pars.set_for_lmax(int(lmax)+500, lens_potential_accuracy=30, lens_margin=2050)
pars.set_accuracy(AccuracyBoost=1.1, lSampleBoost=3.0, lAccuracyBoost=3.0, DoLateRadTruncation=False)
pars.NonLinear = camb.model.NonLinear_both
pars.NonLinearModel.set_params("mead2016")
\end{lstlisting}
We find that the computation of the lensed and delensed $TT, TE, EE, BB$ CMB power spectra plus the CMB lensing power spectrum takes 11 seconds total to run on the NERSC \textit{Perlmutter} machine using the CAMB accuracy settings above.

We also run an MCMC to verify the accuracy of the Fisher estimated biases for our baseline settings. We do this by generating mock data at the accuracy settings of the ``true Universe'' model described above, and running the MCMC with our baseline accuracy settings.  We find consistency between the two methods of bias estimation, and confirm that the parameter biases are well below $1\sigma$ of the expected CMB-HD+BAO parameter errors.

\section{Consistency of Fisher and MCMC Methods}
\label{app:mcmc_vs_fisher}

\begin{figure*}
    \centering
    \includegraphics[width=\textwidth]{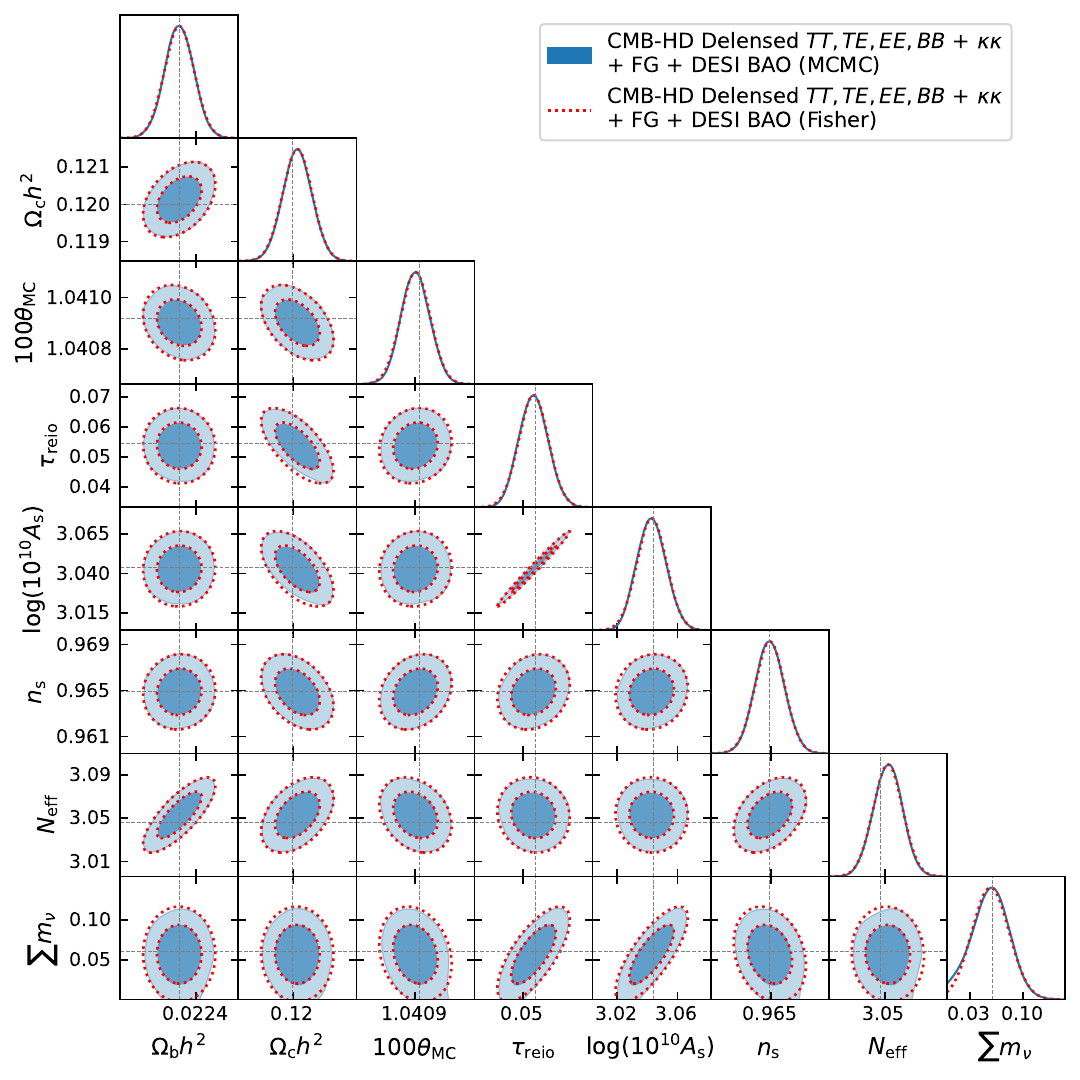}
    \caption{Here we show the parameter constraints for a $\Lambda$CDM + $N_\mathrm{eff}$ + $\sum m_\nu$ model obtained from delensed CMB-HD $TT$, $TE$, $EE$, $BB$ and $\kappa\kappa$ power spectra plus DESI BAO. The Fisher forecasts are shown in red (dotted lines), and the results from running MCMC chains are shown in blue (solid lines/shaded contours). The parameter values used to generate the data are indicated by the grey dashed lines.  We find good agreement between parameter error estimates based on Fisher and MCMC analyses.}
    \label{fig:mcmc_vs_fisher}
\end{figure*}

To verify that our Fisher forecasts accurately predict the expected parameter errors, we create a likelihood and run MCMC chains. In this likelihood, we set the central values of the CMB-HD bandpowers to those of the fiducial model (see footnote~\ref{MCMCbandpowers} in Section~\ref{sec:forecastMethods}), and use the covariance matrix described in Section~\ref{sec:covmat}. In Fig.~\ref{fig:mcmc_vs_fisher} we show as solid lines/shaded contours our MCMC run for the baseline CMB-HD delensed $TT$, $TE$, $EE$, $BB$ and $\kappa\kappa$ power spectra (including foregrounds), combined with mock DESI BAO data.  We see that the MCMC recovers the input values of the data well, which are indicated by the grey dashed lines.  We also overlay the results of the Fisher estimation method as red dotted contours.  We find good agreement between the two methods.  
We also compare the $1\sigma$ parameter errors from each method in Table~\ref{tab:fisher_vs_mcmc}, finding good consistency.  Given the consistency of both methods, we use the Fisher method for all parameter forecasts in this work, unless stated otherwise.

\section{Additional Parameter Forecasts} \label{app:hdparams}

Below we show parameter constraints for CMB-HD in a $\Lambda$CDM model with fixed $N_\mathrm{eff} = 3.046$ and $\sum m_\nu = 0.06$~eV (Table~\ref{tab:hdlcdm}), and in a $\Lambda$CDM + $N_\mathrm{eff}$  model with fixed $\sum m_\nu = 0.06$~eV (Table~\ref{tab:hdlcdmnnu}).  We see that $\sigma(N_\mathrm{eff})$ and $\sigma(n_\mathrm{s})$ do not change when neutrino mass is fixed in the $\Lambda$CDM + $N_\mathrm{eff}$ model compared to when it is free. In the $\Lambda$CDM model, $\sigma(n_\mathrm{s})$ decreases slightly compared to the models where $N_\mathrm{eff}$ is free.  $\sigma(H_0)$ is about a factor of two smaller in the $\Lambda$CDM model compared to the $\Lambda$CDM + $N_\mathrm{eff}$ + $\sum m_\nu$.

In order to see how much improvement is gained from delensing even after adding DESI BAO, we compare the $1\sigma$ parameter uncertainties obtained from lensed or delensed CMB-HD $TT$, $TE$, $EE$, and $BB$ power spectra plus the lensing $\kappa\kappa$ spectrum and DESI BAO data in Table~\ref{tab:lens_vs_delens}.  We find that delensing does improve parameter constraints, with a 20\% improvement for $N_\mathrm{eff}$ which is important for pushing the error well below the 0.027 threshold for a spin-0 particle~\cite{S4forecast}.

Finally, in Table~\ref{tab:params_lmax}, we show the forecasted $1\sigma$ parameter constraints from CMB-HD plus DESI BAO data for different values of the maximum multipole $\ell_\mathrm{max}$ or $L_\mathrm{max}$ used in the analysis.  At higher multipoles the data is more sensitive to non-linear effects in the matter power spectrum, which are more complex to model.  Thus we show the parameter constraints achievable from the linear and semi-linear regimes.  We see that to obtain $\sigma(n_\mathrm{s}) = 0.0013$ and $\sigma(N_\mathrm{eff}) = 0.014$ requires the full range of multipoles out to 20,000.

%%% fisher vs mcmc
\begin{table}[t]
    \begin{center}
    \begin{tabular}{l@{\hskip 1em} l@{\hskip 1em}  l@{\hskip 1em}  l@{\hskip 1em} c}
      \toprule
      \toprule
      Parameter & Fiducial & MCMC & Fisher & Fisher / MCMC 
      \\
       & & $1\sigma$ error & $1\sigma$ error &
      \\
      \midrule
      $\Omega_\mathrm{b} h^2$\dotfill & $0.022370$ & $0.0000258$ & $0.0000263$ & $1.02$
      \\
      $\Omega_\mathrm{c} h^2$\dotfill & $0.12000$ & $0.000410$ & $0.000414$ & $1.01$
      \\
      $\ln(10^{10} A_\mathrm{s})$\dotfill & $3.044$ & $0.00953$ & $0.00974$ & $1.02$
      \\
      $n_\mathrm{s}$\dotfill & $0.9649$ & $0.00133$ & $0.00134$ & $1.01$
      \\
      $\tau$\dotfill & $0.0544$ & $0.00505$ & $0.00515$ & $1.02$
      \\
      $100 \theta_\mathrm{MC}$\dotfill & $1.040920$ & $0.0000583$ & $0.0000598$ & $1.03$
      \\
      $N_\mathrm{eff}$\dotfill & $3.046$ & $0.0139$ & $0.0143$ & $1.03$
      \\
      $\sum m_\nu$ [eV]\dotfill & $0.060$ & $0.0243$ & $0.0244$ & $1.00$
      \\
      \bottomrule
    \end{tabular}
    \caption{Here we show the $1\sigma$ uncertainties for a $\Lambda$CDM + $N_\mathrm{eff}$ + $\sum m_\nu$ model from  delensed CMB-HD $TT$, $TE$, $EE$, $BB$ and $\kappa\kappa$ power spectra plus DESI BAO. The first two columns list the parameter and its fiducial value; the third column lists the marginalized $1\sigma$ uncertainty on each parameter obtained from an MCMC run, and the fourth column lists the uncertainty estimated from a Fisher analysis. In the last column we take the ratio of the two errors, finding good agreement.}
    \label{tab:fisher_vs_mcmc}
    \end{center}
\end{table}

%%% HD LCDM + DESI BAO
\begin{table*}[h]
    \centering
    \begin{tabular}{l@{\hskip 2em} c@{\hskip 2em} l@{\hskip 1.5em} l@{\hskip 1.5em} l@{\hskip 1.5em} l@{\hskip 2em} l}
      \toprule
      \toprule
        &  & \multicolumn{4}{c}{$\ell_\mathrm{max}, ~L_\mathrm{max} = 20,000$}  &  $\ell_\mathrm{max}, L_\mathrm{max} = 10,000$
      \\
      \cmidrule(r){3-6} \cmidrule(r){7-7}
      Parameter & Fiducial & HD Lensed & HD Lensed &  HD Delensed & \textbf{HD Delensed}  & HD Delensed 
      \\
       \multicolumn{1}{l}{$\Lambda$CDM} & & &  + FG & + FG  & \textbf{+ FG + DESI BAO} & + FG + DESI BAO 
      \\
     \midrule
      $\Omega_\mathrm{b} h^2$\dotfill & $0.022370$ & $0.000017$ &  $0.000018$ & $0.000017$ & $0.000017$ & $0.000017$
      \\
      $\Omega_\mathrm{c} h^2$\dotfill & $0.12000$ & $0.00046$ & $0.00048$ & $0.00045$ &  $0.00037$ & $0.00038$
      \\
      $\ln(10^{10} A_\mathrm{s})$\dotfill & $3.0440$ & $0.0078$ & $0.0080$ & $0.0074$ &   $0.0062$ & $0.0063$
      \\
      $n_\mathrm{s}$\dotfill & $0.9649$ & $0.0013$ & $0.0014$ & $0.0013$ & $0.0012$ & $0.0013$
      \\
      $\tau$\dotfill & $0.0544$ & $0.0045$ & $0.0046$ & $0.0043$ & $0.0036$ & $0.0036$
      \\
      $H_0$ [km s$^{-1}$ Mpc$^{-1}$]\dotfill & $67.36$ & $0.17$ & $0.18$ & $0.17$ &  $0.14$ & $0.14$
      \\
      \bottomrule
    \end{tabular}
    \caption{Forecasted cosmological parameter constraints from CMB-HD $TT$, $TE$, $EE$, $BB$ and $\kappa\kappa$ power spectra for a $\Lambda$CDM model. All forecasts include a $\tau$ prior of $\tau = 0.054 \pm 0.007$ from \textit{Planck}~\protect{\cite{planck18params}}. 
    The first two columns list the parameters and their fiducial values. The following two columns list their forecasted marginalized $1\sigma$ uncertainties when using lensed spectra with or without foregrounds in the temperature maps. The fifth column shows the forecast when delensing the CMB spectra, and the sixth column shows the change when including DESI BAO data~\protect{\cite{desi}}.  In each of these cases we use CMB and CMB lensing multipoles out to $\ell_\mathrm{max}, L_\mathrm{max} =20,000$ for CMB-HD. The last column lists the same information as the sixth column, but instead using a maximum multipole of $\ell_\mathrm{max}, L_\mathrm{max} = 10,000$ for both CMB and CMB lensing power spectra.}
    \label{tab:hdlcdm}
\end{table*}

%%% HD LCDM+N_eff + DESI BAO
\begin{table*}[h]
    \centering
    \begin{tabular}{l@{\hskip 2em} c@{\hskip 2em} l@{\hskip 1.5em} l@{\hskip 1.5em} l@{\hskip 1.5em} l@{\hskip 2em} l}
      \toprule
      \toprule
        &  & \multicolumn{4}{c}{ $\ell_\mathrm{max}, L_\mathrm{max} = 20,000$}  &  $\ell_\mathrm{max}, L_\mathrm{max} = 10,000$
      \\
      \cmidrule(r){3-6} \cmidrule(r){7-7}
      Parameter & Fiducial & HD Lensed & HD Lensed &  HD Delensed & \textbf{HD Delensed}  & HD Delensed 
      \\
       \multicolumn{1}{l}{$\Lambda$CDM+$N_\mathrm{eff}$} & & &  + FG & + FG  & \textbf{+ FG + DESI BAO} & + FG + DESI BAO 
      \\
     \midrule
      $\Omega_\mathrm{b} h^2$\dotfill & $0.022370$ & $0.000032$ & $0.000033$ & $0.000027$ &  $0.000026$ & $0.000026$
      \\
      $\Omega_\mathrm{c} h^2$\dotfill & $0.12000$ & $0.00051$ & $0.00052$ & $0.00047$ & $0.00041$ & $0.00041$
      \\
      $\ln(10^{10} A_\mathrm{s})$\dotfill & $3.0440$ & $0.0078$ & $0.0080$ & $0.0075$ &  $0.0062$ & $0.0063$
      \\
      $n_\mathrm{s}$\dotfill & $0.9649$ & $0.0013$ & $0.0015$ & $0.0015$ &  $0.0013$ & $0.0014$
      \\
      $\tau$\dotfill & $0.0544$ & $0.0045$ & $0.0046$ & $0.0043$ &  $0.0036$ & $0.0036$
      \\
      $H_0$ [km s$^{-1}$ Mpc$^{-1}$]\dotfill & $67.36$ & $0.21$ & $0.22$ & $0.21$ &  $0.18$ & $0.19$
      \\
      $N_\mathrm{eff}$\dotfill & $3.046$ & $0.016$ & $0.017$ & $0.015$ &  $0.014$ & $0.015$
      \\
      \bottomrule
    \end{tabular}
    \caption{Forecasted cosmological parameter constraints from CMB-HD $TT$, $TE$, $EE$, $BB$ and $\kappa\kappa$ power spectra on a $\Lambda$CDM + $N_{\rm{eff}}$ model. All forecasts include a $\tau$ prior of $\tau = 0.054 \pm 0.007$ from \textit{Planck}~\protect{\cite{planck18params}}. The first two columns list the parameters and their fiducial values. The following two columns list their forecasted marginalized $1\sigma$ uncertainties when using lensed spectra with or without foregrounds in the temperature maps. The fifth column shows the forecast when delensing the CMB spectra, and the sixth column shows the change when including DESI BAO data~\protect{\cite{desi}}.   In each of these cases we use CMB and CMB lensing multipoles out to $\ell_\mathrm{max}, L_\mathrm{max} =20,000$ for CMB-HD. The last column lists the same information as the sixth column, but instead using a maximum multipole of $\ell_\mathrm{max}, L_\mathrm{max} = 10,000$ for both CMB and CMB lensing power spectra.  }
    \label{tab:hdlcdmnnu}
\end{table*}

% HD lensed vs. delensed
\begin{table*}[h]
    \begin{center}
    \begin{tabular}{l@{\hskip 2em} l@{\hskip 2em} l@{\hskip 1.5em} l}
      \toprule
      \toprule
      Parameter & Fiducial &  HD Lensed & HD Delensed  
      \\
      $\Lambda$CDM+$N_\mathrm{eff}$+$\sum m_\nu$ &  &  + FG + DESI BAO & + FG + DESI BAO  
      \\
      \midrule
      $\Omega_\mathrm{b} h^2$\dotfill & $0.022370$ & $0.000032$ & $0.000026$  
      \\
      $\Omega_\mathrm{c} h^2$\dotfill & $0.12000$ & $0.00045$ & $0.00041$  
      \\
      $\ln(10^{10} A_\mathrm{s})$\dotfill & $3.044$ & $0.010$ & $0.0098$  
      \\
      $n_\mathrm{s}$\dotfill & $0.9649$ & $0.0014$ & $0.0013$  
      \\
      $\tau$\dotfill & $0.0544$ & $0.0055$ & $0.0052$  
      \\
      $H_0$ [km s$^{-1}$ Mpc$^{-1}$]\dotfill & $67.36$ & $0.30$ & $0.29$  
      \\
      $N_\mathrm{eff}$\dotfill & $3.046$ & $0.017$ & $0.014$  
      \\
      $\sum m_\nu$ [eV]\dotfill & $0.06$ & $0.025$ & $0.025$  
      \\
      \bottomrule
    \end{tabular}
    \caption{Here we compare the forecasted $1\sigma$ parameter constraints from lensed (third column) and delensed (fourth column) CMB-HD $TT$, $TE$, $EE$, $BB$ and $\kappa\kappa$ power spectra when including DESI BAO in order to see how much improvement is gained from delensing alone. The parameter names and fiducial values are listed in the first and second columns, respectively. We find that delensing improves parameter constraints even after including DESI BAO, with the most improvement seen for $N_\mathrm{eff}$.}
    \label{tab:lens_vs_delens}
    \end{center}
\end{table*}

%%%%% HD for different lmax
\begin{table*}[h]
    \begin{center}
    \begin{tabular}{l@{\hskip 2em} l@{\hskip 2em} l@{\hskip 1.5em} l@{\hskip 1.5em} l@{\hskip 1.5em} l@{\hskip 2em} l}
      \toprule
      \toprule
      Parameter &  &   \multicolumn{5}{c}{$\ell_\mathrm{max},~L_\mathrm{max}$ for HD Delensed + FG + DESI BAO} 
      \\
      \cmidrule(r){3-7} 
       \multicolumn{1}{l}{$\Lambda$CDM+$N_\mathrm{eff}$+$\sum m_\nu$} & Fiducial & 1000 & 3000 & 5000 & 10,000 & 20,000 
      \\
      \midrule
      $\Omega_\mathrm{b} h^2$\dotfill & $0.022370$ & $0.00016$ & $0.000039$  & $0.000027$ & $0.000026$ & $0.000026$
      \\
      $\Omega_\mathrm{c} h^2$\dotfill & $0.12000$ & $0.0024$ & $0.00054$  & $0.00042$ & $0.00041$ & $0.00041$
      \\
      $\ln(10^{10} A_\mathrm{s})$\dotfill & $3.044$ & $0.014$ & $0.011$  & $0.011$ & $0.010$ & $0.0098$
      \\
      $n_\mathrm{s}$\dotfill & $0.9649$ & $0.0048$ & $0.0023$  & $0.0018$ & $0.0014$ & $0.0013$
      \\
      $\tau$\dotfill & $0.0544$ & $0.0065$ & $0.0060$  & $0.0057$ & $0.0054$ & $0.0052$
      \\
      $H_0$ [km s$^{-1}$ Mpc$^{-1}$]\dotfill & $67.36$ & $0.80$ & $0.32$  & $0.29$ & $0.29$ & $0.29$
      \\
      $N_\mathrm{eff}$\dotfill & $3.046$ & $0.15$ & $0.030$  & $0.018$ & $0.015$ & $0.014$
      \\
      $\sum m_\nu$ [eV]\dotfill & $0.06$ & $0.037$ & $0.029$  & $0.028$ & $0.026$ & $0.025$
      \\
      \bottomrule
    \end{tabular}
    \caption{Shown are the forecasted $1\sigma$ uncertainties from delensed CMB-HD $TT$, $TE$, $EE$, $BB$ and $\kappa\kappa$ spectra plus DESI BAO, when varying the maximum multipole $\ell_\mathrm{max}$ or $L_\mathrm{max}$ for the CMB spectra. The parameter names and fiducial values are listed in the first and second columns, respectively, while the remaining columns list the $1\sigma$ parameter errors for the given maximum multipole. We see that to obtain $\sigma(n_\mathrm{s}) = 0.0013$ and $\sigma(N_\mathrm{eff}) = 0.014$, for example, requires the full range of multipoles out to 20,000.}
    \label{tab:params_lmax}
    \end{center}
\end{table*}

\clearpage

\bibliographystyle{apsrev4-1}
\bibliography{main.bib}

%merlin.mbs apsrev4-1.bst 2010-07-25 4.21a (PWD, AO, DPC) hacked
%Control: key (0)
%Control: author (72) initials jnrlst
%Control: editor formatted (1) identically to author
%Control: production of article title (-1) disabled
%Control: page (0) single
%Control: year (1) truncated
%Control: production of eprint (0) enabled
\begin{thebibliography}{74}%
\makeatletter
\providecommand \@ifxundefined [1]{%
 \@ifx{#1\undefined}
}%
\providecommand \@ifnum [1]{%
 \ifnum #1\expandafter \@firstoftwo
 \else \expandafter \@secondoftwo
 \fi
}%
\providecommand \@ifx [1]{%
 \ifx #1\expandafter \@firstoftwo
 \else \expandafter \@secondoftwo
 \fi
}%
\providecommand \natexlab [1]{#1}%
\providecommand \enquote  [1]{``#1''}%
\providecommand \bibnamefont  [1]{#1}%
\providecommand \bibfnamefont [1]{#1}%
\providecommand \citenamefont [1]{#1}%
\providecommand \href@noop [0]{\@secondoftwo}%
\providecommand \href [0]{\begingroup \@sanitize@url \@href}%
\providecommand \@href[1]{\@@startlink{#1}\@@href}%
\providecommand \@@href[1]{\endgroup#1\@@endlink}%
\providecommand \@sanitize@url [0]{\catcode `\\12\catcode `\$12\catcode
  `\&12\catcode `\#12\catcode `\^12\catcode `\_12\catcode `\%12\relax}%
\providecommand \@@startlink[1]{}%
\providecommand \@@endlink[0]{}%
\providecommand \url  [0]{\begingroup\@sanitize@url \@url }%
\providecommand \@url [1]{\endgroup\@href {#1}{\urlprefix }}%
\providecommand \urlprefix  [0]{URL }%
\providecommand \Eprint [0]{\href }%
\providecommand \doibase [0]{http://dx.doi.org/}%
\providecommand \selectlanguage [0]{\@gobble}%
\providecommand \bibinfo  [0]{\@secondoftwo}%
\providecommand \bibfield  [0]{\@secondoftwo}%
\providecommand \translation [1]{[#1]}%
\providecommand \BibitemOpen [0]{}%
\providecommand \bibitemStop [0]{}%
\providecommand \bibitemNoStop [0]{.\EOS\space}%
\providecommand \EOS [0]{\spacefactor3000\relax}%
\providecommand \BibitemShut  [1]{\csname bibitem#1\endcsname}%
\let\auto@bib@innerbib\@empty
%</preamble>
\bibitem [{\citenamefont {Aghanim}\ \emph {et~al.}(2020)\citenamefont {Aghanim}
  \emph {et~al.}}]{planck18params}%
  \BibitemOpen
  \bibfield  {author} {\bibinfo {author} {\bibfnamefont {N.}~\bibnamefont
  {Aghanim}} \emph {et~al.} (\bibinfo {collaboration} {Planck}),\ }\href
  {\doibase 10.1051/0004-6361/201833910} {\bibfield  {journal} {\bibinfo
  {journal} {\aap}\ }\textbf {\bibinfo {volume} {641}},\ \bibinfo {pages} {A6}
  (\bibinfo {year} {2020})},\ \bibinfo {note} {[Erratum: Astron.Astrophys. 652,
  C4 (2021)]},\ \Eprint {http://arxiv.org/abs/1807.06209} {arXiv:1807.06209
  [astro-ph.CO]} \BibitemShut {NoStop}%
\bibitem [{\citenamefont {Aiola}\ \emph {et~al.}(2020)\citenamefont {Aiola}
  \emph {et~al.}}]{Aiola2020}%
  \BibitemOpen
  \bibfield  {author} {\bibinfo {author} {\bibfnamefont {S.}~\bibnamefont
  {Aiola}} \emph {et~al.} (\bibinfo {collaboration} {ACT}),\ }\href {\doibase
  10.1088/1475-7516/2020/12/047} {\bibfield  {journal} {\bibinfo  {journal}
  {\jcap}\ }\textbf {\bibinfo {volume} {12}},\ \bibinfo {pages} {047} (\bibinfo
  {year} {2020})},\ \Eprint {http://arxiv.org/abs/2007.07288} {arXiv:2007.07288
  [astro-ph.CO]} \BibitemShut {NoStop}%
\bibitem [{\citenamefont {Balkenhol}\ \emph {et~al.}(2023)\citenamefont
  {Balkenhol} \emph {et~al.}}]{SPT-3G:2022hvq}%
  \BibitemOpen
  \bibfield  {author} {\bibinfo {author} {\bibfnamefont {L.}~\bibnamefont
  {Balkenhol}} \emph {et~al.} (\bibinfo {collaboration} {SPT-3G}),\ }\href
  {\doibase 10.1103/PhysRevD.108.023510} {\bibfield  {journal} {\bibinfo
  {journal} {Phys. Rev. D}\ }\textbf {\bibinfo {volume} {108}},\ \bibinfo
  {pages} {023510} (\bibinfo {year} {2023})},\ \Eprint
  {http://arxiv.org/abs/2212.05642} {arXiv:2212.05642 [astro-ph.CO]}
  \BibitemShut {NoStop}%
\bibitem [{\citenamefont {Ade}\ \emph {et~al.}(2019)\citenamefont {Ade} \emph
  {et~al.}}]{SOforecast}%
  \BibitemOpen
  \bibfield  {author} {\bibinfo {author} {\bibfnamefont {P.}~\bibnamefont
  {Ade}} \emph {et~al.} (\bibinfo {collaboration} {Simons Observatory}),\
  }\href {\doibase 10.1088/1475-7516/2019/02/056} {\bibfield  {journal}
  {\bibinfo  {journal} {\jcap}\ }\textbf {\bibinfo {volume} {02}},\ \bibinfo
  {pages} {056} (\bibinfo {year} {2019})},\ \Eprint
  {http://arxiv.org/abs/1808.07445} {arXiv:1808.07445 [astro-ph.CO]}
  \BibitemShut {NoStop}%
\bibitem [{\citenamefont {Abazajian}\ \emph {et~al.}(2019)\citenamefont
  {Abazajian} \emph {et~al.}}]{S4forecast}%
  \BibitemOpen
  \bibfield  {author} {\bibinfo {author} {\bibfnamefont {K.}~\bibnamefont
  {Abazajian}} \emph {et~al.},\ }\href@noop {} {\  (\bibinfo {year} {2019})},\
  \Eprint {http://arxiv.org/abs/1907.04473} {arXiv:1907.04473 [astro-ph.IM]}
  \BibitemShut {NoStop}%
\bibitem [{\citenamefont {Sehgal}\ \emph {et~al.}(2019)\citenamefont {Sehgal}
  \emph {et~al.}}]{HDastro2020}%
  \BibitemOpen
  \bibfield  {author} {\bibinfo {author} {\bibfnamefont {N.}~\bibnamefont
  {Sehgal}} \emph {et~al.},\ }\href@noop {} {\  (\bibinfo {year} {2019})},\
  \Eprint {http://arxiv.org/abs/1906.10134} {arXiv:1906.10134 [astro-ph.CO]}
  \BibitemShut {NoStop}%
\bibitem [{\citenamefont {Aiola}\ \emph {et~al.}(2022)\citenamefont {Aiola}
  \emph {et~al.}}]{HDsnowmass}%
  \BibitemOpen
  \bibfield  {author} {\bibinfo {author} {\bibfnamefont {S.}~\bibnamefont
  {Aiola}} \emph {et~al.} (\bibinfo {collaboration} {CMB-HD}),\ }\href@noop {}
  {\  (\bibinfo {year} {2022})},\ \Eprint {http://arxiv.org/abs/2203.05728}
  {arXiv:2203.05728 [astro-ph.CO]} \BibitemShut {NoStop}%
\bibitem [{\citenamefont {Han}\ \emph {et~al.}(2021)\citenamefont {Han} \emph
  {et~al.}}]{han20delensing}%
  \BibitemOpen
  \bibfield  {author} {\bibinfo {author} {\bibfnamefont {D.}~\bibnamefont
  {Han}} \emph {et~al.} (\bibinfo {collaboration} {ACT}),\ }\href {\doibase
  10.1088/1475-7516/2021/01/031} {\bibfield  {journal} {\bibinfo  {journal}
  {\jcap}\ }\textbf {\bibinfo {volume} {01}},\ \bibinfo {pages} {031} (\bibinfo
  {year} {2021})},\ \Eprint {http://arxiv.org/abs/2007.14405} {arXiv:2007.14405
  [astro-ph.CO]} \BibitemShut {NoStop}%
\bibitem [{\citenamefont {Millea}\ \emph {et~al.}(2021)\citenamefont {Millea}
  \emph {et~al.}}]{Millea:2020iuw}%
  \BibitemOpen
  \bibfield  {author} {\bibinfo {author} {\bibfnamefont {M.}~\bibnamefont
  {Millea}} \emph {et~al.},\ }\href {\doibase 10.3847/1538-4357/ac02bb}
  {\bibfield  {journal} {\bibinfo  {journal} {Astrophys. J.}\ }\textbf
  {\bibinfo {volume} {922}},\ \bibinfo {pages} {259} (\bibinfo {year}
  {2021})},\ \Eprint {http://arxiv.org/abs/2012.01709} {arXiv:2012.01709
  [astro-ph.CO]} \BibitemShut {NoStop}%
\bibitem [{\citenamefont {Carron}\ \emph {et~al.}(2017)\citenamefont {Carron},
  \citenamefont {Lewis},\ and\ \citenamefont {Challinor}}]{Carron:2017vfg}%
  \BibitemOpen
  \bibfield  {author} {\bibinfo {author} {\bibfnamefont {J.}~\bibnamefont
  {Carron}}, \bibinfo {author} {\bibfnamefont {A.}~\bibnamefont {Lewis}}, \
  and\ \bibinfo {author} {\bibfnamefont {A.}~\bibnamefont {Challinor}},\ }\href
  {\doibase 10.1088/1475-7516/2017/05/035} {\bibfield  {journal} {\bibinfo
  {journal} {\jcap}\ }\textbf {\bibinfo {volume} {05}},\ \bibinfo {pages} {035}
  (\bibinfo {year} {2017})},\ \Eprint {http://arxiv.org/abs/1701.01712}
  {arXiv:1701.01712 [astro-ph.CO]} \BibitemShut {NoStop}%
\bibitem [{\citenamefont {Green}\ \emph {et~al.}(2017)\citenamefont {Green},
  \citenamefont {Meyers},\ and\ \citenamefont {van Engelen}}]{Green2017}%
  \BibitemOpen
  \bibfield  {author} {\bibinfo {author} {\bibfnamefont {D.}~\bibnamefont
  {Green}}, \bibinfo {author} {\bibfnamefont {J.}~\bibnamefont {Meyers}}, \
  and\ \bibinfo {author} {\bibfnamefont {A.}~\bibnamefont {van Engelen}},\
  }\href {\doibase 10.1088/1475-7516/2017/12/005} {\bibfield  {journal}
  {\bibinfo  {journal} {\jcap}\ }\textbf {\bibinfo {volume} {12}},\ \bibinfo
  {pages} {005} (\bibinfo {year} {2017})},\ \Eprint
  {http://arxiv.org/abs/1609.08143} {arXiv:1609.08143 [astro-ph.CO]}
  \BibitemShut {NoStop}%
\bibitem [{\citenamefont {Hotinli}\ \emph {et~al.}(2022)\citenamefont
  {Hotinli}, \citenamefont {Meyers}, \citenamefont {Trendafilova},
  \citenamefont {Green},\ and\ \citenamefont {van Engelen}}]{Hotinli2021}%
  \BibitemOpen
  \bibfield  {author} {\bibinfo {author} {\bibfnamefont {S.~C.}\ \bibnamefont
  {Hotinli}}, \bibinfo {author} {\bibfnamefont {J.}~\bibnamefont {Meyers}},
  \bibinfo {author} {\bibfnamefont {C.}~\bibnamefont {Trendafilova}}, \bibinfo
  {author} {\bibfnamefont {D.}~\bibnamefont {Green}}, \ and\ \bibinfo {author}
  {\bibfnamefont {A.}~\bibnamefont {van Engelen}},\ }\href {\doibase
  10.1088/1475-7516/2022/04/020} {\bibfield  {journal} {\bibinfo  {journal}
  {\jcap}\ }\textbf {\bibinfo {volume} {04}},\ \bibinfo {pages} {020} (\bibinfo
  {year} {2022})},\ \Eprint {http://arxiv.org/abs/2111.15036} {arXiv:2111.15036
  [astro-ph.CO]} \BibitemShut {NoStop}%
\bibitem [{\citenamefont {Chung}\ \emph {et~al.}(2020)\citenamefont {Chung},
  \citenamefont {Foreman},\ and\ \citenamefont {van Engelen}}]{Chung2019}%
  \BibitemOpen
  \bibfield  {author} {\bibinfo {author} {\bibfnamefont {E.}~\bibnamefont
  {Chung}}, \bibinfo {author} {\bibfnamefont {S.}~\bibnamefont {Foreman}}, \
  and\ \bibinfo {author} {\bibfnamefont {A.}~\bibnamefont {van Engelen}},\
  }\href {\doibase 10.1103/PhysRevD.101.063534} {\bibfield  {journal} {\bibinfo
   {journal} {Phys. Rev. D}\ }\textbf {\bibinfo {volume} {101}},\ \bibinfo
  {pages} {063534} (\bibinfo {year} {2020})},\ \bibinfo {note} {[Erratum:
  Phys.Rev.D 102, 109903 (2020)]},\ \Eprint {http://arxiv.org/abs/1910.09565}
  {arXiv:1910.09565 [astro-ph.CO]} \BibitemShut {NoStop}%
\bibitem [{\citenamefont {McCarthy}\ \emph {et~al.}(2022)\citenamefont
  {McCarthy}, \citenamefont {Hill},\ and\ \citenamefont
  {Madhavacheril}}]{mccarthy22}%
  \BibitemOpen
  \bibfield  {author} {\bibinfo {author} {\bibfnamefont {F.}~\bibnamefont
  {McCarthy}}, \bibinfo {author} {\bibfnamefont {J.~C.}\ \bibnamefont {Hill}},
  \ and\ \bibinfo {author} {\bibfnamefont {M.~S.}\ \bibnamefont
  {Madhavacheril}},\ }\href {\doibase 10.1103/PhysRevD.105.023517} {\bibfield
  {journal} {\bibinfo  {journal} {Phys. Rev. D}\ }\textbf {\bibinfo {volume}
  {105}},\ \bibinfo {pages} {023517} (\bibinfo {year} {2022})},\ \Eprint
  {http://arxiv.org/abs/2103.05582} {arXiv:2103.05582 [astro-ph.CO]}
  \BibitemShut {NoStop}%
\bibitem [{\citenamefont {Han}\ and\ \citenamefont {Sehgal}(2022)}]{han22}%
  \BibitemOpen
  \bibfield  {author} {\bibinfo {author} {\bibfnamefont {D.}~\bibnamefont
  {Han}}\ and\ \bibinfo {author} {\bibfnamefont {N.}~\bibnamefont {Sehgal}},\
  }\href {\doibase 10.1103/PhysRevD.105.083516} {\bibfield  {journal} {\bibinfo
   {journal} {Phys. Rev. D}\ }\textbf {\bibinfo {volume} {105}},\ \bibinfo
  {pages} {083516} (\bibinfo {year} {2022})},\ \Eprint
  {http://arxiv.org/abs/2112.02109} {arXiv:2112.02109 [astro-ph.CO]}
  \BibitemShut {NoStop}%
\bibitem [{\citenamefont {Aghamousa}\ \emph {et~al.}(2016)\citenamefont
  {Aghamousa} \emph {et~al.}}]{desi}%
  \BibitemOpen
  \bibfield  {author} {\bibinfo {author} {\bibfnamefont {A.}~\bibnamefont
  {Aghamousa}} \emph {et~al.} (\bibinfo {collaboration} {DESI}),\ }\href@noop
  {} {\  (\bibinfo {year} {2016})},\ \Eprint {http://arxiv.org/abs/1611.00036}
  {arXiv:1611.00036 [astro-ph.IM]} \BibitemShut {NoStop}%
\bibitem [{\citenamefont {{Torrado}}\ and\ \citenamefont
  {{Lewis}}(2021)}]{cobaya}%
  \BibitemOpen
  \bibfield  {author} {\bibinfo {author} {\bibfnamefont {J.}~\bibnamefont
  {{Torrado}}}\ and\ \bibinfo {author} {\bibfnamefont {A.}~\bibnamefont
  {{Lewis}}},\ }\href {\doibase 10.1088/1475-7516/2021/05/057} {\bibfield
  {journal} {\bibinfo  {journal} {\jcap}\ }\textbf {\bibinfo {volume} {2021}},\
  \bibinfo {eid} {057} (\bibinfo {year} {2021})},\ \Eprint
  {http://arxiv.org/abs/2005.05290} {arXiv:2005.05290 [astro-ph.IM]}
  \BibitemShut {NoStop}%
\bibitem [{\citenamefont {Lewis}\ \emph {et~al.}(2000)\citenamefont {Lewis},
  \citenamefont {Challinor},\ and\ \citenamefont {Lasenby}}]{CAMBLewis:1999bs}%
  \BibitemOpen
  \bibfield  {author} {\bibinfo {author} {\bibfnamefont {A.}~\bibnamefont
  {Lewis}}, \bibinfo {author} {\bibfnamefont {A.}~\bibnamefont {Challinor}}, \
  and\ \bibinfo {author} {\bibfnamefont {A.}~\bibnamefont {Lasenby}},\ }\href
  {\doibase 10.1086/309179} {\bibfield  {journal} {\bibinfo  {journal}
  {Astrophys. J.}\ }\textbf {\bibinfo {volume} {538}},\ \bibinfo {pages} {473}
  (\bibinfo {year} {2000})},\ \Eprint {http://arxiv.org/abs/astro-ph/9911177}
  {arXiv:astro-ph/9911177} \BibitemShut {NoStop}%
\bibitem [{\citenamefont {Howlett}\ \emph {et~al.}(2012)\citenamefont
  {Howlett}, \citenamefont {Lewis}, \citenamefont {Hall},\ and\ \citenamefont
  {Challinor}}]{CAMBHowlett:2012mh}%
  \BibitemOpen
  \bibfield  {author} {\bibinfo {author} {\bibfnamefont {C.}~\bibnamefont
  {Howlett}}, \bibinfo {author} {\bibfnamefont {A.}~\bibnamefont {Lewis}},
  \bibinfo {author} {\bibfnamefont {A.}~\bibnamefont {Hall}}, \ and\ \bibinfo
  {author} {\bibfnamefont {A.}~\bibnamefont {Challinor}},\ }\href {\doibase
  10.1088/1475-7516/2012/04/027} {\bibfield  {journal} {\bibinfo  {journal}
  {\jcap}\ }\textbf {\bibinfo {volume} {1204}},\ \bibinfo {pages} {027}
  (\bibinfo {year} {2012})},\ \Eprint {http://arxiv.org/abs/1201.3654}
  {arXiv:1201.3654 [astro-ph.CO]} \BibitemShut {NoStop}%
%%CITATION = ARXIV:1201.3654;%%
\bibitem [{\citenamefont {Mandal}\ \emph {et~al.}(2022)\citenamefont {Mandal},
  \citenamefont {Sehgal},\ and\ \citenamefont {Namikawa}}]{Mandal:2022tqu}%
  \BibitemOpen
  \bibfield  {author} {\bibinfo {author} {\bibfnamefont {S.}~\bibnamefont
  {Mandal}}, \bibinfo {author} {\bibfnamefont {N.}~\bibnamefont {Sehgal}}, \
  and\ \bibinfo {author} {\bibfnamefont {T.}~\bibnamefont {Namikawa}},\ }\href
  {\doibase 10.1103/PhysRevD.105.063537} {\bibfield  {journal} {\bibinfo
  {journal} {Phys. Rev. D}\ }\textbf {\bibinfo {volume} {105}},\ \bibinfo
  {pages} {063537} (\bibinfo {year} {2022})},\ \Eprint
  {http://arxiv.org/abs/2201.02204} {arXiv:2201.02204 [astro-ph.CO]}
  \BibitemShut {NoStop}%
\bibitem [{\citenamefont {Mead}\ \emph {et~al.}(2015)\citenamefont {Mead},
  \citenamefont {Peacock}, \citenamefont {Heymans}, \citenamefont {Joudaki},\
  and\ \citenamefont {Heavens}}]{Mead2015}%
  \BibitemOpen
  \bibfield  {author} {\bibinfo {author} {\bibfnamefont {A.}~\bibnamefont
  {Mead}}, \bibinfo {author} {\bibfnamefont {J.}~\bibnamefont {Peacock}},
  \bibinfo {author} {\bibfnamefont {C.}~\bibnamefont {Heymans}}, \bibinfo
  {author} {\bibfnamefont {S.}~\bibnamefont {Joudaki}}, \ and\ \bibinfo
  {author} {\bibfnamefont {A.}~\bibnamefont {Heavens}},\ }\href {\doibase
  10.1093/mnras/stv2036} {\bibfield  {journal} {\bibinfo  {journal} {\mnras}\
  }\textbf {\bibinfo {volume} {454}},\ \bibinfo {pages} {1958} (\bibinfo {year}
  {2015})},\ \Eprint {http://arxiv.org/abs/1505.07833} {arXiv:1505.07833
  [astro-ph.CO]} \BibitemShut {NoStop}%
\bibitem [{\citenamefont {Mead}\ \emph {et~al.}(2016)\citenamefont {Mead},
  \citenamefont {Heymans}, \citenamefont {Lombriser}, \citenamefont {Peacock},
  \citenamefont {Steele},\ and\ \citenamefont {Winther}}]{Mead2016}%
  \BibitemOpen
  \bibfield  {author} {\bibinfo {author} {\bibfnamefont {A.}~\bibnamefont
  {Mead}}, \bibinfo {author} {\bibfnamefont {C.}~\bibnamefont {Heymans}},
  \bibinfo {author} {\bibfnamefont {L.}~\bibnamefont {Lombriser}}, \bibinfo
  {author} {\bibfnamefont {J.}~\bibnamefont {Peacock}}, \bibinfo {author}
  {\bibfnamefont {O.}~\bibnamefont {Steele}}, \ and\ \bibinfo {author}
  {\bibfnamefont {H.}~\bibnamefont {Winther}},\ }\href {\doibase
  10.1093/mnras/stw681} {\bibfield  {journal} {\bibinfo  {journal} {\mnras}\
  }\textbf {\bibinfo {volume} {459}},\ \bibinfo {pages} {1468} (\bibinfo {year}
  {2016})},\ \Eprint {http://arxiv.org/abs/1602.02154} {arXiv:1602.02154
  [astro-ph.CO]} \BibitemShut {NoStop}%
\bibitem [{\citenamefont {{Zeldovich}}\ and\ \citenamefont
  {{Sunyaev}}(1969)}]{SZ1969}%
  \BibitemOpen
  \bibfield  {author} {\bibinfo {author} {\bibfnamefont {Y.~B.}\ \bibnamefont
  {{Zeldovich}}}\ and\ \bibinfo {author} {\bibfnamefont {R.~A.}\ \bibnamefont
  {{Sunyaev}}},\ }\href {\doibase 10.1007/BF00661821} {\bibfield  {journal}
  {\bibinfo  {journal} {Astrophysics and Space Science}\ }\textbf {\bibinfo
  {volume} {4}},\ \bibinfo {pages} {301} (\bibinfo {year} {1969})}\BibitemShut
  {NoStop}%
\bibitem [{\citenamefont {{Sunyaev}}\ and\ \citenamefont
  {{Zeldovich}}(1970)}]{SZ1970}%
  \BibitemOpen
  \bibfield  {author} {\bibinfo {author} {\bibfnamefont {R.~A.}\ \bibnamefont
  {{Sunyaev}}}\ and\ \bibinfo {author} {\bibfnamefont {Y.~B.}\ \bibnamefont
  {{Zeldovich}}},\ }\href {\doibase 10.1007/BF00653471} {\bibfield  {journal}
  {\bibinfo  {journal} {Astrophysics and Space Science}\ }\textbf {\bibinfo
  {volume} {7}},\ \bibinfo {pages} {3} (\bibinfo {year} {1970})}\BibitemShut
  {NoStop}%
\bibitem [{\citenamefont {{Sunyaev}}\ and\ \citenamefont
  {{Zeldovich}}(1972)}]{SZ1972}%
  \BibitemOpen
  \bibfield  {author} {\bibinfo {author} {\bibfnamefont {R.~A.}\ \bibnamefont
  {{Sunyaev}}}\ and\ \bibinfo {author} {\bibfnamefont {Y.~B.}\ \bibnamefont
  {{Zeldovich}}},\ }\href@noop {} {\bibfield  {journal} {\bibinfo  {journal}
  {Comments on Astrophysics and Space Physics}\ }\textbf {\bibinfo {volume}
  {4}},\ \bibinfo {pages} {173} (\bibinfo {year} {1972})}\BibitemShut {NoStop}%
\bibitem [{\citenamefont {Foreman}\ \emph {et~al.}(2023)\citenamefont
  {Foreman}, \citenamefont {Hotinli}, \citenamefont {Madhavacheril},
  \citenamefont {van Engelen},\ and\ \citenamefont
  {Kreisch}}]{Foreman:2022ves}%
  \BibitemOpen
  \bibfield  {author} {\bibinfo {author} {\bibfnamefont {S.}~\bibnamefont
  {Foreman}}, \bibinfo {author} {\bibfnamefont {S.~C.}\ \bibnamefont
  {Hotinli}}, \bibinfo {author} {\bibfnamefont {M.~S.}\ \bibnamefont
  {Madhavacheril}}, \bibinfo {author} {\bibfnamefont {A.}~\bibnamefont {van
  Engelen}}, \ and\ \bibinfo {author} {\bibfnamefont {C.~D.}\ \bibnamefont
  {Kreisch}},\ }\href {\doibase 10.1103/PhysRevD.107.083502} {\bibfield
  {journal} {\bibinfo  {journal} {Phys. Rev. D}\ }\textbf {\bibinfo {volume}
  {107}},\ \bibinfo {pages} {083502} (\bibinfo {year} {2023})},\ \Eprint
  {http://arxiv.org/abs/2209.03973} {arXiv:2209.03973 [astro-ph.CO]}
  \BibitemShut {NoStop}%
\bibitem [{\citenamefont {Hu}(2001)}]{hu2001mappingDM}%
  \BibitemOpen
  \bibfield  {author} {\bibinfo {author} {\bibfnamefont {W.}~\bibnamefont
  {Hu}},\ }\href {\doibase 10.1086/323253} {\bibfield  {journal} {\bibinfo
  {journal} {\apjl}\ }\textbf {\bibinfo {volume} {557}},\ \bibinfo {pages}
  {L79} (\bibinfo {year} {2001})},\ \Eprint
  {http://arxiv.org/abs/astro-ph/0105424} {arXiv:astro-ph/0105424} \BibitemShut
  {NoStop}%
\bibitem [{\citenamefont {Hu}\ and\ \citenamefont
  {Okamoto}(2002)}]{huokamoto2002}%
  \BibitemOpen
  \bibfield  {author} {\bibinfo {author} {\bibfnamefont {W.}~\bibnamefont
  {Hu}}\ and\ \bibinfo {author} {\bibfnamefont {T.}~\bibnamefont {Okamoto}},\
  }\href {\doibase 10.1086/341110} {\bibfield  {journal} {\bibinfo  {journal}
  {Astrophys. J.}\ }\textbf {\bibinfo {volume} {574}},\ \bibinfo {pages} {566}
  (\bibinfo {year} {2002})},\ \Eprint {http://arxiv.org/abs/astro-ph/0111606}
  {arXiv:astro-ph/0111606} \BibitemShut {NoStop}%
\bibitem [{\citenamefont {Okamoto}\ and\ \citenamefont
  {Hu}(2003)}]{OkamotoHu2003}%
  \BibitemOpen
  \bibfield  {author} {\bibinfo {author} {\bibfnamefont {T.}~\bibnamefont
  {Okamoto}}\ and\ \bibinfo {author} {\bibfnamefont {W.}~\bibnamefont {Hu}},\
  }\href {\doibase 10.1103/PhysRevD.67.083002} {\bibfield  {journal} {\bibinfo
  {journal} {Phys. Rev. D}\ }\textbf {\bibinfo {volume} {67}},\ \bibinfo
  {pages} {083002} (\bibinfo {year} {2003})},\ \Eprint
  {http://arxiv.org/abs/astro-ph/0301031} {arXiv:astro-ph/0301031} \BibitemShut
  {NoStop}%
\bibitem [{\citenamefont {Hu}\ \emph {et~al.}(2007)\citenamefont {Hu},
  \citenamefont {DeDeo},\ and\ \citenamefont {Vale}}]{hdv2007}%
  \BibitemOpen
  \bibfield  {author} {\bibinfo {author} {\bibfnamefont {W.}~\bibnamefont
  {Hu}}, \bibinfo {author} {\bibfnamefont {S.}~\bibnamefont {DeDeo}}, \ and\
  \bibinfo {author} {\bibfnamefont {C.}~\bibnamefont {Vale}},\ }\href {\doibase
  10.1088/1367-2630/9/12/441} {\bibfield  {journal} {\bibinfo  {journal} {New
  J. Phys.}\ }\textbf {\bibinfo {volume} {9}},\ \bibinfo {pages} {441}
  (\bibinfo {year} {2007})},\ \Eprint {http://arxiv.org/abs/astro-ph/0701276}
  {arXiv:astro-ph/0701276} \BibitemShut {NoStop}%
\bibitem [{\citenamefont {Kesden}\ \emph
  {et~al.}(2003{\natexlab{a}})\citenamefont {Kesden}, \citenamefont {Cooray},\
  and\ \citenamefont {Kamionkowski}}]{Kesden2003}%
  \BibitemOpen
  \bibfield  {author} {\bibinfo {author} {\bibfnamefont {M.~H.}\ \bibnamefont
  {Kesden}}, \bibinfo {author} {\bibfnamefont {A.}~\bibnamefont {Cooray}}, \
  and\ \bibinfo {author} {\bibfnamefont {M.}~\bibnamefont {Kamionkowski}},\
  }\href {\doibase 10.1103/PhysRevD.67.123507} {\bibfield  {journal} {\bibinfo
  {journal} {Phys. Rev. D}\ }\textbf {\bibinfo {volume} {67}},\ \bibinfo
  {pages} {123507} (\bibinfo {year} {2003}{\natexlab{a}})},\ \Eprint
  {http://arxiv.org/abs/astro-ph/0302536} {arXiv:astro-ph/0302536} \BibitemShut
  {NoStop}%
\bibitem [{\citenamefont {{Namikawa}}\ \emph {et~al.}(2013)\citenamefont
  {{Namikawa}}, \citenamefont {{Hanson}},\ and\ \citenamefont
  {{Takahashi}}}]{Namikawa2013}%
  \BibitemOpen
  \bibfield  {author} {\bibinfo {author} {\bibfnamefont {T.}~\bibnamefont
  {{Namikawa}}}, \bibinfo {author} {\bibfnamefont {D.}~\bibnamefont
  {{Hanson}}}, \ and\ \bibinfo {author} {\bibfnamefont {R.}~\bibnamefont
  {{Takahashi}}},\ }\href {\doibase 10.1093/mnras/stt195} {\bibfield  {journal}
  {\bibinfo  {journal} {\mnras}\ }\textbf {\bibinfo {volume} {431}},\ \bibinfo
  {pages} {609} (\bibinfo {year} {2013})},\ \Eprint
  {http://arxiv.org/abs/1209.0091} {arXiv:1209.0091 [astro-ph.CO]} \BibitemShut
  {NoStop}%
\bibitem [{\citenamefont {Madhavacheril}\ \emph {et~al.}(2020)\citenamefont
  {Madhavacheril}, \citenamefont {Smith}, \citenamefont {Sherwin},\ and\
  \citenamefont {Naess}}]{Madhavacheril:2020ido}%
  \BibitemOpen
  \bibfield  {author} {\bibinfo {author} {\bibfnamefont {M.~S.}\ \bibnamefont
  {Madhavacheril}}, \bibinfo {author} {\bibfnamefont {K.~M.}\ \bibnamefont
  {Smith}}, \bibinfo {author} {\bibfnamefont {B.~D.}\ \bibnamefont {Sherwin}},
  \ and\ \bibinfo {author} {\bibfnamefont {S.}~\bibnamefont {Naess}},\ }\href
  {\doibase 10.1088/1475-7516/2021/05/028} {\  (\bibinfo {year} {2020}),\
  10.1088/1475-7516/2021/05/028},\ \Eprint {http://arxiv.org/abs/2011.02475}
  {arXiv:2011.02475 [astro-ph.CO]} \BibitemShut {NoStop}%
\bibitem [{\citenamefont {Nguyen}\ \emph {et~al.}(2019)\citenamefont {Nguyen},
  \citenamefont {Sehgal},\ and\ \citenamefont {Madhavacheril}}]{Nguyen2017}%
  \BibitemOpen
  \bibfield  {author} {\bibinfo {author} {\bibfnamefont {H.~N.}\ \bibnamefont
  {Nguyen}}, \bibinfo {author} {\bibfnamefont {N.}~\bibnamefont {Sehgal}}, \
  and\ \bibinfo {author} {\bibfnamefont {M.}~\bibnamefont {Madhavacheril}},\
  }\href {\doibase 10.1103/PhysRevD.99.023502} {\bibfield  {journal} {\bibinfo
  {journal} {Phys. Rev. D}\ }\textbf {\bibinfo {volume} {99}},\ \bibinfo
  {pages} {023502} (\bibinfo {year} {2019})},\ \Eprint
  {http://arxiv.org/abs/1710.03747} {arXiv:1710.03747 [astro-ph.CO]}
  \BibitemShut {NoStop}%
\bibitem [{\citenamefont {{Hanson}}\ \emph {et~al.}(2011)\citenamefont
  {{Hanson}}, \citenamefont {{Challinor}}, \citenamefont {{Efstathiou}},\ and\
  \citenamefont {{Bielewicz}}}]{Hanson2011}%
  \BibitemOpen
  \bibfield  {author} {\bibinfo {author} {\bibfnamefont {D.}~\bibnamefont
  {{Hanson}}}, \bibinfo {author} {\bibfnamefont {A.}~\bibnamefont
  {{Challinor}}}, \bibinfo {author} {\bibfnamefont {G.}~\bibnamefont
  {{Efstathiou}}}, \ and\ \bibinfo {author} {\bibfnamefont {P.}~\bibnamefont
  {{Bielewicz}}},\ }\href {\doibase 10.1103/PhysRevD.83.043005} {\bibfield
  {journal} {\bibinfo  {journal} {Phys. Rev. D}\ }\textbf {\bibinfo {volume}
  {83}},\ \bibinfo {eid} {043005} (\bibinfo {year} {2011})},\ \Eprint
  {http://arxiv.org/abs/1008.4403} {arXiv:1008.4403 [astro-ph.CO]} \BibitemShut
  {NoStop}%
\bibitem [{\citenamefont {Horowitz}\ \emph {et~al.}(2019)\citenamefont
  {Horowitz}, \citenamefont {Ferraro},\ and\ \citenamefont
  {Sherwin}}]{Horowitz:2017iql}%
  \BibitemOpen
  \bibfield  {author} {\bibinfo {author} {\bibfnamefont {B.}~\bibnamefont
  {Horowitz}}, \bibinfo {author} {\bibfnamefont {S.}~\bibnamefont {Ferraro}}, \
  and\ \bibinfo {author} {\bibfnamefont {B.~D.}\ \bibnamefont {Sherwin}},\
  }\href {\doibase 10.1093/mnras/stz566} {\bibfield  {journal} {\bibinfo
  {journal} {Mon. Not. Roy. Astron. Soc.}\ }\textbf {\bibinfo {volume} {485}},\
  \bibinfo {pages} {3919} (\bibinfo {year} {2019})},\ \Eprint
  {http://arxiv.org/abs/1710.10236} {arXiv:1710.10236 [astro-ph.CO]}
  \BibitemShut {NoStop}%
\bibitem [{\citenamefont {Hadzhiyska}\ \emph {et~al.}(2019)\citenamefont
  {Hadzhiyska}, \citenamefont {Sherwin}, \citenamefont {Madhavacheril},\ and\
  \citenamefont {Ferraro}}]{Hadzhiyska2019}%
  \BibitemOpen
  \bibfield  {author} {\bibinfo {author} {\bibfnamefont {B.}~\bibnamefont
  {Hadzhiyska}}, \bibinfo {author} {\bibfnamefont {B.~D.}\ \bibnamefont
  {Sherwin}}, \bibinfo {author} {\bibfnamefont {M.}~\bibnamefont
  {Madhavacheril}}, \ and\ \bibinfo {author} {\bibfnamefont {S.}~\bibnamefont
  {Ferraro}},\ }\href {\doibase 10.1103/PhysRevD.100.023547} {\bibfield
  {journal} {\bibinfo  {journal} {Phys. Rev. D}\ }\textbf {\bibinfo {volume}
  {100}},\ \bibinfo {pages} {023547} (\bibinfo {year} {2019})},\ \Eprint
  {http://arxiv.org/abs/1905.04217} {arXiv:1905.04217 [astro-ph.CO]}
  \BibitemShut {NoStop}%
\bibitem [{\citenamefont {Chan}\ \emph {et~al.}(2023)\citenamefont {Chan},
  \citenamefont {Hlo\v{z}ek}, \citenamefont {Meyers},\ and\ \citenamefont {van
  Engelen}}]{Chan2023}%
  \BibitemOpen
  \bibfield  {author} {\bibinfo {author} {\bibfnamefont {V.~C.}\ \bibnamefont
  {Chan}}, \bibinfo {author} {\bibfnamefont {R.}~\bibnamefont {Hlo\v{z}ek}},
  \bibinfo {author} {\bibfnamefont {J.}~\bibnamefont {Meyers}}, \ and\ \bibinfo
  {author} {\bibfnamefont {A.}~\bibnamefont {van Engelen}},\ }\href@noop {} {\
  (\bibinfo {year} {2023})},\ \Eprint {http://arxiv.org/abs/2302.13350}
  {arXiv:2302.13350 [astro-ph.CO]} \BibitemShut {NoStop}%
\bibitem [{\citenamefont {Seljak}\ and\ \citenamefont
  {Hirata}(2004)}]{Seljak:2003pn}%
  \BibitemOpen
  \bibfield  {author} {\bibinfo {author} {\bibfnamefont {U.}~\bibnamefont
  {Seljak}}\ and\ \bibinfo {author} {\bibfnamefont {C.~M.}\ \bibnamefont
  {Hirata}},\ }\href {\doibase 10.1103/PhysRevD.69.043005} {\bibfield
  {journal} {\bibinfo  {journal} {Phys. Rev. D}\ }\textbf {\bibinfo {volume}
  {69}},\ \bibinfo {pages} {043005} (\bibinfo {year} {2004})},\ \Eprint
  {http://arxiv.org/abs/astro-ph/0310163} {arXiv:astro-ph/0310163} \BibitemShut
  {NoStop}%
\bibitem [{\citenamefont {{Smith}}\ \emph {et~al.}(2012)\citenamefont
  {{Smith}}, \citenamefont {{Hanson}}, \citenamefont {{LoVerde}}, \citenamefont
  {{Hirata}},\ and\ \citenamefont {{Zahn}}}]{Smith2012}%
  \BibitemOpen
  \bibfield  {author} {\bibinfo {author} {\bibfnamefont {K.~M.}\ \bibnamefont
  {{Smith}}}, \bibinfo {author} {\bibfnamefont {D.}~\bibnamefont {{Hanson}}},
  \bibinfo {author} {\bibfnamefont {M.}~\bibnamefont {{LoVerde}}}, \bibinfo
  {author} {\bibfnamefont {C.~M.}\ \bibnamefont {{Hirata}}}, \ and\ \bibinfo
  {author} {\bibfnamefont {O.}~\bibnamefont {{Zahn}}},\ }\href {\doibase
  10.1088/1475-7516/2012/06/014} {\bibfield  {journal} {\bibinfo  {journal}
  {\jcap}\ }\textbf {\bibinfo {volume} {2012}},\ \bibinfo {eid} {014} (\bibinfo
  {year} {2012})},\ \Eprint {http://arxiv.org/abs/1010.0048} {arXiv:1010.0048
  [astro-ph.CO]} \BibitemShut {NoStop}%
\bibitem [{\citenamefont {Simard}\ \emph {et~al.}(2015)\citenamefont {Simard},
  \citenamefont {Hanson},\ and\ \citenamefont {Holder}}]{Simard:2014aqa}%
  \BibitemOpen
  \bibfield  {author} {\bibinfo {author} {\bibfnamefont {G.}~\bibnamefont
  {Simard}}, \bibinfo {author} {\bibfnamefont {D.}~\bibnamefont {Hanson}}, \
  and\ \bibinfo {author} {\bibfnamefont {G.}~\bibnamefont {Holder}},\ }\href
  {\doibase 10.1088/0004-637X/807/2/166} {\bibfield  {journal} {\bibinfo
  {journal} {Astrophys. J.}\ }\textbf {\bibinfo {volume} {807}},\ \bibinfo
  {pages} {166} (\bibinfo {year} {2015})},\ \Eprint
  {http://arxiv.org/abs/1410.0691} {arXiv:1410.0691 [astro-ph.CO]} \BibitemShut
  {NoStop}%
\bibitem [{\citenamefont {Sehgal}\ \emph {et~al.}(2017)\citenamefont {Sehgal},
  \citenamefont {Madhavacheril}, \citenamefont {Sherwin},\ and\ \citenamefont
  {van Engelen}}]{Sehgal:2016eag}%
  \BibitemOpen
  \bibfield  {author} {\bibinfo {author} {\bibfnamefont {N.}~\bibnamefont
  {Sehgal}}, \bibinfo {author} {\bibfnamefont {M.~S.}\ \bibnamefont
  {Madhavacheril}}, \bibinfo {author} {\bibfnamefont {B.}~\bibnamefont
  {Sherwin}}, \ and\ \bibinfo {author} {\bibfnamefont {A.}~\bibnamefont {van
  Engelen}},\ }\href {\doibase 10.1103/PhysRevD.95.103512} {\bibfield
  {journal} {\bibinfo  {journal} {Phys. Rev. D}\ }\textbf {\bibinfo {volume}
  {95}},\ \bibinfo {pages} {103512} (\bibinfo {year} {2017})},\ \Eprint
  {http://arxiv.org/abs/1612.03898} {arXiv:1612.03898 [astro-ph.CO]}
  \BibitemShut {NoStop}%
\bibitem [{\citenamefont {{Benoit-L{\'e}vy}}\ \emph {et~al.}(2012)\citenamefont
  {{Benoit-L{\'e}vy}}, \citenamefont {{Smith}},\ and\ \citenamefont
  {{Hu}}}]{BenoitLevySmithHu2012}%
  \BibitemOpen
  \bibfield  {author} {\bibinfo {author} {\bibfnamefont {A.}~\bibnamefont
  {{Benoit-L{\'e}vy}}}, \bibinfo {author} {\bibfnamefont {K.~M.}\ \bibnamefont
  {{Smith}}}, \ and\ \bibinfo {author} {\bibfnamefont {W.}~\bibnamefont
  {{Hu}}},\ }\href {\doibase 10.1103/PhysRevD.86.123008} {\bibfield  {journal}
  {\bibinfo  {journal} {Phys. Rev. D}\ }\textbf {\bibinfo {volume} {86}},\
  \bibinfo {eid} {123008} (\bibinfo {year} {2012})},\ \Eprint
  {http://arxiv.org/abs/1205.0474} {arXiv:1205.0474 [astro-ph.CO]} \BibitemShut
  {NoStop}%
\bibitem [{\citenamefont {Peloton}\ \emph {et~al.}(2017)\citenamefont
  {Peloton}, \citenamefont {Schmittfull}, \citenamefont {Lewis}, \citenamefont
  {Carron},\ and\ \citenamefont {Zahn}}]{Peloton2017}%
  \BibitemOpen
  \bibfield  {author} {\bibinfo {author} {\bibfnamefont {J.}~\bibnamefont
  {Peloton}}, \bibinfo {author} {\bibfnamefont {M.}~\bibnamefont
  {Schmittfull}}, \bibinfo {author} {\bibfnamefont {A.}~\bibnamefont {Lewis}},
  \bibinfo {author} {\bibfnamefont {J.}~\bibnamefont {Carron}}, \ and\ \bibinfo
  {author} {\bibfnamefont {O.}~\bibnamefont {Zahn}},\ }\href {\doibase
  10.1103/PhysRevD.95.043508} {\bibfield  {journal} {\bibinfo  {journal} {Phys.
  Rev. D}\ }\textbf {\bibinfo {volume} {95}},\ \bibinfo {pages} {043508}
  (\bibinfo {year} {2017})},\ \Eprint {http://arxiv.org/abs/1611.01446}
  {arXiv:1611.01446 [astro-ph.CO]} \BibitemShut {NoStop}%
\bibitem [{\citenamefont {Schmittfull}\ \emph {et~al.}(2013)\citenamefont
  {Schmittfull}, \citenamefont {Challinor}, \citenamefont {Hanson},\ and\
  \citenamefont {Lewis}}]{Schmittfull2013}%
  \BibitemOpen
  \bibfield  {author} {\bibinfo {author} {\bibfnamefont {M.~M.}\ \bibnamefont
  {Schmittfull}}, \bibinfo {author} {\bibfnamefont {A.}~\bibnamefont
  {Challinor}}, \bibinfo {author} {\bibfnamefont {D.}~\bibnamefont {Hanson}}, \
  and\ \bibinfo {author} {\bibfnamefont {A.}~\bibnamefont {Lewis}},\ }\href
  {\doibase 10.1103/PhysRevD.88.063012} {\bibfield  {journal} {\bibinfo
  {journal} {Phys. Rev. D}\ }\textbf {\bibinfo {volume} {88}},\ \bibinfo
  {pages} {063012} (\bibinfo {year} {2013})},\ \Eprint
  {http://arxiv.org/abs/1308.0286} {arXiv:1308.0286 [astro-ph.CO]} \BibitemShut
  {NoStop}%
\bibitem [{\citenamefont {Manzotti}\ \emph {et~al.}(2014)\citenamefont
  {Manzotti}, \citenamefont {Hu},\ and\ \citenamefont
  {Benoit-L\'evy}}]{Manzotti2014}%
  \BibitemOpen
  \bibfield  {author} {\bibinfo {author} {\bibfnamefont {A.}~\bibnamefont
  {Manzotti}}, \bibinfo {author} {\bibfnamefont {W.}~\bibnamefont {Hu}}, \ and\
  \bibinfo {author} {\bibfnamefont {A.}~\bibnamefont {Benoit-L\'evy}},\ }\href
  {\doibase 10.1103/PhysRevD.90.023003} {\bibfield  {journal} {\bibinfo
  {journal} {Phys. Rev. D}\ }\textbf {\bibinfo {volume} {90}},\ \bibinfo
  {pages} {023003} (\bibinfo {year} {2014})},\ \Eprint
  {http://arxiv.org/abs/1401.7992} {arXiv:1401.7992 [astro-ph.CO]} \BibitemShut
  {NoStop}%
\bibitem [{\citenamefont {Kesden}\ \emph
  {et~al.}(2003{\natexlab{b}})\citenamefont {Kesden}, \citenamefont {Cooray},\
  and\ \citenamefont {Kamionkowski}}]{Kesden:2003cc}%
  \BibitemOpen
  \bibfield  {author} {\bibinfo {author} {\bibfnamefont {M.~H.}\ \bibnamefont
  {Kesden}}, \bibinfo {author} {\bibfnamefont {A.}~\bibnamefont {Cooray}}, \
  and\ \bibinfo {author} {\bibfnamefont {M.}~\bibnamefont {Kamionkowski}},\
  }\href {\doibase 10.1103/PhysRevD.67.123507} {\bibfield  {journal} {\bibinfo
  {journal} {Phys. Rev. D}\ }\textbf {\bibinfo {volume} {67}},\ \bibinfo
  {pages} {123507} (\bibinfo {year} {2003}{\natexlab{b}})},\ \Eprint
  {http://arxiv.org/abs/astro-ph/0302536} {arXiv:astro-ph/0302536} \BibitemShut
  {NoStop}%
\bibitem [{\citenamefont {Eisenstein}\ \emph {et~al.}(2005)\citenamefont
  {Eisenstein} \emph {et~al.}}]{EisensteinSDSS2005}%
  \BibitemOpen
  \bibfield  {author} {\bibinfo {author} {\bibfnamefont {D.~J.}\ \bibnamefont
  {Eisenstein}} \emph {et~al.} (\bibinfo {collaboration} {SDSS}),\ }\href
  {\doibase 10.1086/466512} {\bibfield  {journal} {\bibinfo  {journal}
  {Astrophys. J.}\ }\textbf {\bibinfo {volume} {633}},\ \bibinfo {pages} {560}
  (\bibinfo {year} {2005})},\ \Eprint {http://arxiv.org/abs/astro-ph/0501171}
  {arXiv:astro-ph/0501171} \BibitemShut {NoStop}%
\bibitem [{\citenamefont {Allison}\ \emph {et~al.}(2015)\citenamefont
  {Allison}, \citenamefont {Caucal}, \citenamefont {Calabrese}, \citenamefont
  {Dunkley},\ and\ \citenamefont {Louis}}]{Allison2015}%
  \BibitemOpen
  \bibfield  {author} {\bibinfo {author} {\bibfnamefont {R.}~\bibnamefont
  {Allison}}, \bibinfo {author} {\bibfnamefont {P.}~\bibnamefont {Caucal}},
  \bibinfo {author} {\bibfnamefont {E.}~\bibnamefont {Calabrese}}, \bibinfo
  {author} {\bibfnamefont {J.}~\bibnamefont {Dunkley}}, \ and\ \bibinfo
  {author} {\bibfnamefont {T.}~\bibnamefont {Louis}},\ }\href {\doibase
  10.1103/PhysRevD.92.123535} {\bibfield  {journal} {\bibinfo  {journal} {Phys.
  Rev. D}\ }\textbf {\bibinfo {volume} {92}},\ \bibinfo {pages} {123535}
  (\bibinfo {year} {2015})},\ \Eprint {http://arxiv.org/abs/1509.07471}
  {arXiv:1509.07471 [astro-ph.CO]} \BibitemShut {NoStop}%
\bibitem [{\citenamefont {Dodelson}(2003)}]{Dodelson2003}%
  \BibitemOpen
  \bibfield  {author} {\bibinfo {author} {\bibfnamefont {S.}~\bibnamefont
  {Dodelson}},\ }\href@noop {} {\emph {\bibinfo {title} {{Modern Cosmology}}}}\
  (\bibinfo  {publisher} {Academic Press},\ \bibinfo {address} {Amsterdam},\
  \bibinfo {year} {2003})\BibitemShut {NoStop}%
\bibitem [{\citenamefont {Albrecht}\ \emph {et~al.}(2006)\citenamefont
  {Albrecht} \emph {et~al.}}]{DarkEnergyTaskForce}%
  \BibitemOpen
  \bibfield  {author} {\bibinfo {author} {\bibfnamefont {A.}~\bibnamefont
  {Albrecht}} \emph {et~al.},\ }\href@noop {} {\  (\bibinfo {year} {2006})},\
  \Eprint {http://arxiv.org/abs/astro-ph/0609591} {arXiv:astro-ph/0609591}
  \BibitemShut {NoStop}%
\bibitem [{\citenamefont {Huterer}\ and\ \citenamefont
  {Takada}(2005)}]{Huterer2004}%
  \BibitemOpen
  \bibfield  {author} {\bibinfo {author} {\bibfnamefont {D.}~\bibnamefont
  {Huterer}}\ and\ \bibinfo {author} {\bibfnamefont {M.}~\bibnamefont
  {Takada}},\ }\href {\doibase 10.1016/j.astropartphys.2005.02.006} {\bibfield
  {journal} {\bibinfo  {journal} {Astropart. Phys.}\ }\textbf {\bibinfo
  {volume} {23}},\ \bibinfo {pages} {369} (\bibinfo {year} {2005})},\ \Eprint
  {http://arxiv.org/abs/astro-ph/0412142} {arXiv:astro-ph/0412142} \BibitemShut
  {NoStop}%
\bibitem [{\citenamefont {Amara}\ and\ \citenamefont
  {Refregier}(2008)}]{Amara2007}%
  \BibitemOpen
  \bibfield  {author} {\bibinfo {author} {\bibfnamefont {A.}~\bibnamefont
  {Amara}}\ and\ \bibinfo {author} {\bibfnamefont {A.}~\bibnamefont
  {Refregier}},\ }\href {\doibase 10.1111/j.1365-2966.2008.13880.x} {\bibfield
  {journal} {\bibinfo  {journal} {\mnras}\ }\textbf {\bibinfo {volume} {391}},\
  \bibinfo {pages} {228} (\bibinfo {year} {2008})},\ \Eprint
  {http://arxiv.org/abs/0710.5171} {arXiv:0710.5171 [astro-ph]} \BibitemShut
  {NoStop}%
\bibitem [{\citenamefont {Bernal}\ \emph {et~al.}(2020)\citenamefont {Bernal},
  \citenamefont {Bellomo}, \citenamefont {Raccanelli},\ and\ \citenamefont
  {Verde}}]{Bernal2020}%
  \BibitemOpen
  \bibfield  {author} {\bibinfo {author} {\bibfnamefont {J.~L.}\ \bibnamefont
  {Bernal}}, \bibinfo {author} {\bibfnamefont {N.}~\bibnamefont {Bellomo}},
  \bibinfo {author} {\bibfnamefont {A.}~\bibnamefont {Raccanelli}}, \ and\
  \bibinfo {author} {\bibfnamefont {L.}~\bibnamefont {Verde}},\ }\href
  {\doibase 10.1088/1475-7516/2020/10/017} {\bibfield  {journal} {\bibinfo
  {journal} {\jcap}\ }\textbf {\bibinfo {volume} {10}},\ \bibinfo {pages} {017}
  (\bibinfo {year} {2020})},\ \Eprint {http://arxiv.org/abs/2005.09666}
  {arXiv:2005.09666 [astro-ph.CO]} \BibitemShut {NoStop}%
\bibitem [{\citenamefont {Hogg}\ and\ \citenamefont
  {Foreman-Mackey}(2018)}]{Hogg2017MCMC}%
  \BibitemOpen
  \bibfield  {author} {\bibinfo {author} {\bibfnamefont {D.~W.}\ \bibnamefont
  {Hogg}}\ and\ \bibinfo {author} {\bibfnamefont {D.}~\bibnamefont
  {Foreman-Mackey}},\ }\href {\doibase 10.3847/1538-4365/aab76e} {\bibfield
  {journal} {\bibinfo  {journal} {\apjs}\ }\textbf {\bibinfo {volume} {236}},\
  \bibinfo {pages} {11} (\bibinfo {year} {2018})},\ \Eprint
  {http://arxiv.org/abs/1710.06068} {arXiv:1710.06068 [astro-ph.IM]}
  \BibitemShut {NoStop}%
\bibitem [{\citenamefont {Lewis}\ and\ \citenamefont
  {Bridle}(2002)}]{cobayaMCMClewisbrindle}%
  \BibitemOpen
  \bibfield  {author} {\bibinfo {author} {\bibfnamefont {A.}~\bibnamefont
  {Lewis}}\ and\ \bibinfo {author} {\bibfnamefont {S.}~\bibnamefont {Bridle}},\
  }\href {\doibase 10.1103/PhysRevD.66.103511} {\bibfield  {journal} {\bibinfo
  {journal} {Phys. Rev. D}\ }\textbf {\bibinfo {volume} {66}},\ \bibinfo
  {pages} {103511} (\bibinfo {year} {2002})},\ \Eprint
  {http://arxiv.org/abs/astro-ph/0205436} {arXiv:astro-ph/0205436} \BibitemShut
  {NoStop}%
\bibitem [{\citenamefont {Lewis}(2013)}]{cobayaMCMClewis}%
  \BibitemOpen
  \bibfield  {author} {\bibinfo {author} {\bibfnamefont {A.}~\bibnamefont
  {Lewis}},\ }\href {\doibase 10.1103/PhysRevD.87.103529} {\bibfield  {journal}
  {\bibinfo  {journal} {Phys. Rev. D}\ }\textbf {\bibinfo {volume} {87}},\
  \bibinfo {pages} {103529} (\bibinfo {year} {2013})},\ \Eprint
  {http://arxiv.org/abs/1304.4473} {arXiv:1304.4473 [astro-ph.CO]} \BibitemShut
  {NoStop}%
\bibitem [{\citenamefont {Gelman}\ and\ \citenamefont
  {Rubin}(1992)}]{GelmanRubin1992}%
  \BibitemOpen
  \bibfield  {author} {\bibinfo {author} {\bibfnamefont {A.}~\bibnamefont
  {Gelman}}\ and\ \bibinfo {author} {\bibfnamefont {D.~B.}\ \bibnamefont
  {Rubin}},\ }\href {\doibase 10.1214/ss/1177011136} {\bibfield  {journal}
  {\bibinfo  {journal} {Statistical Science}\ }\textbf {\bibinfo {volume}
  {7}},\ \bibinfo {pages} {457} (\bibinfo {year} {1992})}\BibitemShut {NoStop}%
\bibitem [{\citenamefont {Tr\"oster}\ \emph {et~al.}(2022)\citenamefont
  {Tr\"oster} \emph {et~al.}}]{Troster2021}%
  \BibitemOpen
  \bibfield  {author} {\bibinfo {author} {\bibfnamefont {T.}~\bibnamefont
  {Tr\"oster}} \emph {et~al.},\ }\href {\doibase 10.1051/0004-6361/202142197}
  {\bibfield  {journal} {\bibinfo  {journal} {\aap}\ }\textbf {\bibinfo
  {volume} {660}},\ \bibinfo {pages} {A27} (\bibinfo {year} {2022})},\ \Eprint
  {http://arxiv.org/abs/2109.04458} {arXiv:2109.04458 [astro-ph.CO]}
  \BibitemShut {NoStop}%
\bibitem [{\citenamefont {Mead}\ \emph {et~al.}(2021)\citenamefont {Mead},
  \citenamefont {Brieden}, \citenamefont {Tr\"oster},\ and\ \citenamefont
  {Heymans}}]{Mead2020}%
  \BibitemOpen
  \bibfield  {author} {\bibinfo {author} {\bibfnamefont {A.}~\bibnamefont
  {Mead}}, \bibinfo {author} {\bibfnamefont {S.}~\bibnamefont {Brieden}},
  \bibinfo {author} {\bibfnamefont {T.}~\bibnamefont {Tr\"oster}}, \ and\
  \bibinfo {author} {\bibfnamefont {C.}~\bibnamefont {Heymans}},\ }\href
  {\doibase 10.1093/mnras/stab082} {\bibfield  {journal} {\bibinfo  {journal}
  {\mnras}\ }\textbf {\bibinfo {volume} {502}},\ \bibinfo {pages} {1401}
  (\bibinfo {year} {2021})},\ \Eprint {http://arxiv.org/abs/2009.01858}
  {arXiv:2009.01858 [astro-ph.CO]} \BibitemShut {NoStop}%
\bibitem [{\citenamefont {McCarthy}\ \emph {et~al.}(2021)\citenamefont
  {McCarthy}, \citenamefont {Foreman},\ and\ \citenamefont {van
  Engelen}}]{McCarthy2020}%
  \BibitemOpen
  \bibfield  {author} {\bibinfo {author} {\bibfnamefont {F.}~\bibnamefont
  {McCarthy}}, \bibinfo {author} {\bibfnamefont {S.}~\bibnamefont {Foreman}}, \
  and\ \bibinfo {author} {\bibfnamefont {A.}~\bibnamefont {van Engelen}},\
  }\href {\doibase 10.1103/PhysRevD.103.103538} {\bibfield  {journal} {\bibinfo
   {journal} {Phys. Rev. D}\ }\textbf {\bibinfo {volume} {103}},\ \bibinfo
  {pages} {103538} (\bibinfo {year} {2021})},\ \Eprint
  {http://arxiv.org/abs/2011.06582} {arXiv:2011.06582 [astro-ph.CO]}
  \BibitemShut {NoStop}%
\bibitem [{\citenamefont {Wallisch}(2018)}]{Wallisch2018}%
  \BibitemOpen
  \bibfield  {author} {\bibinfo {author} {\bibfnamefont {B.}~\bibnamefont
  {Wallisch}},\ }\emph {\bibinfo {title} {{Cosmological Probes of Light
  Relics}}},\ \href {\doibase 10.17863/CAM.30368} {Ph.D. thesis},\ \bibinfo
  {school} {Cambridge U.} (\bibinfo {year} {2018}),\ \Eprint
  {http://arxiv.org/abs/1810.02800} {arXiv:1810.02800 [astro-ph.CO]}
  \BibitemShut {NoStop}%
\bibitem [{\citenamefont {Green}\ \emph {et~al.}(2019)\citenamefont {Green}
  \emph {et~al.}}]{Green2019}%
  \BibitemOpen
  \bibfield  {author} {\bibinfo {author} {\bibfnamefont {D.}~\bibnamefont
  {Green}} \emph {et~al.},\ }\href@noop {} {\bibfield  {journal} {\bibinfo
  {journal} {Bull. Am. Astron. Soc.}\ }\textbf {\bibinfo {volume} {51}},\
  \bibinfo {pages} {159} (\bibinfo {year} {2019})},\ \Eprint
  {http://arxiv.org/abs/1903.04763} {arXiv:1903.04763 [astro-ph.CO]}
  \BibitemShut {NoStop}%
\bibitem [{\citenamefont {Peccei}(2008)}]{Peccei2006}%
  \BibitemOpen
  \bibfield  {author} {\bibinfo {author} {\bibfnamefont {R.~D.}\ \bibnamefont
  {Peccei}},\ }\href {\doibase 10.1007/978-3-540-73518-2_1} {\bibfield
  {journal} {\bibinfo  {journal} {Lect. Notes Phys.}\ }\textbf {\bibinfo
  {volume} {741}},\ \bibinfo {pages} {3} (\bibinfo {year} {2008})},\ \Eprint
  {http://arxiv.org/abs/hep-ph/0607268} {arXiv:hep-ph/0607268} \BibitemShut
  {NoStop}%
\bibitem [{\citenamefont {Marsh}(2018)}]{Marsh2017}%
  \BibitemOpen
  \bibfield  {author} {\bibinfo {author} {\bibfnamefont {D.~J.~E.}\
  \bibnamefont {Marsh}},\ }in\ \href {\doibase
  10.3204/DESY-PROC-2017-02/marsh_david} {\emph {\bibinfo {booktitle} {{13th
  Patras Workshop on Axions, WIMPs and WISPs}}}}\ (\bibinfo {year} {2018})\
  pp.\ \bibinfo {pages} {59--74},\ \Eprint {http://arxiv.org/abs/1712.03018}
  {arXiv:1712.03018 [hep-ph]} \BibitemShut {NoStop}%
\bibitem [{\citenamefont {Di~Luzio}\ \emph {et~al.}(2020)\citenamefont
  {Di~Luzio}, \citenamefont {Giannotti}, \citenamefont {Nardi},\ and\
  \citenamefont {Visinelli}}]{DiLuzio2020}%
  \BibitemOpen
  \bibfield  {author} {\bibinfo {author} {\bibfnamefont {L.}~\bibnamefont
  {Di~Luzio}}, \bibinfo {author} {\bibfnamefont {M.}~\bibnamefont {Giannotti}},
  \bibinfo {author} {\bibfnamefont {E.}~\bibnamefont {Nardi}}, \ and\ \bibinfo
  {author} {\bibfnamefont {L.}~\bibnamefont {Visinelli}},\ }\href {\doibase
  10.1016/j.physrep.2020.06.002} {\bibfield  {journal} {\bibinfo  {journal}
  {Phys. Rept.}\ }\textbf {\bibinfo {volume} {870}},\ \bibinfo {pages} {1}
  (\bibinfo {year} {2020})},\ \Eprint {http://arxiv.org/abs/2003.01100}
  {arXiv:2003.01100 [hep-ph]} \BibitemShut {NoStop}%
\bibitem [{\citenamefont {Baumann}\ \emph {et~al.}(2016)\citenamefont
  {Baumann}, \citenamefont {Green},\ and\ \citenamefont
  {Wallisch}}]{Baumann2016}%
  \BibitemOpen
  \bibfield  {author} {\bibinfo {author} {\bibfnamefont {D.}~\bibnamefont
  {Baumann}}, \bibinfo {author} {\bibfnamefont {D.}~\bibnamefont {Green}}, \
  and\ \bibinfo {author} {\bibfnamefont {B.}~\bibnamefont {Wallisch}},\ }\href
  {\doibase 10.1103/PhysRevLett.117.171301} {\bibfield  {journal} {\bibinfo
  {journal} {Phys. Rev. Lett.}\ }\textbf {\bibinfo {volume} {117}},\ \bibinfo
  {pages} {171301} (\bibinfo {year} {2016})},\ \Eprint
  {http://arxiv.org/abs/1604.08614} {arXiv:1604.08614 [astro-ph.CO]}
  \BibitemShut {NoStop}%
\bibitem [{\citenamefont {Blum}\ \emph {et~al.}(2014)\citenamefont {Blum},
  \citenamefont {D'Agnolo}, \citenamefont {Lisanti},\ and\ \citenamefont
  {Safdi}}]{Blum2014}%
  \BibitemOpen
  \bibfield  {author} {\bibinfo {author} {\bibfnamefont {K.}~\bibnamefont
  {Blum}}, \bibinfo {author} {\bibfnamefont {R.~T.}\ \bibnamefont {D'Agnolo}},
  \bibinfo {author} {\bibfnamefont {M.}~\bibnamefont {Lisanti}}, \ and\
  \bibinfo {author} {\bibfnamefont {B.~R.}\ \bibnamefont {Safdi}},\ }\href
  {\doibase 10.1016/j.physletb.2014.07.059} {\bibfield  {journal} {\bibinfo
  {journal} {Phys. Lett. B}\ }\textbf {\bibinfo {volume} {737}},\ \bibinfo
  {pages} {30} (\bibinfo {year} {2014})},\ \Eprint
  {http://arxiv.org/abs/1401.6460} {arXiv:1401.6460 [hep-ph]} \BibitemShut
  {NoStop}%
\bibitem [{\citenamefont {Hannestad}\ \emph {et~al.}(2007)\citenamefont
  {Hannestad}, \citenamefont {Mirizzi}, \citenamefont {Raffelt},\ and\
  \citenamefont {Wong}}]{Hannestad2007}%
  \BibitemOpen
  \bibfield  {author} {\bibinfo {author} {\bibfnamefont {S.}~\bibnamefont
  {Hannestad}}, \bibinfo {author} {\bibfnamefont {A.}~\bibnamefont {Mirizzi}},
  \bibinfo {author} {\bibfnamefont {G.~G.}\ \bibnamefont {Raffelt}}, \ and\
  \bibinfo {author} {\bibfnamefont {Y.~Y.~Y.}\ \bibnamefont {Wong}},\ }\href
  {\doibase 10.1088/1475-7516/2007/08/015} {\bibfield  {journal} {\bibinfo
  {journal} {\jcap}\ }\textbf {\bibinfo {volume} {08}},\ \bibinfo {pages} {015}
  (\bibinfo {year} {2007})},\ \Eprint {http://arxiv.org/abs/0706.4198}
  {arXiv:0706.4198 [astro-ph]} \BibitemShut {NoStop}%
\bibitem [{\citenamefont {Di~Valentino}\ \emph {et~al.}(2016)\citenamefont
  {Di~Valentino}, \citenamefont {Giusarma}, \citenamefont {Lattanzi},
  \citenamefont {Mena}, \citenamefont {Melchiorri},\ and\ \citenamefont
  {Silk}}]{DiValentino2015}%
  \BibitemOpen
  \bibfield  {author} {\bibinfo {author} {\bibfnamefont {E.}~\bibnamefont
  {Di~Valentino}}, \bibinfo {author} {\bibfnamefont {E.}~\bibnamefont
  {Giusarma}}, \bibinfo {author} {\bibfnamefont {M.}~\bibnamefont {Lattanzi}},
  \bibinfo {author} {\bibfnamefont {O.}~\bibnamefont {Mena}}, \bibinfo {author}
  {\bibfnamefont {A.}~\bibnamefont {Melchiorri}}, \ and\ \bibinfo {author}
  {\bibfnamefont {J.}~\bibnamefont {Silk}},\ }\href {\doibase
  10.1016/j.physletb.2015.11.025} {\bibfield  {journal} {\bibinfo  {journal}
  {Phys. Lett. B}\ }\textbf {\bibinfo {volume} {752}},\ \bibinfo {pages} {182}
  (\bibinfo {year} {2016})},\ \Eprint {http://arxiv.org/abs/1507.08665}
  {arXiv:1507.08665 [astro-ph.CO]} \BibitemShut {NoStop}%
\bibitem [{\citenamefont {Schaan}\ and\ \citenamefont
  {Ferraro}(2019)}]{Schaan:2018tup}%
  \BibitemOpen
  \bibfield  {author} {\bibinfo {author} {\bibfnamefont {E.}~\bibnamefont
  {Schaan}}\ and\ \bibinfo {author} {\bibfnamefont {S.}~\bibnamefont
  {Ferraro}},\ }\href {\doibase 10.1103/PhysRevLett.122.181301} {\bibfield
  {journal} {\bibinfo  {journal} {Phys. Rev. Lett.}\ }\textbf {\bibinfo
  {volume} {122}},\ \bibinfo {pages} {181301} (\bibinfo {year} {2019})},\
  \Eprint {http://arxiv.org/abs/1804.06403} {arXiv:1804.06403 [astro-ph.CO]}
  \BibitemShut {NoStop}%
\bibitem [{\citenamefont {Legrand}\ and\ \citenamefont
  {Carron}(2022)}]{Legrand:2021qdu}%
  \BibitemOpen
  \bibfield  {author} {\bibinfo {author} {\bibfnamefont {L.}~\bibnamefont
  {Legrand}}\ and\ \bibinfo {author} {\bibfnamefont {J.}~\bibnamefont
  {Carron}},\ }\href {\doibase 10.1103/PhysRevD.105.123519} {\bibfield
  {journal} {\bibinfo  {journal} {Phys. Rev. D}\ }\textbf {\bibinfo {volume}
  {105}},\ \bibinfo {pages} {123519} (\bibinfo {year} {2022})},\ \Eprint
  {http://arxiv.org/abs/2112.05764} {arXiv:2112.05764 [astro-ph.CO]}
  \BibitemShut {NoStop}%
\bibitem [{\citenamefont {Sailer}\ \emph {et~al.}(2023)\citenamefont {Sailer},
  \citenamefont {Ferraro},\ and\ \citenamefont {Schaan}}]{Sailer:2022jwt}%
  \BibitemOpen
  \bibfield  {author} {\bibinfo {author} {\bibfnamefont {N.}~\bibnamefont
  {Sailer}}, \bibinfo {author} {\bibfnamefont {S.}~\bibnamefont {Ferraro}}, \
  and\ \bibinfo {author} {\bibfnamefont {E.}~\bibnamefont {Schaan}},\ }\href
  {\doibase 10.1103/PhysRevD.107.023504} {\bibfield  {journal} {\bibinfo
  {journal} {Phys. Rev. D}\ }\textbf {\bibinfo {volume} {107}},\ \bibinfo
  {pages} {023504} (\bibinfo {year} {2023})},\ \Eprint
  {http://arxiv.org/abs/2211.03786} {arXiv:2211.03786 [astro-ph.CO]}
  \BibitemShut {NoStop}%
\bibitem [{\citenamefont {Legrand}\ and\ \citenamefont
  {Carron}(2023)}]{Legrand:2023jne}%
  \BibitemOpen
  \bibfield  {author} {\bibinfo {author} {\bibfnamefont {L.}~\bibnamefont
  {Legrand}}\ and\ \bibinfo {author} {\bibfnamefont {J.}~\bibnamefont
  {Carron}},\ }\href@noop {} {\  (\bibinfo {year} {2023})},\ \Eprint
  {http://arxiv.org/abs/2304.02584} {arXiv:2304.02584 [astro-ph.CO]}
  \BibitemShut {NoStop}%
\end{thebibliography}%

\end{document}